\documentclass{jfm}

\usepackage{lineno}

\usepackage{float}
\usepackage{amsmath}

\usepackage{hyperref}
\usepackage[hyphenbreaks]{breakurl}

\usepackage[usenames,dvipsnames,svgnames,table]{xcolor}
\usepackage{latexsym}
\usepackage{graphicx}
\usepackage{natbib,hyperref}
\usepackage[percent]{overpic}
\usepackage{mathrsfs}  
\usepackage{relsize}
\usepackage{enumitem}
\usepackage{lipsum}
\usepackage{placeins}

\ifCUPmtlplainloaded \else
  \checkfont{eurm10}
  \iffontfound
    \IfFileExists{upmath.sty}
      {\typeout{^^JFound AMS Euler Roman fonts on the system,
                   using the 'upmath' package.^^J}%
       \usepackage{upmath}}
      {\typeout{^^JFound AMS Euler Roman fonts on the system, but you
                   dont seem to have the}%
       \typeout{'upmath' package installed. JFM.cls can take advantage
                 of these fonts,^^Jif you use 'upmath' package.^^J}%
      }
  \else
  \fi
\fi

\ifCUPmtlplainloaded \else
  \checkfont{msam10}
  \iffontfound
    \IfFileExists{amssymb.sty}
      {\typeout{^^JFound AMS Symbol fonts on the system, using the
                'amssymb' package.^^J}%
       \usepackage{amssymb}%
       \let\le=\leqslant  \let\leq=\leqslant
       \let\ge=\geqslant  
      }{}
  \fi
\fi

\ifCUPmtlplainloaded \else
  \IfFileExists{amsbsy.sty}
    {\typeout{^^JFound the 'amsbsy' package on the system, using it.^^J}%
     \usepackage{amsbsy}}
    {}
\fi

\newcommand\tpu{\tilde{u}}

\newcommand\hu{\hat{u}}

\newcommand\hv{\hat{v}}

\newcommand\hp{\hat{p}}

\newcommand\Rey{\mbox{\textit{Re}}}  

\newsavebox{\astrutbox}
\sbox{\astrutbox}{\rule[-5pt]{0pt}{20pt}}

\newcommand\p{\ensuremath{\partial}}

\newcommand{\SJS}[1]{{\color{red}#1}}

\title[Influence of localised smooth steps on the instability of a boundary layer]{Influence of localised smooth steps on the instability of a boundary layer}

\author[H. Xu, J.-E. W. Lombard, S. J. Sherwin ]%
{
Hui Xu,\ns
Jean-Eloi W. Lombard,\ns
Spencer J. Sherwin
}

\affiliation{
Department of Aeronautics, Imperial College London,\\ 180 Queen’s Gate, London SW7 2AZ, UK\\[\affilskip]
}

\pubyear{2010}
\volume{650}
\pagerange{119--126}
\date{?; revised ?; accepted ?. - To be entered by editorial office}
\begin{document}

\maketitle

\begin{abstract}
We consider a smooth forward facing step defined by the Gauss error function of height 4-30\% and four times the width of the local boundary layer thickness $\delta_{99}$. The boundary layer flow over a smooth forward-facing stepped plate is studied with particular emphasis on stabilisation and destabilisation of the Tollmien-Schlichting (TS)  waves and subsequently on transition. The interaction between TS waves at a range of frequencies and a base flow over a single/two forward facing smooth steps is conducted by linear analysis. The results indicate that for a high frequency TS wave, the amplitude of the TS wave is attenuated in the unstable regime of the neutral stability curve corresponding to a flat plate boundary layer. Furthermore, it is observed that two smooth forward facing steps lead to a more acute reduction of the amplitude of the TS wave. When the height of a step is increased to more than 20\% of the local boundary layer thickness for a fixed width parameter, the TS wave is amplified and thereby a destabilisation effect is introduced.  Therefore, stabilisation or destabilisation effect of a smooth step is typically dependent on its shape parameters. To validate the results of the linear stability analysis, where a high-frequency TS wave is damped by the forward facing smooth steps direct numerical simulation (DNS) is performed. The results of the DNS correlate favorably with the linear analysis and show that for the investigated high frequency TS wave, the K-type transition process is altered whereas the onset of the H-type transition is postponed. The results of the DNS suggest that for a high-frequency perturbation $\mathcal{F}=150$ and in the absence of other external perturbations, two forward facing steps of height 5\% and 12\% of the boundary layer thickness delayed H-type transition scenario and completely suppresses it for the K-type transition.
\end{abstract}

\begin{keywords}
Boundary layers, instability, Navier-Stokes equations.
\end{keywords}

\section{Introduction}\label{sec:1}
\subsection{Motivation behind the study of steps in boundary layers}

In environments with low levels of  disturbances, transition to turbulence is initiated by the exponential amplification of the Tollmien-Schlichting (TS) waves followed by the growth of secondary instabilities. Breakdown to turbulence generally occurs when the amplitude of the primary instability typically reaches $10\%$ of the free-stream velocity magnitude \citep{herbert1988,cossu2002}. The classical process of  laminar-turbulent  transition is subdivided into three stages: receptivity, linear eigenmode growth and non-linear breakdown to turbulence. A long-standing goal of laminar flow control (LFC) is the development of drag-reduction mechanisms by delaying the onset of transition. The process of laminar to turbulent transition has been shown to be influenced  by many factors such as surface roughness elements, slits, surface waviness and steps. These surface imperfections can significantly influence the the laminar-turbulent transition by influencing the growth of TS waves in accordance  with linear stability theory and then non-linear breakdown along with three dimensional effects \citep{kachanov1994}. 
Since the existence of TS  waves was confirmed by \cite{schubauer1948},  numerous studies aiming to  stabilise or destabilise the  TS modes have been carried out in order to explore and explain different paths to transition. If the growth of the TS waves is reduced or completely suppressed and, providing no other instability mechanism comes into play, it has been suggested transition could be postponed or even eliminated \citep{davies1996}. Despite roughness elements being traditionally seen as an impediment to the stability of the flat plate boundary layer recent research has shown this might not always be the case. For example, \cite{shahinfar2012} showed that classical vortex generators, known for their
efficiency in delaying, or even inhibiting, boundary layer separation can be equally effective
in delaying transition. An array of miniature vortex generator (MVGs) 
was shown experimentally to strongly damp TS waves at a frequency $\mathcal{F}=\omega\nu/U_\infty^2 \times 10^6=102$ and delay the onset of transition. Similar results were
obtained for $\mathcal{F}=$135 and 178.
\cite{downs2014} found that TS wave $\mathcal{F} \in \{100, 110, 120, 130 \}$ amplitudes over spanwise periodic
surface patterns can be reduced and demonstrated substantial delays in the
onset of transition when TS waves are forced with large amplitudes.

Over the past two decades, most investigations on topics of laminar-turbulent transition in a boundary layer have focused on two kinds of problems: the receptivity mechanism \citep{wu2001b,wu2001,saric2002,ruban2013} and stabilisation/destabilisation of TS waves\citep{cossu2002,cossu2004,fransson2005,fransson2006,garzon2013}. Receptivity is the initial stage of the natural transition process, first highlighted by \cite{morkovin1969a}, where environmental disturbances, such as acoustic waves or vorticity, are transformed into small scale perturbations within the boundary layer \citep{morkovin1969}. The aim of these studies is to assess the initial condition of the disturbance amplitude, frequency, and phase within the boundary\citep{morkovin1969,saric2002}.  So far, the receptivity mechanism of isolated small height roughness is well understood from theoretical, numerical and experimental points of view\citep{gaster1965,murdock1980,goldstein1983,kerschen1989,kerschen1990,dietz1999,wu2001,saric2002}. 
The receptivity
mechanism shows that the deviation on the length scale of eigenmodes from a smooth
surface can excite TS waves by interacting with free-stream disturbances or acoustic
noise. From a theoretical point of view, \cite{ruban1984,Goldstein1985,duck1996} studied the interactions of free-stream disturbances with an isolated steady hump within the viscous sublayer of a triple-deck region.
However, theoretical studies of the interaction between the TS wave and a distorted base flow have received less attention. More recently, the important theoretical work on the interaction of isolated roughness with either acoustic or vortical freestream disturbances was investigated by \cite{wu2006} within the framework of the triple-deck theory.  For distributed roughness \cite{corke1986} further inferred that the faster growth of TS waves on the rough wall was not attributable to the destabilization effect of roughness, such as an inflectional instability, but claimed that the growth was due to the continual excitation of TS waves on rough wall by free-stream turbulence. As we know, from a theoretical point of view, the understanding of the effect of isolated localised roughness on receptivity mechanism is  better than that on growth of the TS wave \citep{wu2001b,wu2006}.

In this paper, we investigate the effect of a smooth forward facing step on the growth properties of TS wave excited by forcing the boundary layer at different unstable non-dimensional frequencies. The amplitude of the forcing was chosen such that the velocity profiles of the resulting TS waves are well resolved but weak enough that no secondary instabilities are introduced. For the domain of the flat plate considered in the test cases presented here the unstable frequencies span $\mathcal{F}\in[27, 250]$ for displacement thickness Reynolds number $Re_{\delta^*} \in [320, 1500]$. We denote frequencies as high for $\mathcal{F} \in [100, 250]$. These high-frequencies are of particular interest because they can lead to transition towards the leading edge of the flat plate \citep{downs2014}. Note, the difference between the smooth step considered in this paper and a traditional sharp step is that the geometry for a smooth step is described by no less than two parameters (height and width) whereas for a traditional sharp step only one parameter (height) is required.  \cite{nenni1966} first explicitly gave a critical height for sharp forward-facing steps corresponding to a Reynolds number, defined according to the step height, of $Re_{H,\rm critical}=1800$. As discussed by \cite{edelmann2015}, the research on the influence of steps on the stability of the boundary layer can be divided into two approaches: one focuses on finding a critical step height $\Rey_{H,\rm{critical}}$ \citep{nenni1966} whereas the other is based on the idea that the effect of a  protuberance can be incorporated in the $e^N$  method \citep{perraud2004,wang2005,crouch2006,edelmann2013,edelmann2015}.  It is worth mentioning 
that \cite{wu2006} showed that as the TS wave propagates
through and is scattered by the mean-flow distortion induced by the roughness, it acquires a different amplitude down stream. They introduced the concept of a transmission coefficient. A further numerical study by \cite{xu2016a} confirms the localised isolated roughness has a local stabilising effect but overall a destabilising effect. Additionally an alternate expression of the transmission coefficient is introduced  which  can be incorporated into  the $e^N$ method. Recently, based on the second approach, \cite{edelmann2015} found that generally, for transonic flows, sharp forward-facing steps led to an enhanced amplification of disturbances. They also found sub- and super-sonic results showed significant differences in the generation mechanism of the separation bubbles. A different phenomenon was found for incompressible flows when investigated numerically, by using an immersed boundary technique \citep{worner2003}.  The authors observed that for a nondimensional frequency $\mathcal{F}=49.34$, the amplitude of the TS wave is reduced throughout the domain considered by the forward-facing step. They attribute this stabilising effect to the thinner boundary layer evolving on the step in comparison to the boundary layer without a step. They also claimed that  when a small separation zone appears in front of the step it has no influence on the TS wave.

The presence of separation bubbles gives rise to a destabilising effect on a boundary layer.  Generally, the separated shear layer will undergo rapid transition to turbulence and, even at rather small Reynolds numbers, separation provokes an increase in velocity perturbations and laminar flow breakdown, taking place in the separation region or close to it. The first observation of laminar separation bubbles were done by \cite{jones1938} and the structure of a time-averaged bubble was given by \cite{horton1968} and the interested reader can find a detailed review of the experimental work on the subject in \cite{young1966}. \cite{hammond1998} found that a separation bubble could become absolutely unstable for peak reversed flow velocity in excess of 30\% of the free-stream velocity.  \cite{theofilis2000} report the shape of the globally unstable mode in recirculation bubble. A separation bubble can have important impact on the global stability of a boundary layer. In the following, we shall only focus on amplification of the TS wave by a separation bubble. Numerically, with laminar separation bubbles, \cite{rist1993} suggested a three-dimensional oblique mode breakdown rather than a secondary instability of finite-amplitude two-dimensional waves. \cite{xu2016a} investigated the behavior of TS waves undergoing  small-scale localised distortions and found even a small separation bubble can amplify a TS wave. When a sharp forward-facing step of sufficient height is present in a boundary layer, a separation bubble can easily be generated. In particular, the effective transformation or scattering of the freestream disturbances to the TS waves occurs preferably alongside sudden changes of the mean flow (e.g. over the leading edge, separation region or suction slits). The external acoustics, surface vibrations and vortical disturbances in the form of localised flow modulations or freestream turbulence are those which most frequently contribute to the boundary layer receptivity as reviewed by \cite{nishioka1986,kozlov1990,saric1990, bippes1999}. Additionally, a smooth step is less receptive than a sharp step \citep{kachanov1979}.  Another benefit of using a smooth step is to circumvent biglobal instability \citep{hammond1998}.

\begin{figure}
\vskip4mm
\centerline{
\begin{overpic}[width=0.7\textwidth]{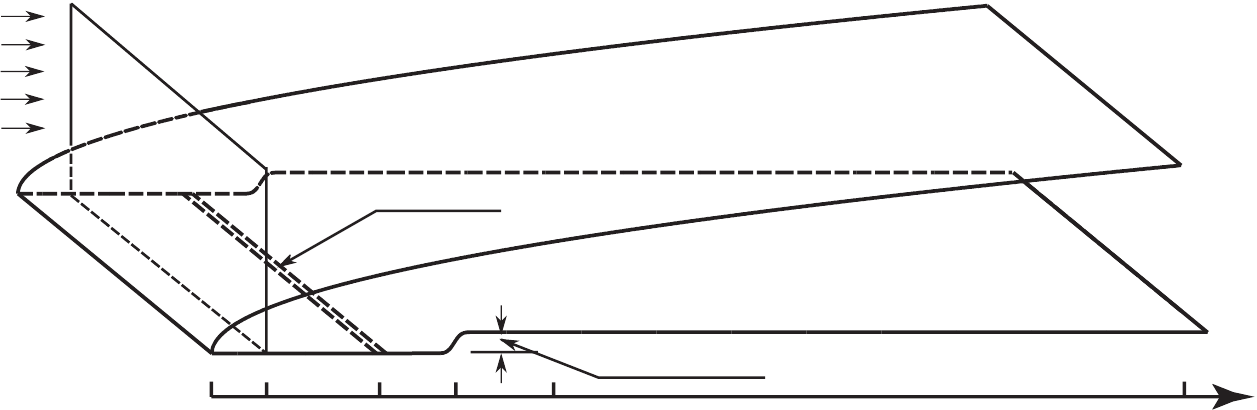}
\put(-10,28){$U_\infty$}
\put( 28,16.1){{ Disturbance strip}}
\put( 57,3){{$h$}}
\put( 18,-2){{$\Rey_{\delta_*^i}$}}
\put( 33,-2){{$\Rey_{\delta_*^{c}}$}}
\put( 90,-2){{$\Rey_{\delta_*}=1500$}}
\end{overpic}
}
 \caption{Overview of the computational setup with the Blasius boundary layer profile at the inflow and the disturbance position.}
\label{fig:domains}
\end{figure}
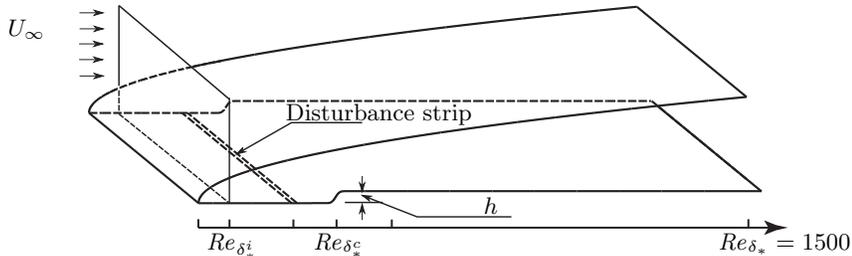
\subsection{Problem definition}
The smooth step considered in this paper, shown in figure \ref{fig:domains}, is located in the unstable regime of the neutral stability curve close to the leading edge. This work mostly considers medium and high frequencies $\mathcal{F} \in \{100, 140, 150, 160 \}$ TS waves but also the case of a low frequency forcing $\mathcal{F} =49.34$ is investigated for closer comparison with \cite{worner2003}. The low frequency forcing is particularly interesting because it offers, in the context of the neutral stability curve of the zero pressure gradient flat plate, a much larger unstable regime compatible with the so called critical amplification factor $N=8$ \citep{edelmann2015}. Recently, \cite{downs2014} studied TS wave growth over spanwise-periodic surface patterns excited at $\mathcal{F} \in \{100, 110, 120, 130 \}$. They report that TS waves excited by high frequencies and large amplitudes ($A^{\rm int, I}_{\rm TS} < 0.48\%U_\infty$), producing well resolved profiles without triggering secondary instabilities,  can be reduced by spanwise-periodic surface patterns compared to the flat plate case. Therefore, over a smooth step, understanding growth properties of the TS waves with high frequencies  is pertinent. The linear analysis shows that in the presence of a single smooth step the TS wave can be attenuated below a critical height and smoothness but amplified above this critical height. Furthermore, the linear stability investigation of two isolated smooth steps instead of a single step with the same geometrical configuration revealed further reduction of the amplitude  of the TS wave can be obtained compared to a single smooth step when the same step parameters bar location are used. Again, past a critical height, two forward facing smooth steps can have a destabilising effect on the TS mode.  Furthermore, a smooth step can tolerate a higher height scale compared with a sharp step and does not introduce separation bubbles. In order to validate the stabilising effect of the smooth forward facing steps seen in the linear stability analysis, full non-linear direct numerical simulations of both the  K- and H-type  transition scenarios are conducted. 
These two transition scenarios are particularly compelling because they exhibit a long region of linear growth particularly suitable for the investigation of the effect of the smooth forward facing step in the boundary layer on the TS wave \citep{moin2013}. The results from the direct numerical simulation confirm the findings from the linear analysis. For both K- and H-type  transition scenarios the forward facing smooth steps configuration has a stabilising effect even avoiding transition for the K-type scenario.

The paper is organised as follows. In \S 2, we give fundamental definitions and describe the numerical tools employed in \S 3.1. In \S 3.2 and \S 3.3 and 3.4 we present results of linear analysis for high, medium and low frequency perturbations. \S 3.4, the results from the DNS are presented and discussed. A further discussion is then given in \S 4 and subsequently, the work is concluded.

\section{Mathematical formulations}\label{sec:2}
\subsection{Fully nonlinear and linearised Navier-Stokes equations}
The non-dimensional momentum and continuity  equations governing unsteady viscous flow with constant density are given as follows
\begin{equation}\label{ns}
\begin{array}{rll}
\p_t u_i-\Rey^{-1}\p^2_j u_i + u_j\p_j u_i + \p_i p &=& 0,\\
\p_j u_j &=&0,
\end{array}
\end{equation}
where $u_i$ is one component of velocity field along the $i$th direction, $\p_t$ denotes the derivative with respect to time, $\p_j$  is the $j$th direction spatial derivative , $\Rey$ is the Reynolds number defined by $LU_\infty/\nu$ where $L$ is the distance from the leading edge, $\nu$ is the kinematic viscosity and $p$ is the pressure. For a two-dimensional (2D) problem, $i=1,2$ and $(u_1,u_2)=(u,v)$ and for a  three-dimensional (3D) problem, $i=1,2,3$ and $(u_1,u_2,u_3)=(u,v,w)$.

Considering a steady state $\bar{u}_i$ of (\ref{ns}) about which a small perturbation $\tilde{u}_i$, such that $u_i=\bar{u}_i+\tilde{u}_i$ and dropping the second order terms in $\tilde{u}_i$, (\ref{ns}) can be linearised as follows
\begin{equation}\label{lns}
\begin{array}{rll}
\p_t \tilde{u}_i-\Rey^{-1}\p^2_j \tilde{u}_i + \bar{u}_j\p_j \tilde{u}_i + \tilde{u}_j\p_j \bar{u}_i + \p_i \tilde{p} &=& 0,\\
\p_j \tilde{u}_j &=&0.
\end{array}
\end{equation}
With suitable boundary conditions, in a linear regime, the system (\ref{lns}) can be used to exactly simulate  evolution of a small perturbation $\tilde{u}_i$ in a boundary layer.

In the flat-plate simulations undertaken, with the assumptions of relatively large $\Rey$  and no pressure gradient  the base flow can be approximated by the well-known Blasius equation
\begin{equation}
f^{\prime\prime\prime}(\eta)+\frac{1}{2}f(\eta)f^{\prime\prime}(\eta)=0,
\end{equation}
subject to the following boundary conditions
\begin{equation}
f(\eta)=f^\prime(\eta)=0 \mbox{~at~} \eta=0, f^\prime=1 \mbox{~at~} \eta\rightarrow\infty,
\end{equation}
where the prime denotes the derivative with respect to the similarity variable $\eta$. Specifically in the above, the dimensionless variables are defined by 
\begin{equation}
f=\Psi/\sqrt{\nu U_\infty x} \mbox{~and~} \eta=y/\sqrt{U_\infty/(\nu x)},
\end{equation}
where $\Psi$ is the stream function. The streamwise and vertical velocity profiles of the Blasius boundary layer can be calculated by
\begin{equation}\label{bv}
\bar{u}_B=U_\infty f^\prime(\eta) \mbox{~and~} \bar{v}_B=\frac{1}{2}\sqrt{\frac{\nu U_\infty}{x}}(\eta f^\prime(\eta)-f(\eta)).
\end{equation}

Under the assumption of streamwise parallel flow in two dimensions, the perturbation assumes the normal form
\begin{equation}\label{lnsmode}
(\tilde{u},\tilde{v}, \tilde{p}) =(\hu,\hv,\hp)\exp({\rm i}(\alpha x-\omega t)) + {\rm c.c.},
\end{equation}
where $\alpha$ and $\omega$ denote wave-number and frequency of a perturbation, respectively. The mode $(\hu,\hv,\hp)$ in (\ref{lnsmode}) generally can be obtained by solving the well-known Orr-Sommerfeld (O-S) equations, the solution of which for eigenvalues and eigenfunctions has been well studied \citep{stuart1963,schlichting1968,drazin1981}. When a surface imperfection occurs, the same notations $\hu$ and $\hv$ are used to denote the TS mode.

Generally, for an unstable frequency $\omega\in\mathbb{R}^{+}$, assuming that the TS mode is dependent on both $x$ and $y$, the TS wave envelope is defined by the absolute maximum amplitude of the TS  wave as follows
\begin{equation}
A(x) = \max\left\{\left|{\tpu}(x,\eta,t)\right|: \forall \eta\in[0, \infty), \forall t\in\mathbb{R}^+\right\}.
\end{equation}

\subsection{Definitions correlated with a surface imperfection}

In order to rescale the step, we introduce a reference boundary layer thickness $\delta_{99}=4.91x_cRe_{x_c}^{-1/2}$ and displacement thickness scales $\delta^*|_{x_c}=1.7208x_cRe_{x_c}^{-1/2}$, defined according to a flat plate boundary layer, where $x_c$ is  the distance from the leading edge to the centre position of a surface imperfection and $Re_{x_c}=U_\infty x_c/\nu$. We also let  $\Rey_{\delta^*}=U_\infty\delta^*/\nu$ be the displacement Reynolds number. Now, we consider a forward-step-like surface imperfection, which is defined by
\begin{equation}\label{eq:hs}
f_s(X,\hat{h})=\frac{\hat{h}}{2}\cdot\left(1+{\rm erf}\left(\frac{X}{\sqrt{2}\hat{d}}\right)\right)
\end{equation}
where $\hat{d}$ and $\hat{h} (>0)$ are the streamwise width scale and the normal direction length scale defined by the corresponding physical scales $d$ and $h$  as
\begin{equation}
\hat{h}=h/\delta_{99}, \hat{d}=d/\delta_{99}, 
\end{equation} 
and $X$ is a streamwise local coordinate defined as follows  
\begin{equation}
X=(x-x_c)/\delta_{99}.
\end{equation}
For multiple smooth steps,  the wall profile is formally defined  by 
\begin{equation}\label{eq:ms}
\sum_{i=0}^n f_s(X-X_i,\hat{h}),
\end{equation}
where $X_i$ ($X_0=0$) denotes the relative centre position of each individual step with respect to the first step and $n+1$ is the number of steps.

In order to characterise or quantify geometrical steepness of a continuous function $f(x)$, assuming at least $f(x)\in C^1$, we introduce the following quantity
\begin{equation}\label{steep}
\gamma(x) = \frac{\max |\p_x f(x)|}{\sqrt{\max|\p_x f(x)|^2+\varepsilon^{-2}}},
\end{equation}
where $\varepsilon$ is a smooth parameter. It is clear that $\gamma\in[0,1)$.
For a smooth step, $\varepsilon$ is defined by the ratio  $\hat{h}/\hat{d}=h/d\in [0, \infty)$ and the formula (\ref{steep}) can be explained as follows
\begin{equation}
\gamma(X)=\left\{
\begin{array}{rl}
1,&\hat{h}/\hat{d}\rightarrow \infty \mbox{~for~} \hat{h}\neq0\\
0,&\hat{h}/\hat{d}\rightarrow 0 \\
\end{array}
\right.,\mbox{~for~} f(X)=f_s(X,\hat{h}), X\in [-\hat{d}/2,\hat{d}/2].
\end{equation}
When $\gamma=1$ the step is a sharp whereas for $\gamma=0$ the smooth step tends to a flat plate.

\section{Numerical approach and results}

\subsection{Numerical strategy}
A spectral/{\it hp} element discretisation, implemented in the Nektar++ package, is  used in this work to solve the linear as well as  nonlinear Navier-Stokes equations.  A stiffly stable splitting scheme is adopted which decouples the velocity and pressure fields and time integration is achieved by  a  second-order accurate implicit-explicit scheme \citep{karniadakis1991,nektar}.  

For 2D simulations, a convergence study by {\it p}-type refinement is performed to demonstrate mesh independence is achieved throughout this study. For 3D calculations, provided the same 2D mesh in the  $x-y$ plane is used with the addition of a hybrid Fourier-Spectral/{\it hp} discretisation in the third direction yielding a hybrid Fourier/spectral/{\it hp} discretisation of the full 3D incompressible Navier-Stokes equations.

\begin{figure}
  \centerline{
\begin{overpic}[width=0.96\textwidth]{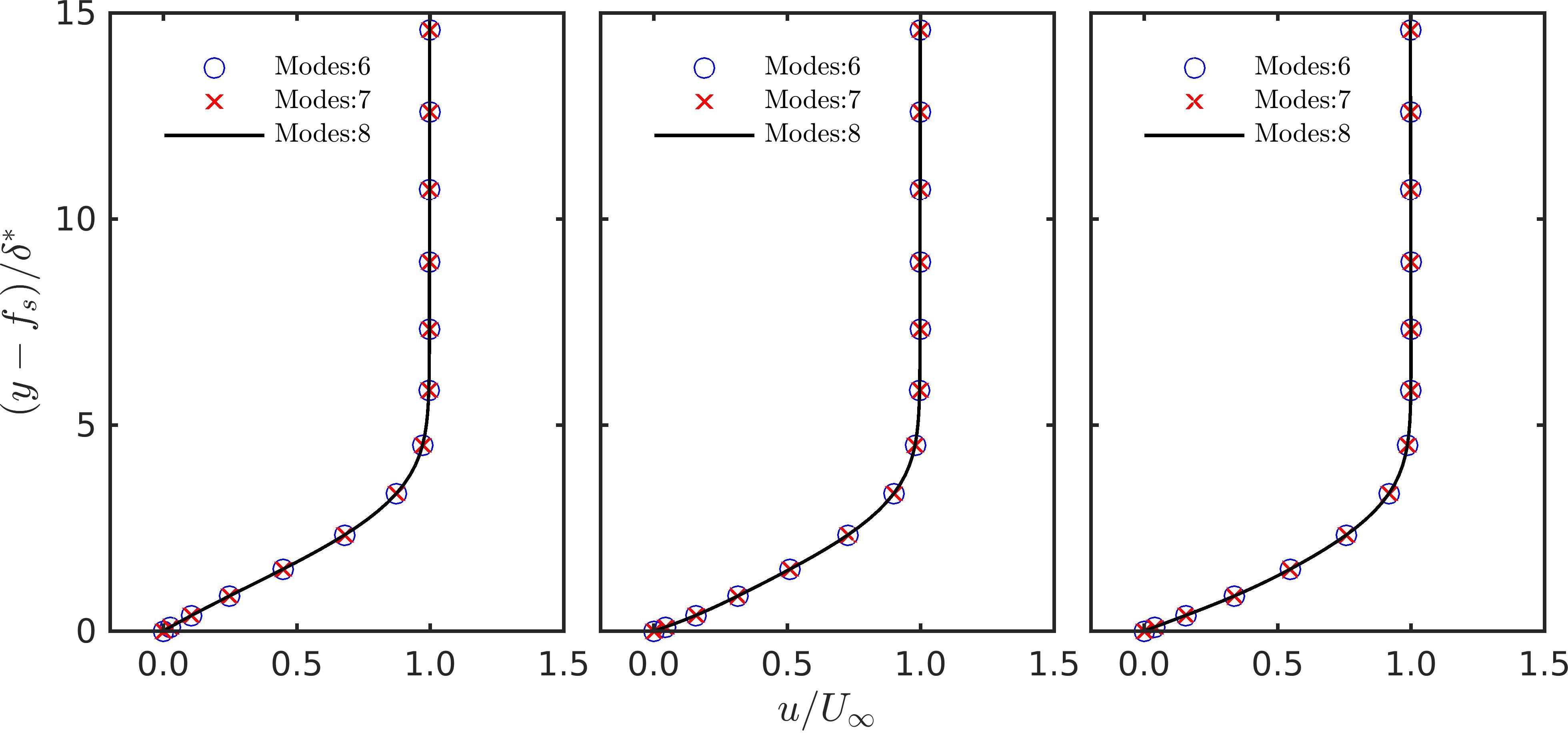}
\put(10,43){(a)}
\put(41,43){(b)}
\put(72,43){(c)}
\end{overpic}
}
 \caption{Streamwise velocity profiles at 3 different streamwise locations: (a) $Re_{\delta^*}=821$, (b) $Re_{\delta^*}=866$ and (c) $Re_{\delta^*}=897$. The physical parameters corresponding to each case are from Case E in Table \ref{tab:121106}.}
\label{fig:uind}
\end{figure}

\begin{figure}
 \centerline{
\begin{overpic}[width=0.96\textwidth]{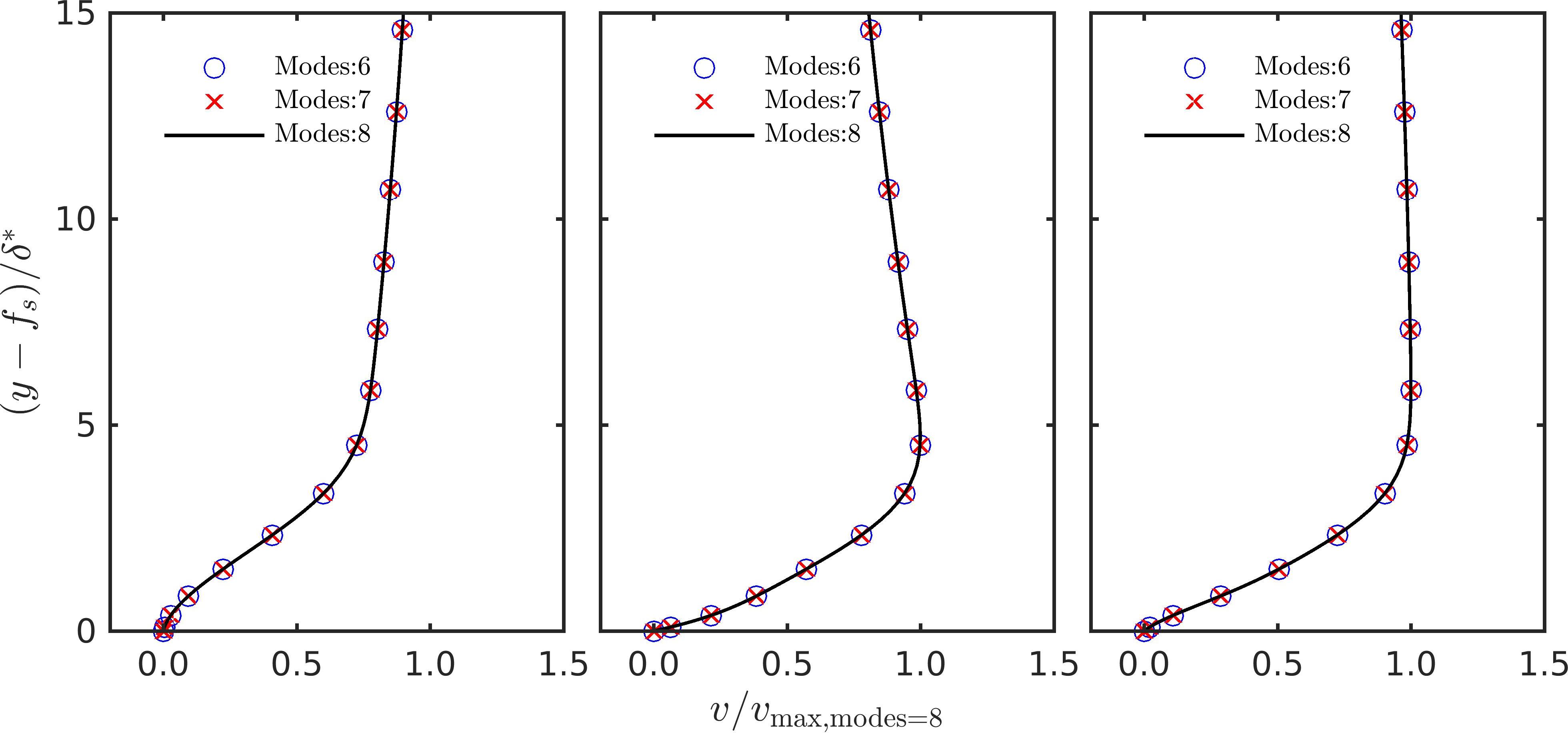}
\put(10,43){(a)}
\put(41,43){(b)}
\put(72,43){(c)}
\end{overpic}
}
 \caption{Wall-normal velocity profiles at 3 different streamwise locations: (a) $Re_{\delta^*}=821$, (b) $Re_{\delta^*}=866$ and (c) $Re_{\delta^*}=897$. The physical parameters corresponding to each case are from Case E in Table \ref{tab:121106}.}
\label{fig:vind}
\end{figure}

To obtain the best fidelity in representing the curved surface of the smooth step, high-order curved elements are defined by means of  an analytical mapping. The governing equations are then discretised in each curved element by seventh-order polynomials. The choice of polynomial order is guaranteed by the mesh independence study. Figures \ref{fig:uind} and \ref{fig:vind} compare  the horizontal and vertical velocity profiles over a smooth step in $\Rey_{\delta^*}=866$ are given at different positions for polynomial order ranging from six to eight. In the whole domain, the $L^2$ relative error of velocity fields is lower than $10^{-6}$, which is consistent with the convergence tolerance of the base flow generation  defined by
$$\label{eq:cc}
{\left\|{ (\p_t^d u,\p_t^d v)}\right\|_{0}}/{\|(u,v)\|_{0}}\cdot T_c<10^{-6},
$$
where $\|\cdot\|_0$ means the standard $L^2$ norm, $\p_t^d$ denotes the discrete temporal derivative and $T_c$ is the convective time scale. Once 2D steady base flows are generated by using the non-linear Navier-Stokes equations (NSEs), the TS waves are simulated by the linearised Navier-Stokes equations (LNSEs). As discussed below, for base flow generation, the inlet position is located sufficiently far from the first step to allow the base flows to recover the Blasius profile.  Following the experimental methodology used by \cite{downs2014}, the TS waves are excited by periodic suction and blowing on wall.


\subsection{Linear analysis in a narrow unstable regime of the neutral stability curve}
\begin{figure}
\vskip0.5cm
\centerline{
\begin{overpic}[width=0.46\textwidth]{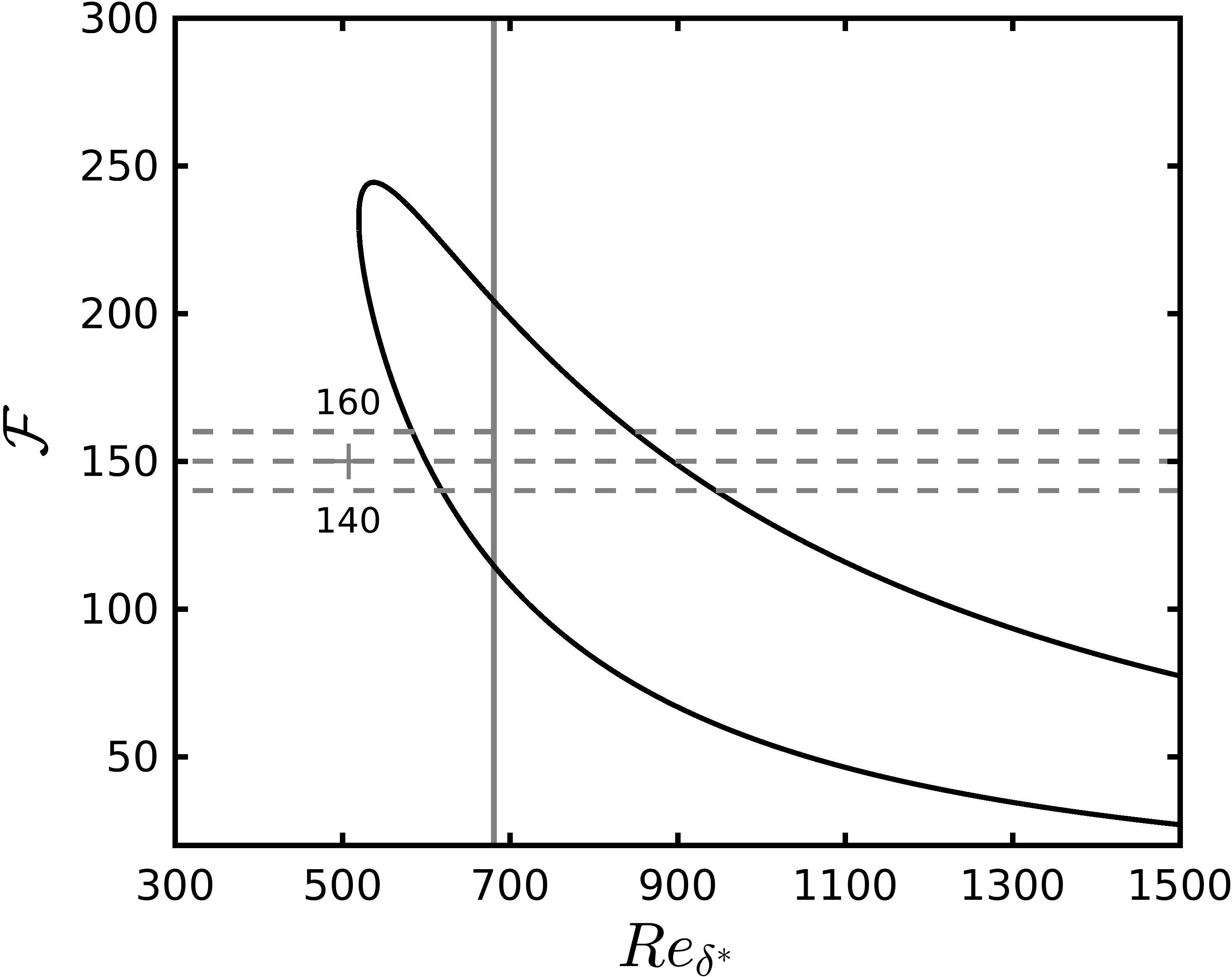}
\put(18,72){(a)}
\end{overpic}
\begin{overpic}[width=0.46\textwidth]{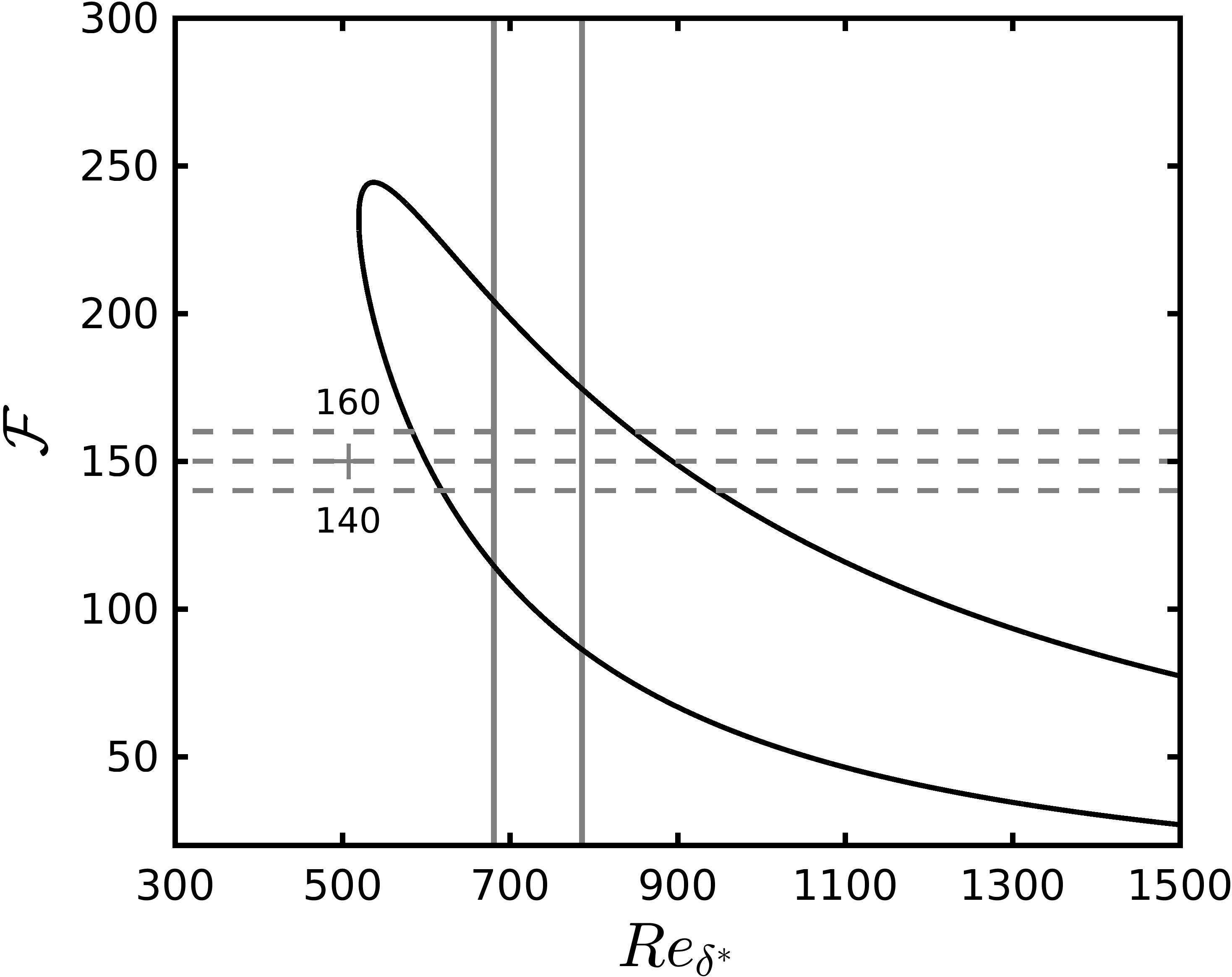}
\put(18,72){(b)}
\end{overpic}
}
 \caption{Positions of exciters (+) and one single smooth steps (a) and two smooth isolated steps (b). Three different frequencies corresponding to steps' positions are given by horizontal dashed lines.}
\label{fig:neutral}
\end{figure}

\begin{figure}


\centerline{
\begin{overpic}[height=0.15\textheight]{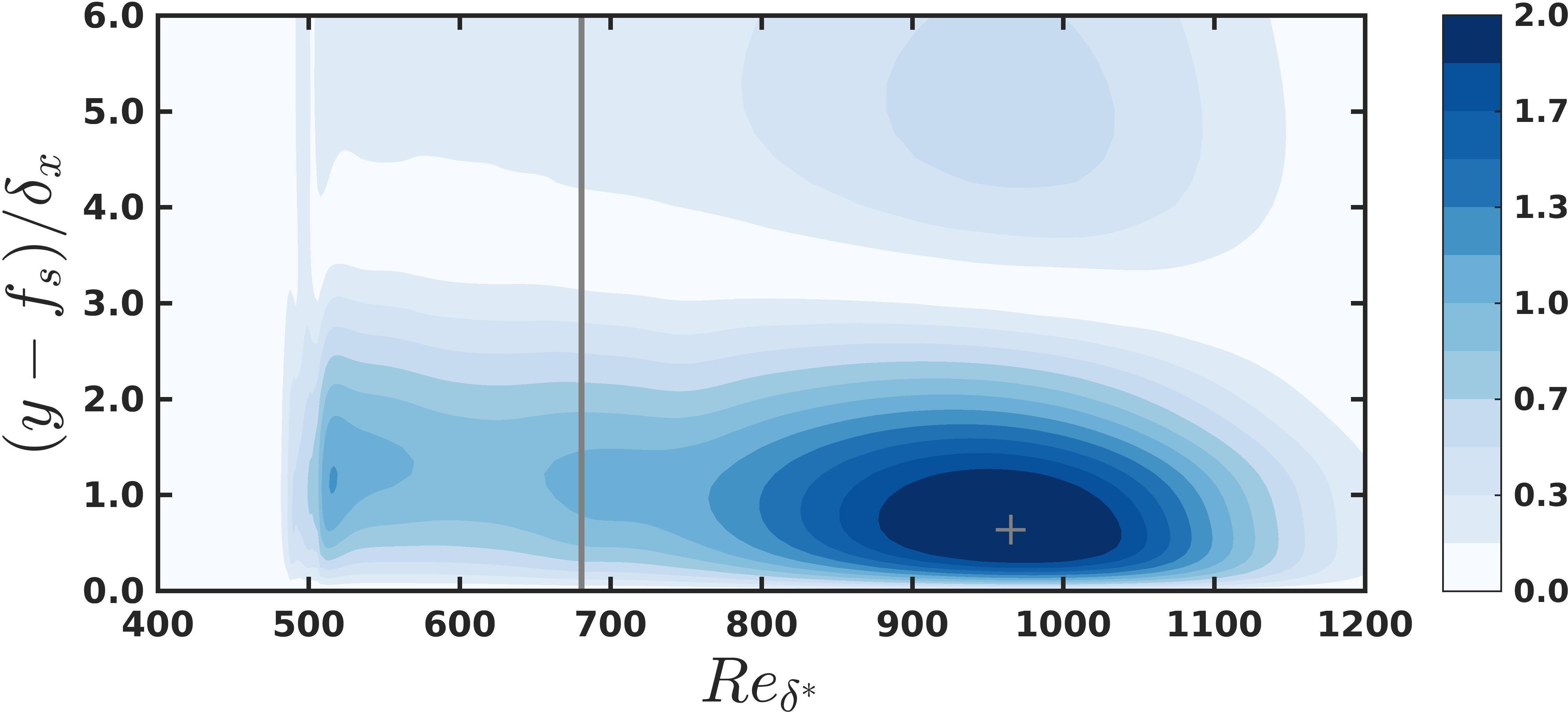}
\put(12,39){(a$_1$)}
\put(40,39){$\hat{h}=5.48\% \quad \mathcal{F}=140$}
\end{overpic}
\begin{overpic}[height=0.15\textheight]{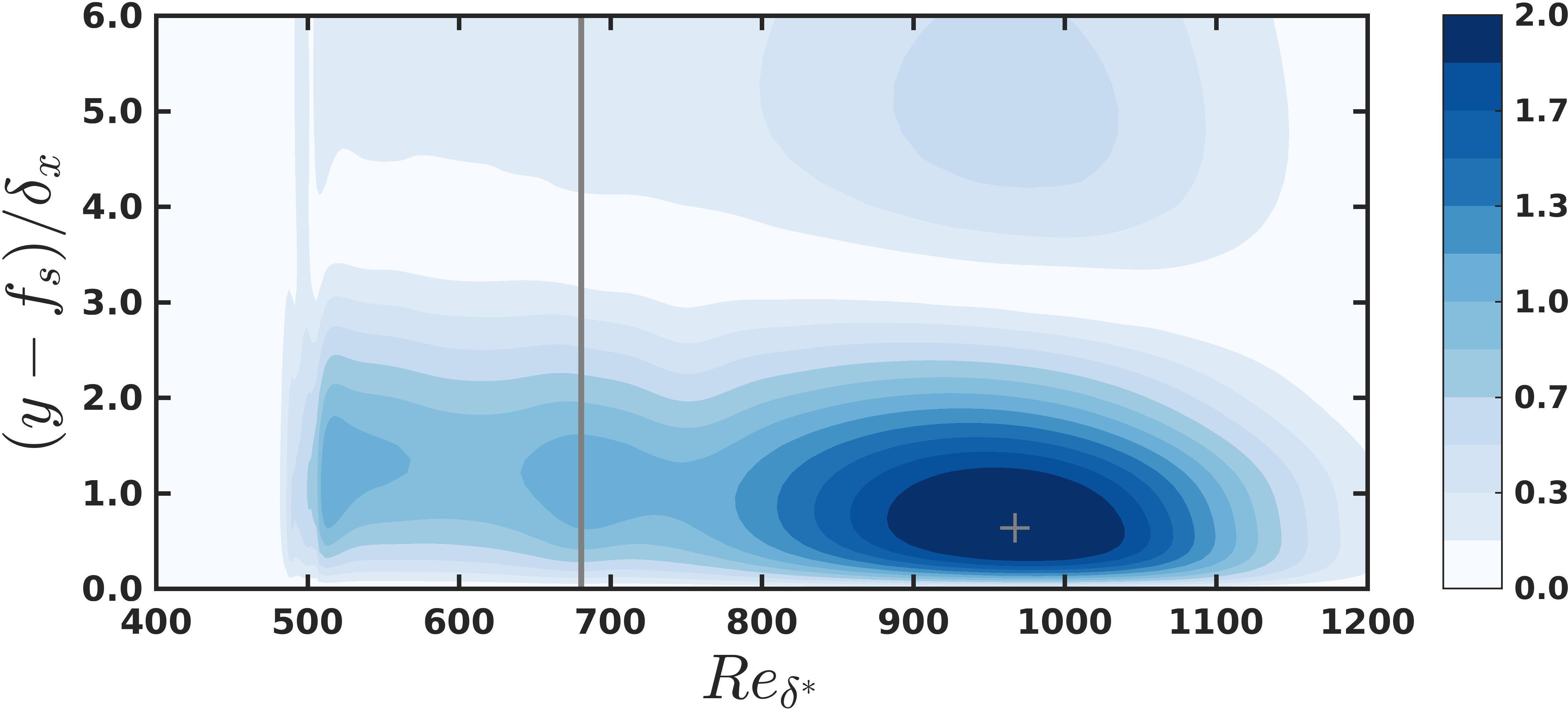}
\put(12,40){(b$_1$)}
\put(40,39){$\hat{h}=10.96\% \quad \mathcal{F}=140$}
\end{overpic}
}
\centerline{
\begin{overpic}[height=0.15\textheight]{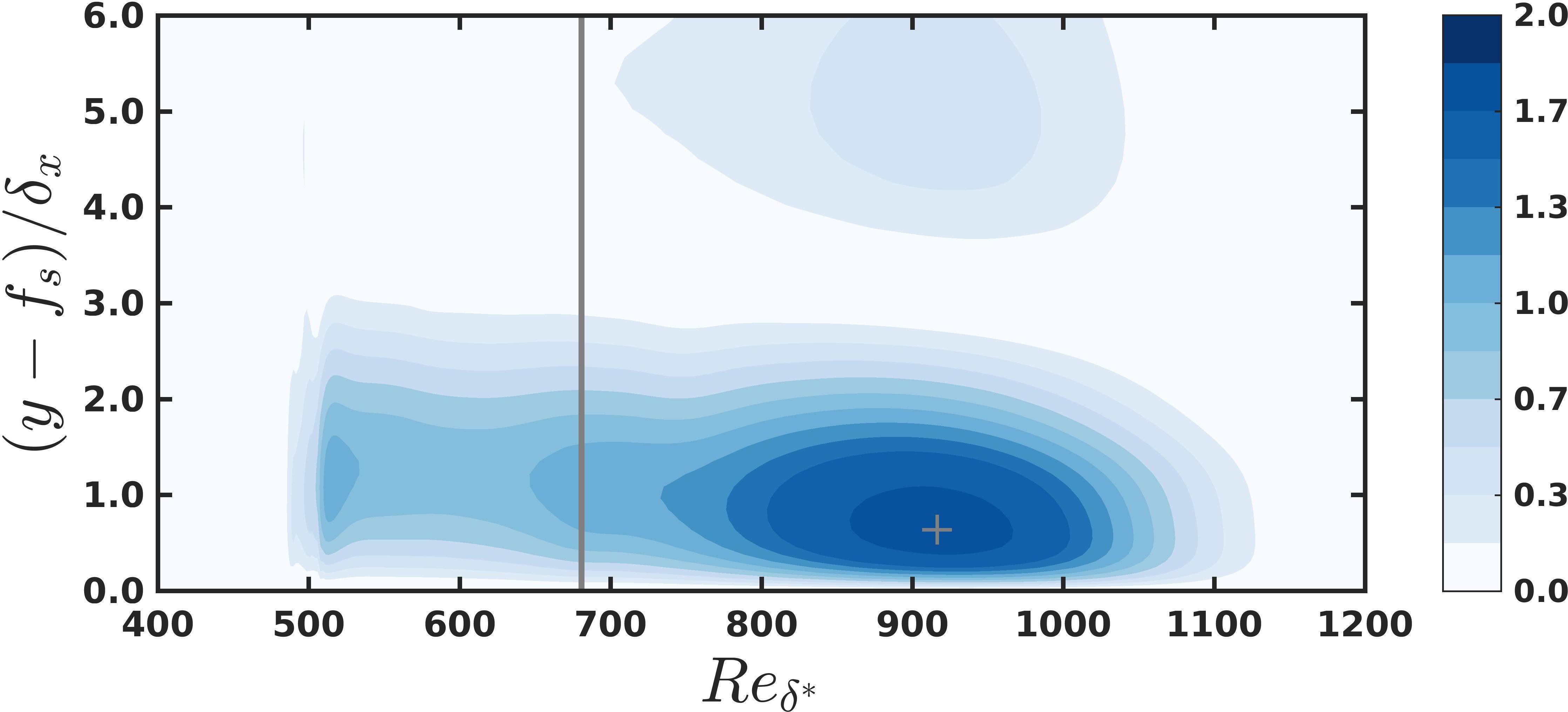}
\put(12,40){(a$_2$)}
\put(40,39){$\hat{h}=5.48\% \quad \mathcal{F}=150$}
\end{overpic}
\begin{overpic}[height=0.15\textheight]{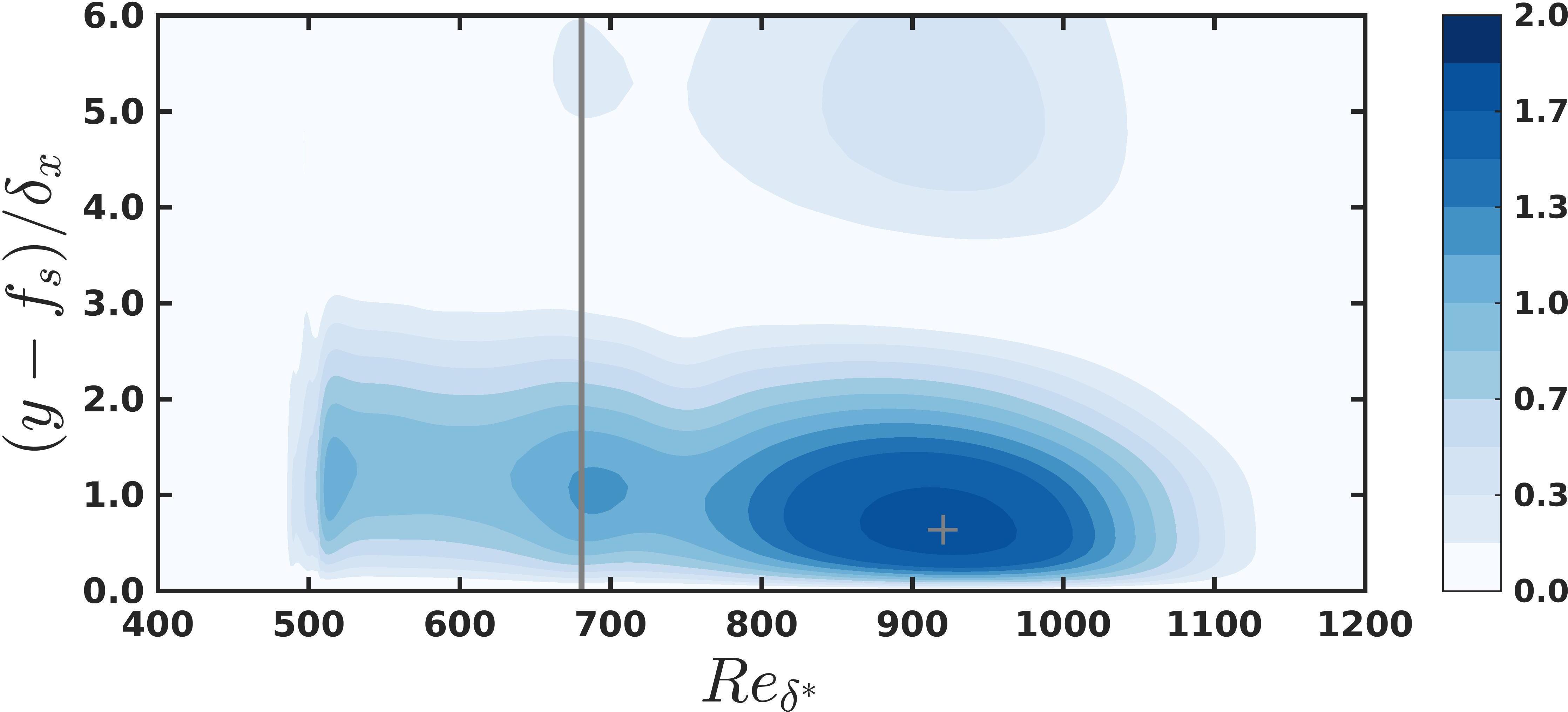}
\put(12,40){(b$_2$)}
\put(40,39){$\hat{h}=10.96\% \quad \mathcal{F}=150$}
\end{overpic}
}
\centerline{
\begin{overpic}[height=0.15\textheight]{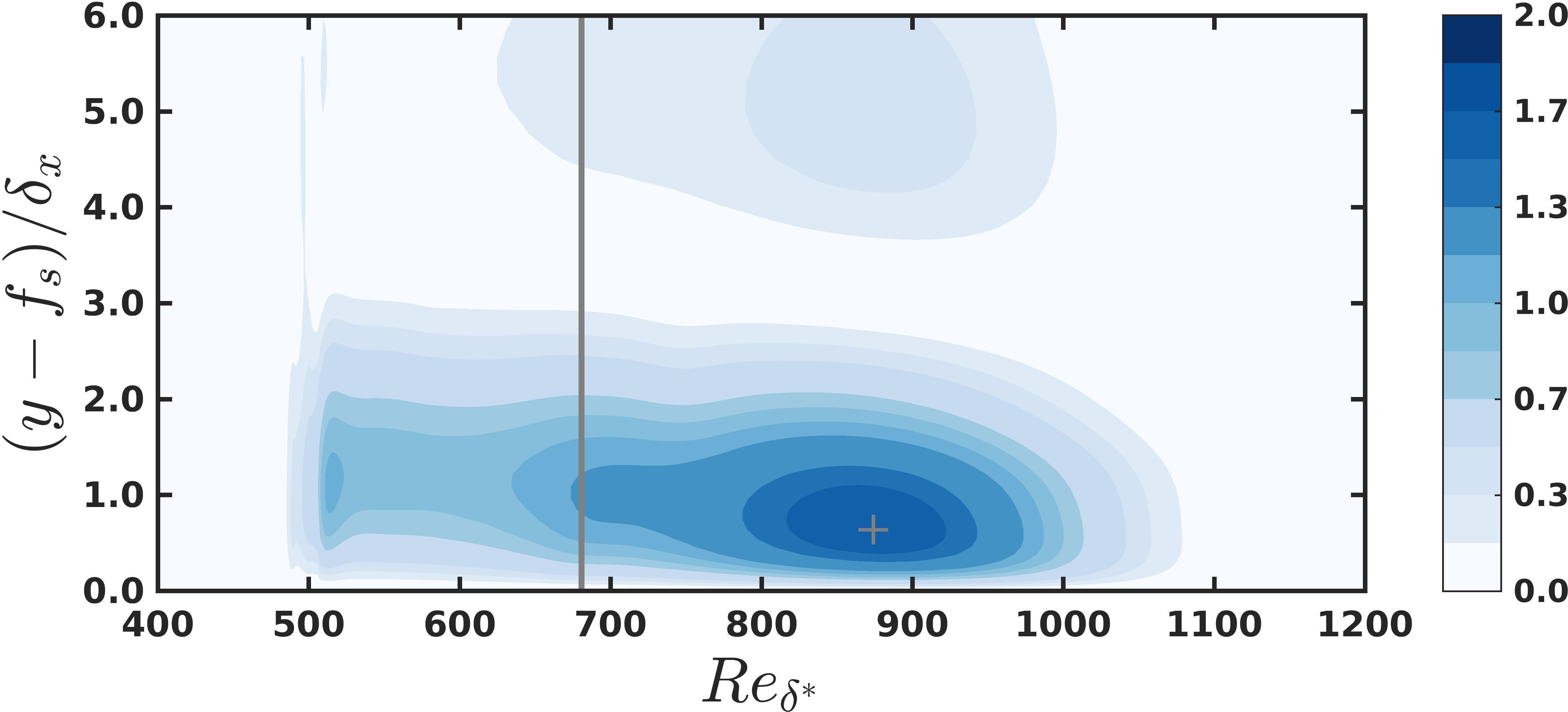}
\put(12,40){(a$_3$)}
\put(40,39){$\hat{h}=5.48\% \quad \mathcal{F}=160$}
\end{overpic}
\begin{overpic}[height=0.15\textheight]{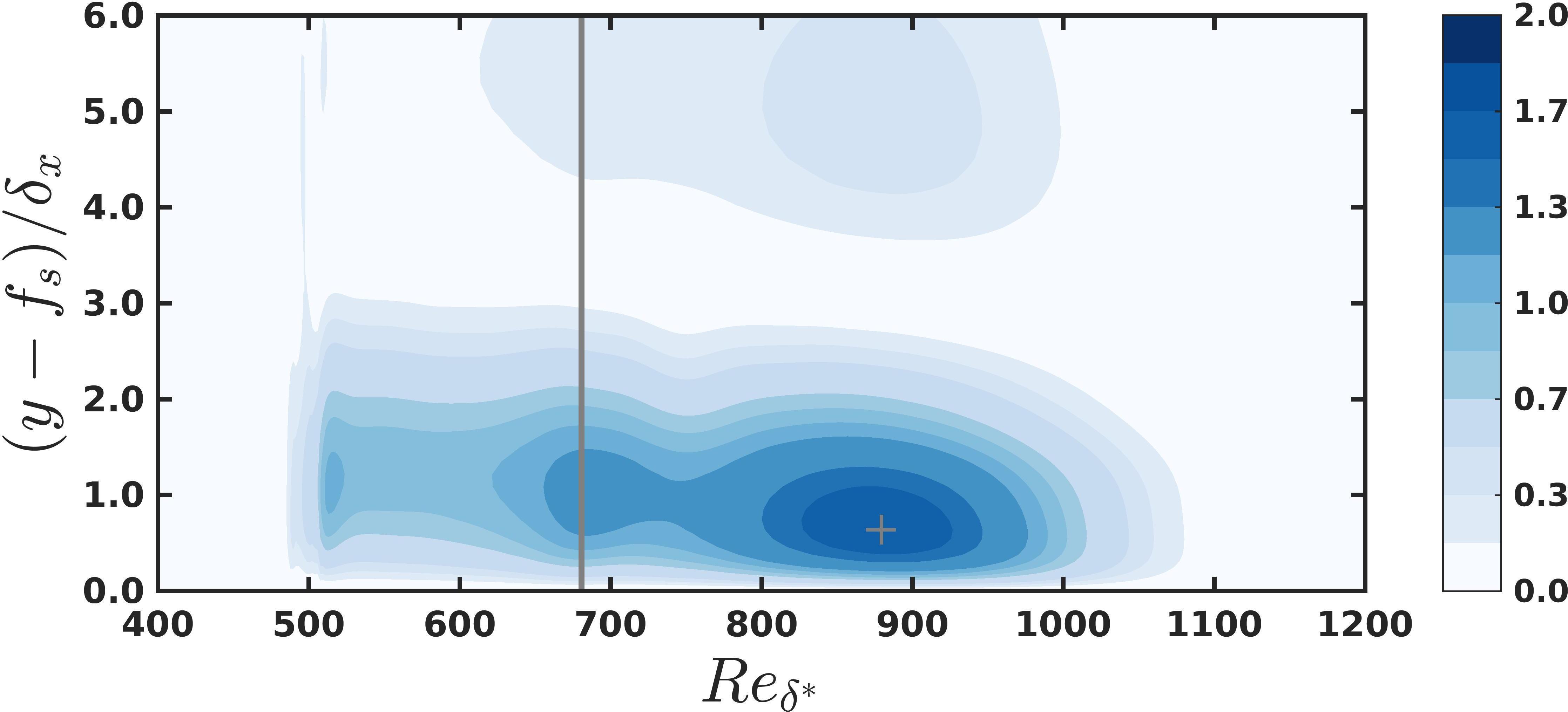}
\put(12,40){(b$_3$)}
\put(40,39){$\hat{h}=10.96\% \quad \mathcal{F}=160$}
\end{overpic}
}
\centerline{
\begin{overpic}[height=0.15\textheight]{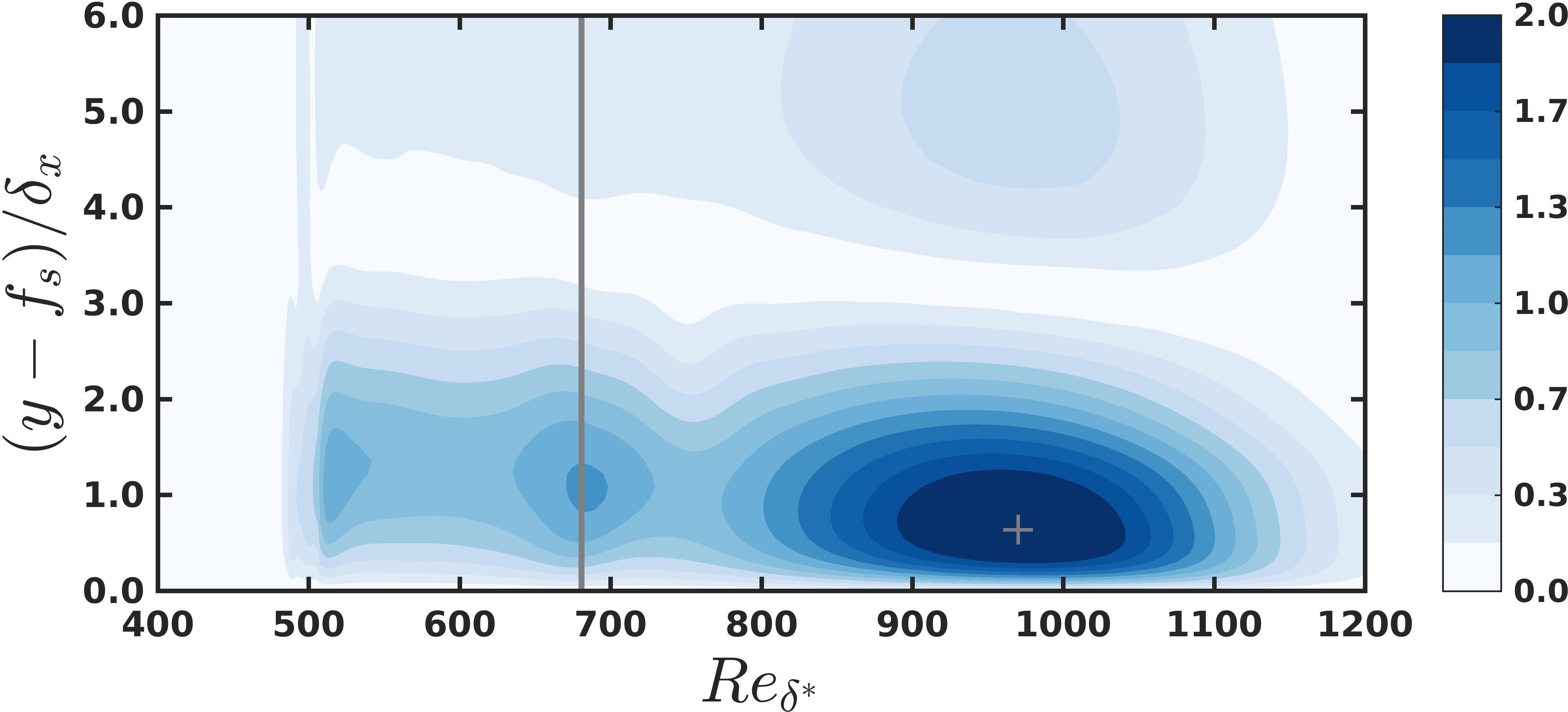}
\put(12,40){(c$_1$)}
\put(40,39){$\hat{h}=20\% \quad \mathcal{F}=140$}
\end{overpic}
\begin{overpic}[height=0.15\textheight]{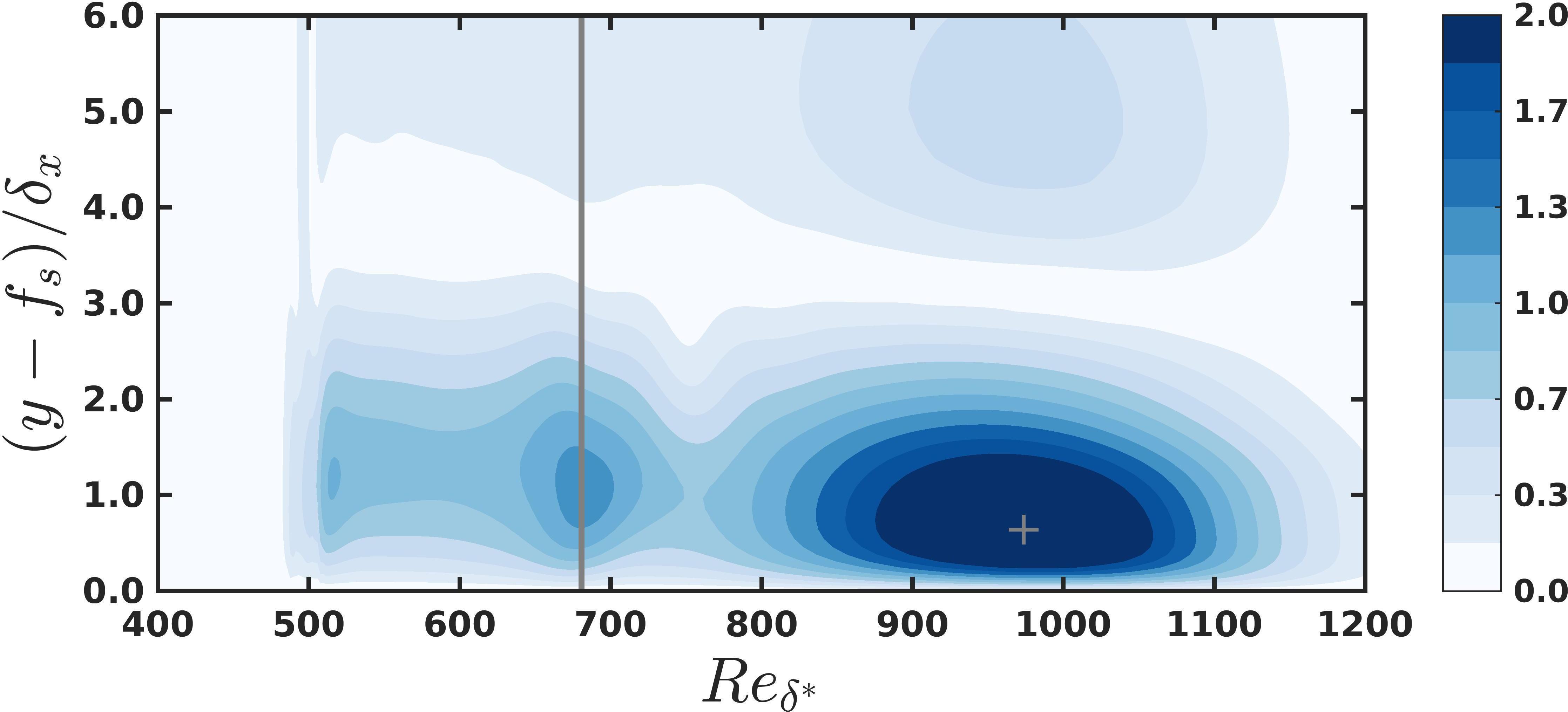}
\put(12,40){(d$_1$)}
\put(40,39){$\hat{h}=30\% \quad \mathcal{F}=140$}
\end{overpic}
}
\centerline{
\begin{overpic}[height=0.15\textheight]{pic/071114ts/C1_U_h02}
\put(12,40){(c$_2$)}
\put(40,39){$\hat{h}=20\% \quad \mathcal{F}=150$}
\end{overpic}
\begin{overpic}[height=0.15\textheight]{pic/071114ts/C1_U_h03}
\put(40,39){$\hat{h}=30\% \quad \mathcal{F}=150$}
\put(12,40){(d$_2$)}
\end{overpic}
}
\centerline{
\begin{overpic}[height=0.15\textheight]{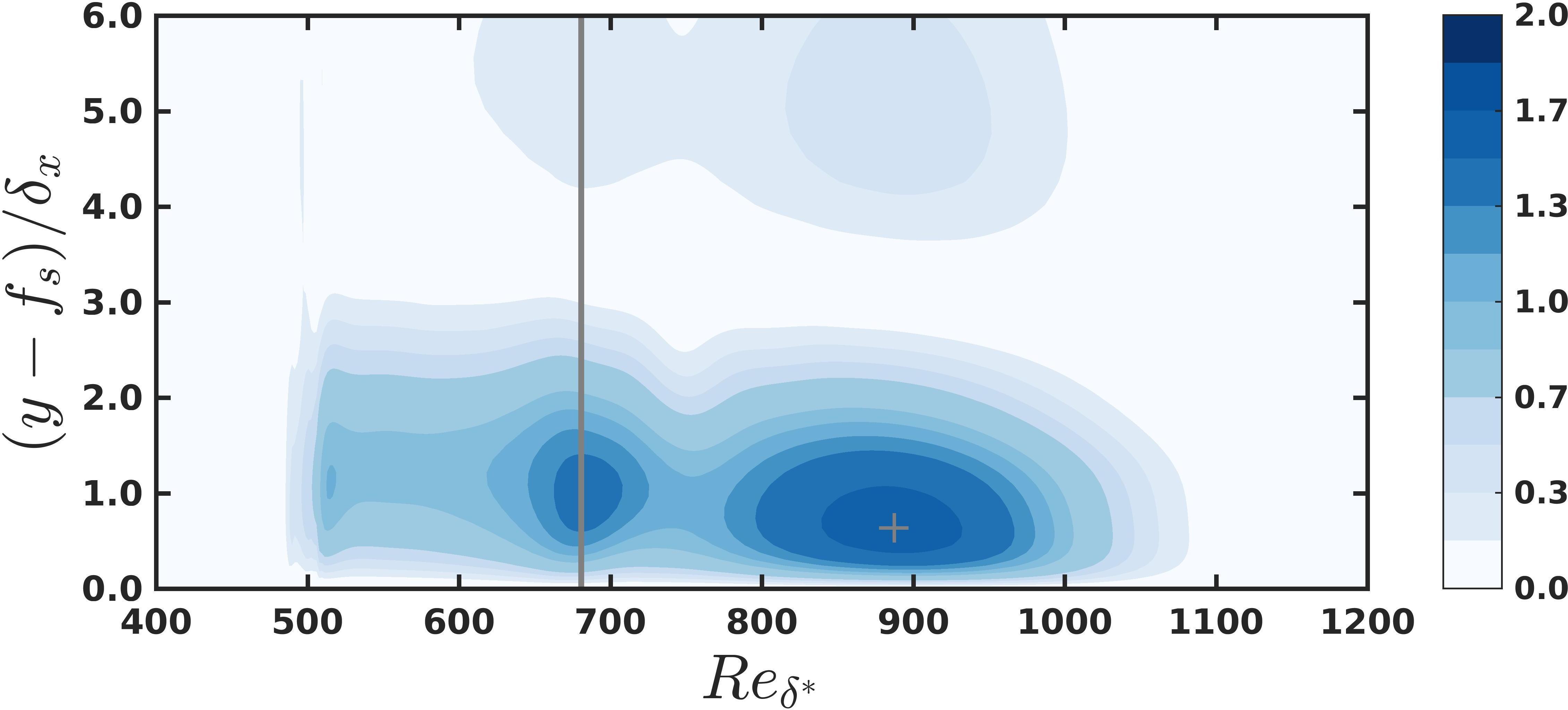}
\put(12,40){(c$_3$)}
\put(40,39){$\hat{h}=20\% \quad \mathcal{F}=160$}
\end{overpic}
\begin{overpic}[height=0.15\textheight]{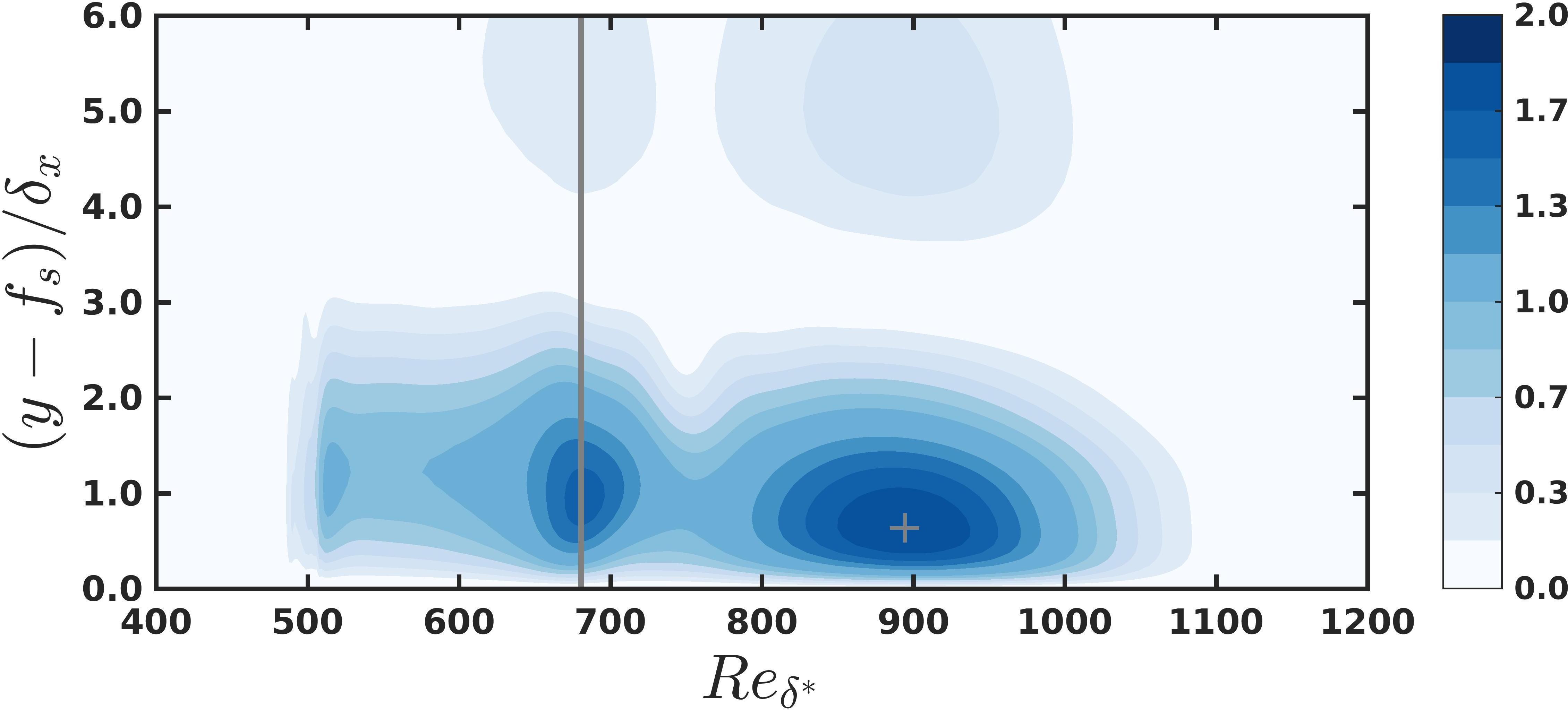}
\put(12,40){(d$_3$)}
\put(40,39){$\hat{h}=30\% \quad \mathcal{F}=160$}
\end{overpic}
}

 \caption{Contour plots of $|\tilde{u}|/A_0$ and the physical parameters corresponding to (a$_\#$), (b$_\#$), (c$_\#$) and (d$_\#$) are from Case A, B, C  and D in Table \ref{tab:071114} where $\#$ denotes the number 1, 2 or 3, which \SJS{corresponds to frequency} $\mathcal{F}_\#$. \SJS{The} step position is determined by $\Rey_{\delta_*^{\rm c_1}}$. The grey cross + indicates the location of the maximum amplitude of the TS-wave and the vertical grey line represents the location of the forward facing smooth step.}
\label{fig:071114}
\end{figure}

In order to detect the effect of smooth steps on the stability of the boundary layer, three different high frequency perturbations $\mathcal{F}$ ($\in$\{140, 150 160\}) are excited within the boundary, up-stream of the unstable region, by periodic blowing and suction (see figure \ref{fig:neutral}). Hereafter, for each perturbation frequency studied, let $A_0$ indicate a reference maximum TS mode amplitude at the lower branch of the neutral stability curve in a flat plate boundary layer. The contours of $|\tilde{u}|/A_0$ are given in figure \ref{fig:071114} for  three different non-dimensional frequencies ($\mathcal{F} \in \{140, 150, 160 \}$) for four different single smooth steps of different height and smoothness all located in $\Rey_{\delta_*^{c_1}}=680$. A summary of the parameters of these computations can be found in table \ref{tab:071114}.

First, we observe that around a smooth step, the TS mode is energised and subsequently weakened (figure \ref{fig:071114}(a1-d3)). By {\it energising} we mean that around the step, there exists a local maximum of $|\tilde{u}|/A_0$. Secondly, a higher step height  $\hat{h}$ gives a stronger local maximum. Thirdly, increasing the height of the step moves the location of maximum amplitude of the TS mode downstream. For example for the an excitation frequency of $\mathcal{F}=150$ the maximum is located at $Re_{\delta^*}=920$ for $\hat{h} =5.48\%$ (figure \ref{fig:071114} (a2)) and at $Re_{\delta^*}=950$ for $\hat{h} =30\%$ (figure \ref{fig:071114} (d2)).

\begin{figure}
  \centerline{
\begin{overpic}[height=0.15\textheight]{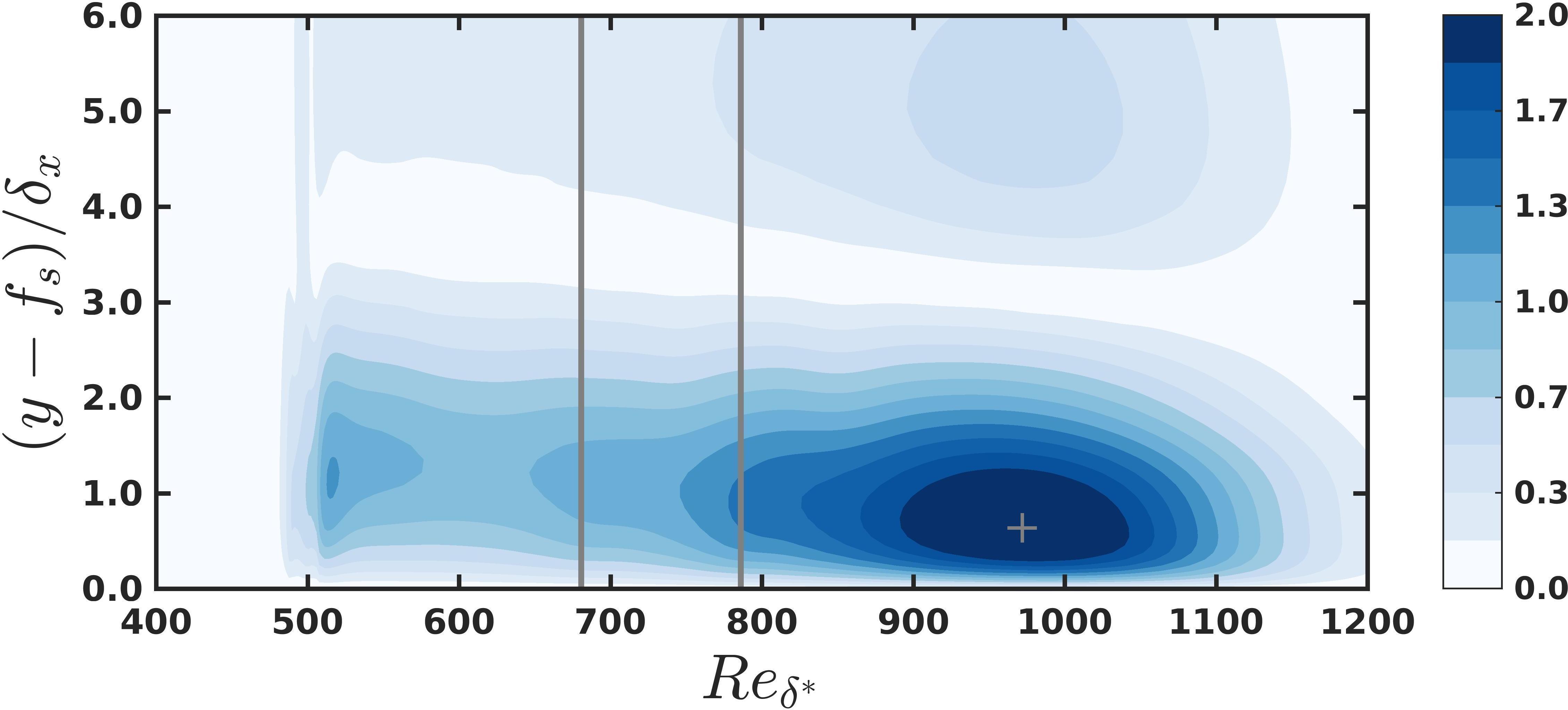}
\put(12,40){(a$_1$)}
\put(40,39){$\hat{h}=5\% \quad \mathcal{F}=140$}
\end{overpic}
\begin{overpic}[height=0.15\textheight]{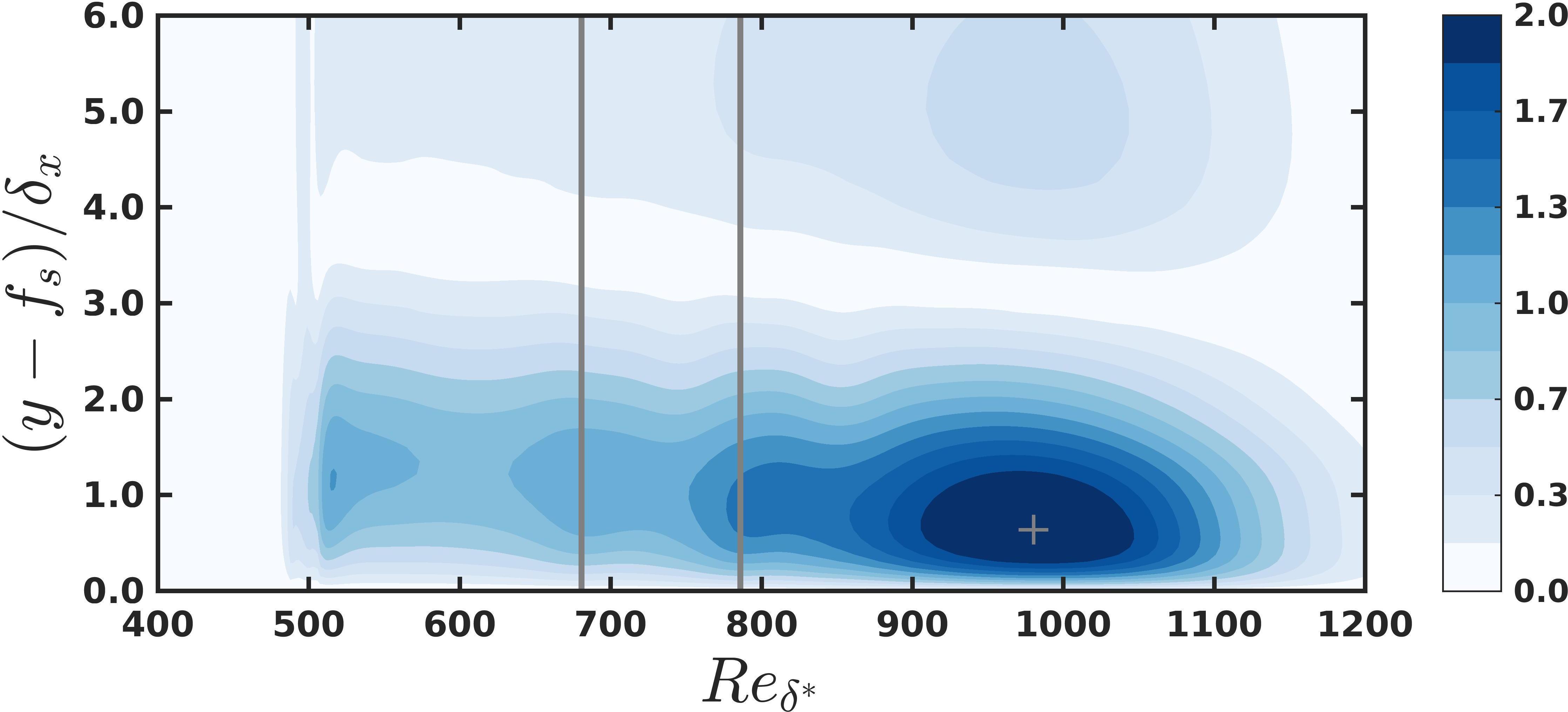}
\put(40,39){$\hat{h}=10\% \quad \mathcal{F}=140$}
\put(12,40){(b$_1$)}
\end{overpic}
}
  \centerline{
\begin{overpic}[height=0.15\textheight]{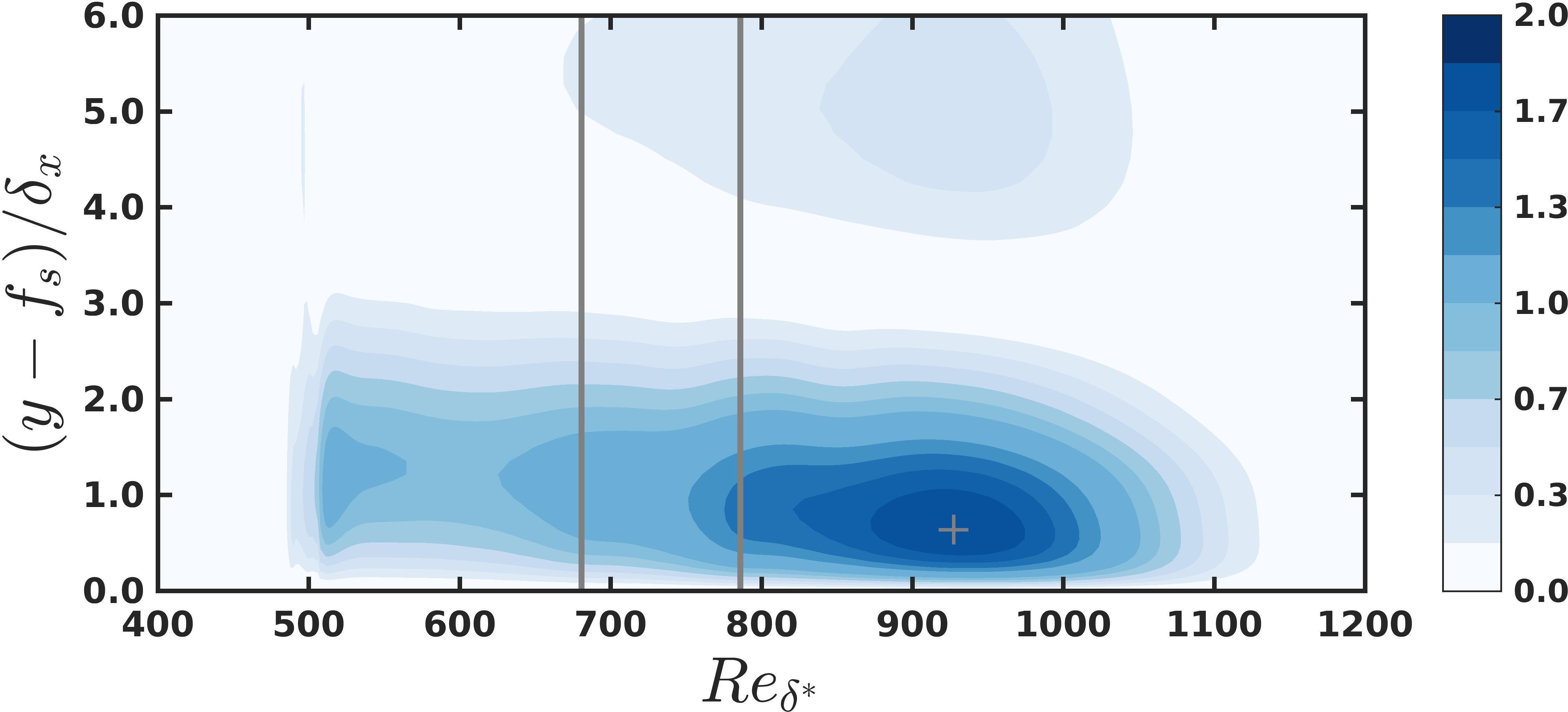}
\put(12,40){(a$_2$)}
\put(40,39){$\hat{h}=5\% \quad \mathcal{F}=150$}
\end{overpic}
\begin{overpic}[height=0.15\textheight]{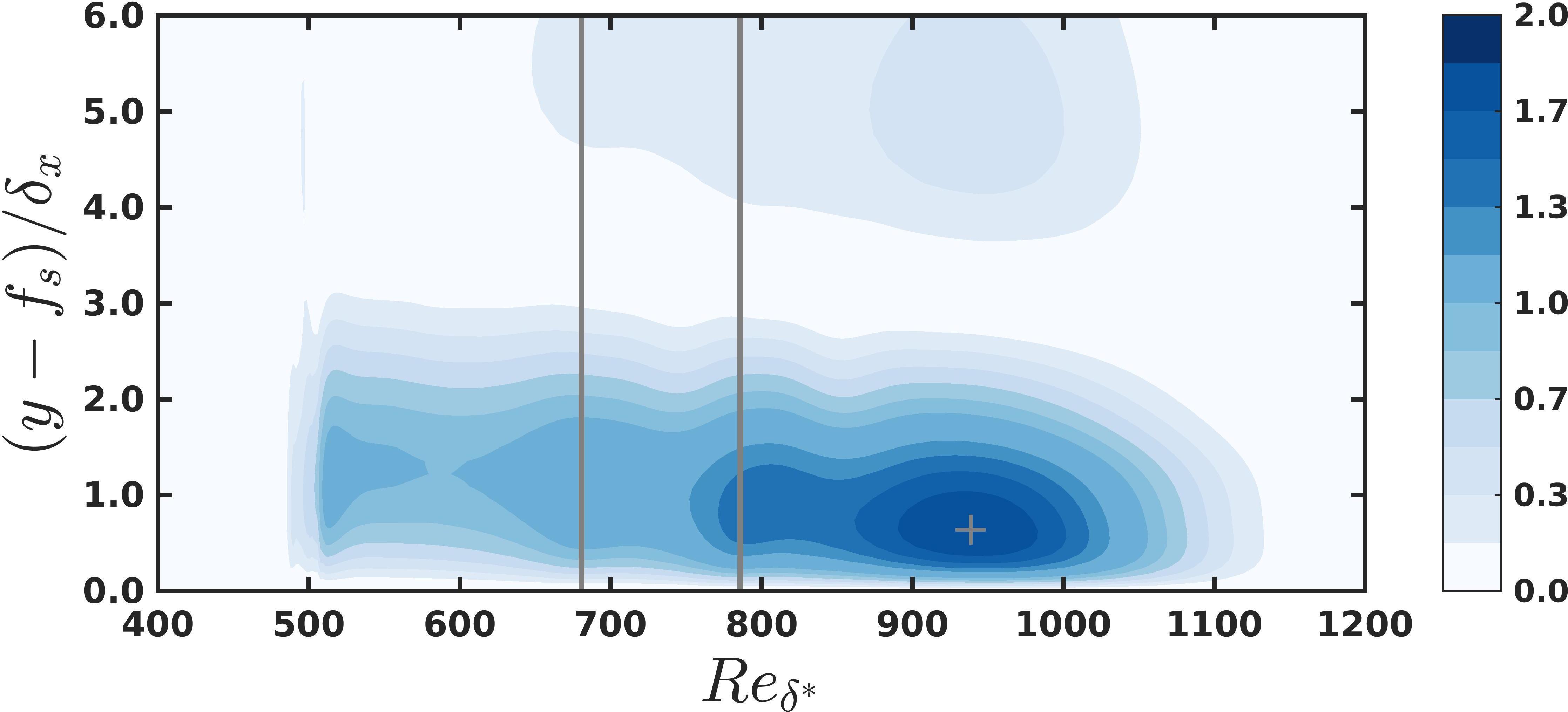}
\put(12,40){(b$_2$)}
\put(40,39){$\hat{h}=10\% \quad \mathcal{F}=150$}
\end{overpic}
}
  \centerline{
\begin{overpic}[height=0.15\textheight]{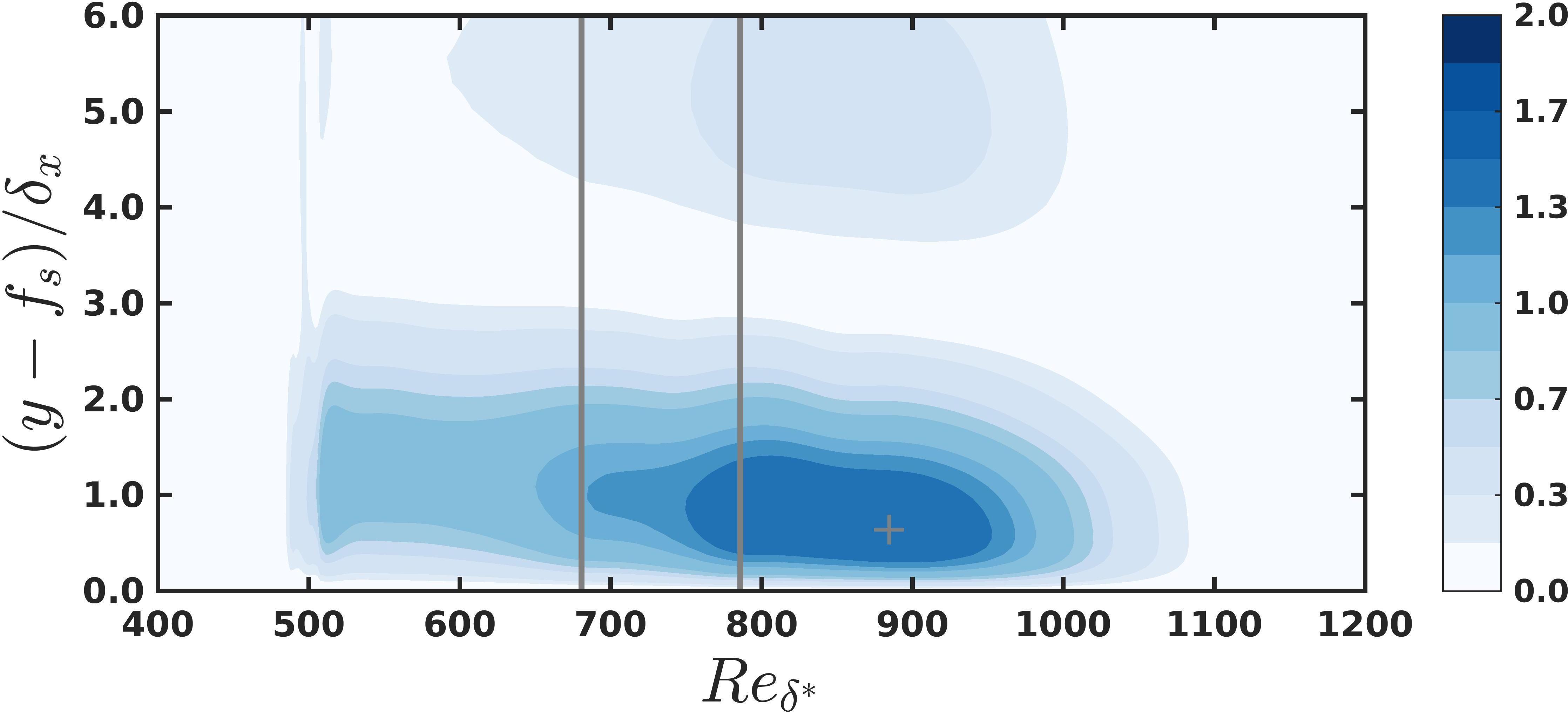}
\put(12,40){(a$_3$)}
\put(40,39){$\hat{h}=5\% \quad \mathcal{F}=160$}
\end{overpic}
\begin{overpic}[height=0.15\textheight]{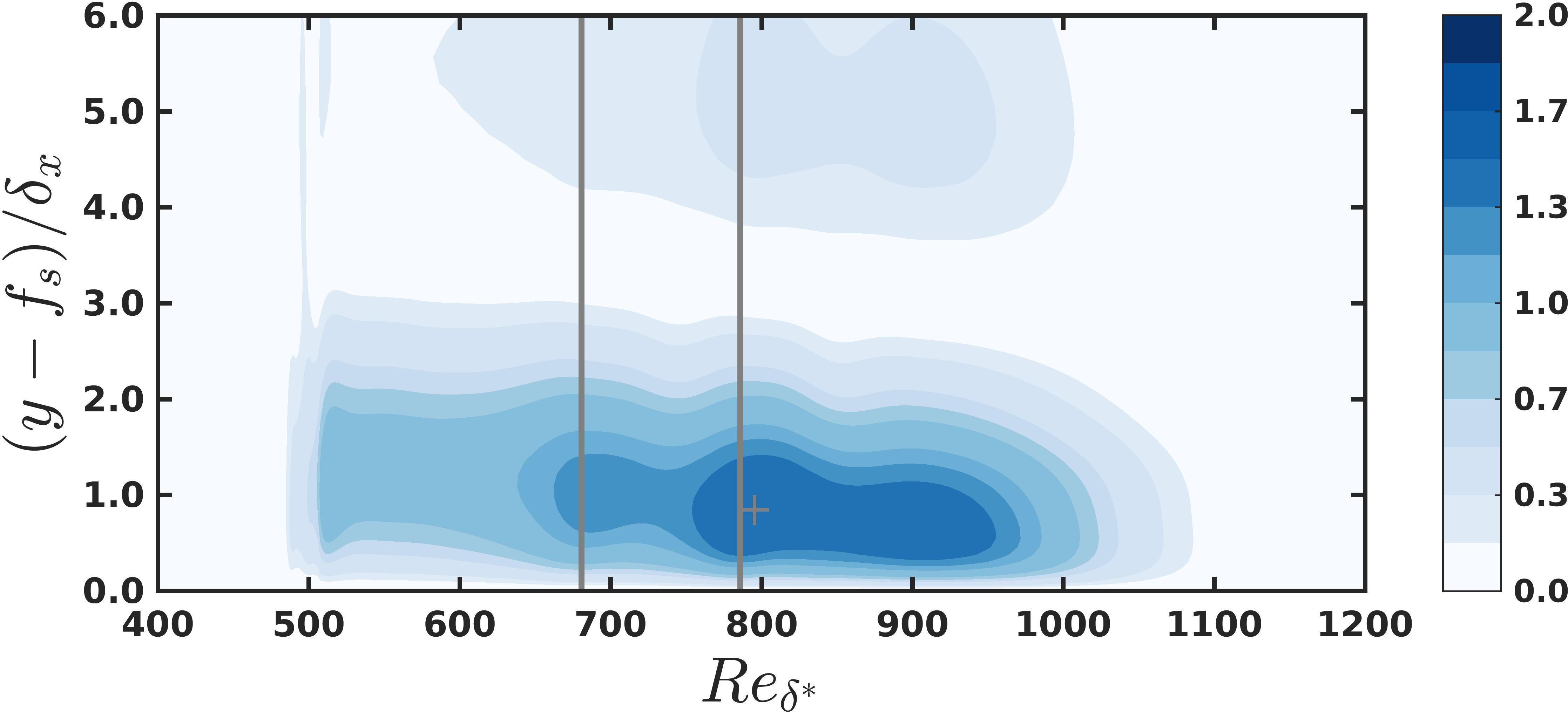}
\put(40,39){$\hat{h}=10\% \quad \mathcal{F}=160$}
\put(12,40){(b$_3$)}
\end{overpic}
}
 \centerline{
\begin{overpic}[height=0.15\textheight]{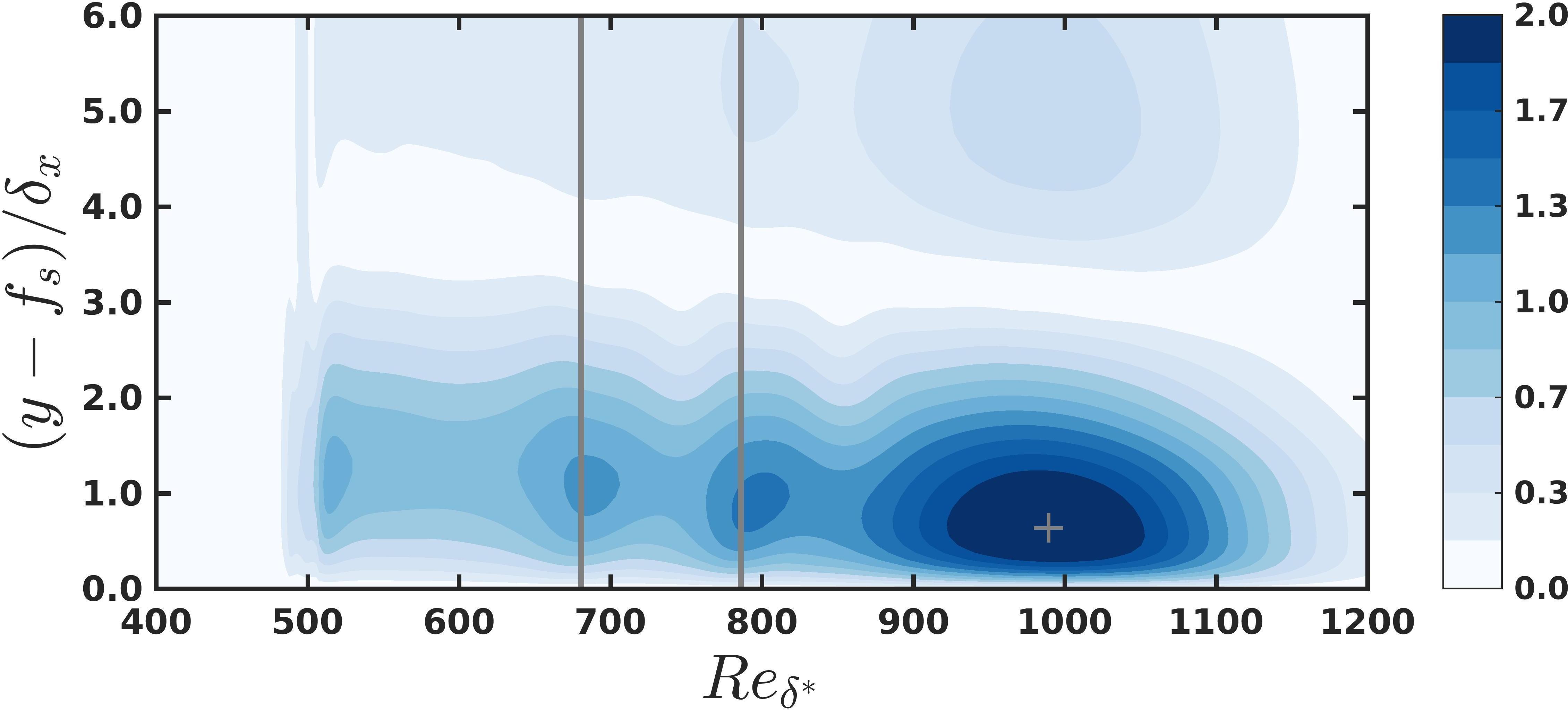}
\put(12,40){(c$_1$)}
\put(40,39){$\hat{h}=20\% \quad \mathcal{F}=140$}
\end{overpic}
\begin{overpic}[height=0.15\textheight]{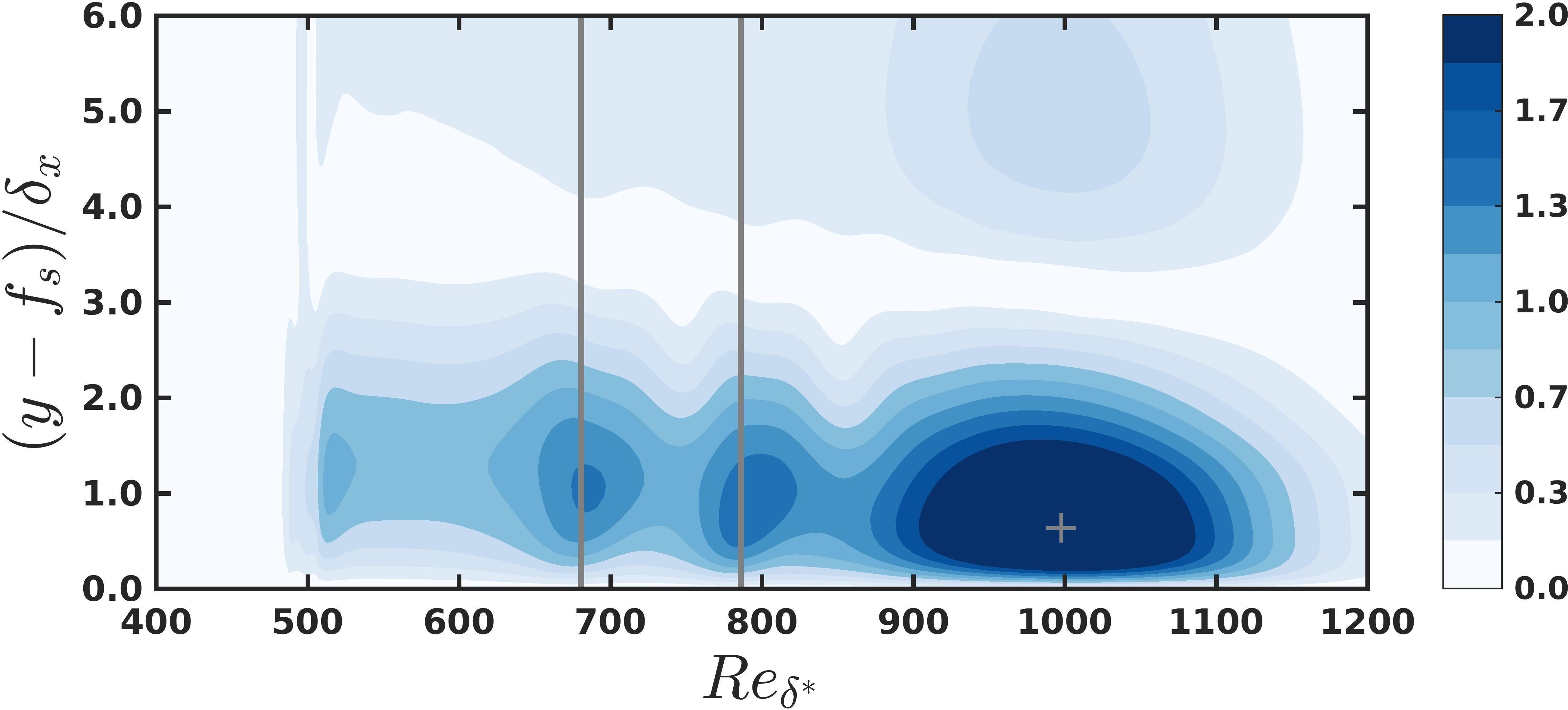}
\put(12,40){(d$_1$)}
\put(40,39){$\hat{h}=30\% \quad \mathcal{F}=140$}
\end{overpic}
}
  \centerline{
\begin{overpic}[height=0.15\textheight]{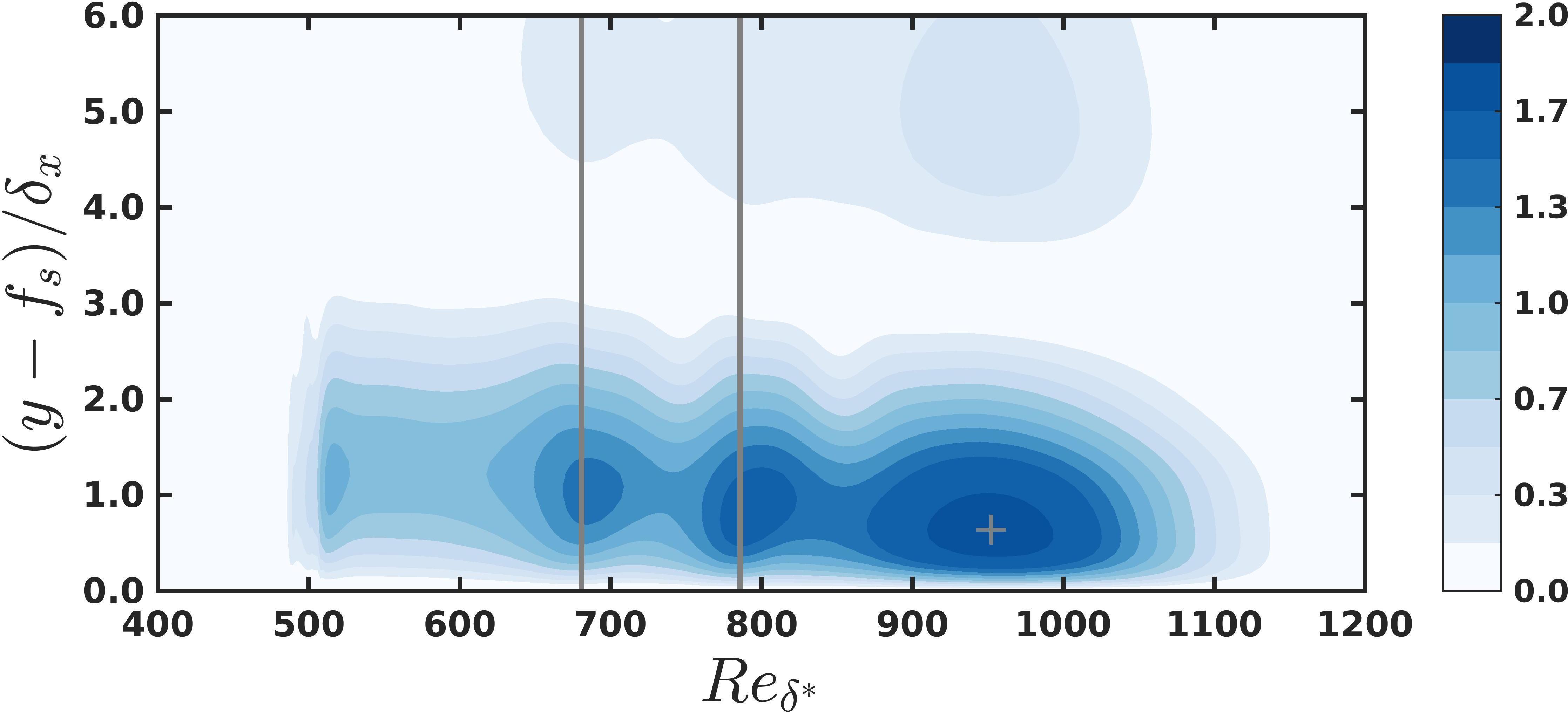}
\put(12,40){(c$_2$)}
\put(40,39){$\hat{h}=20\% \quad \mathcal{F}=150$}
\end{overpic}
\begin{overpic}[height=0.15\textheight]{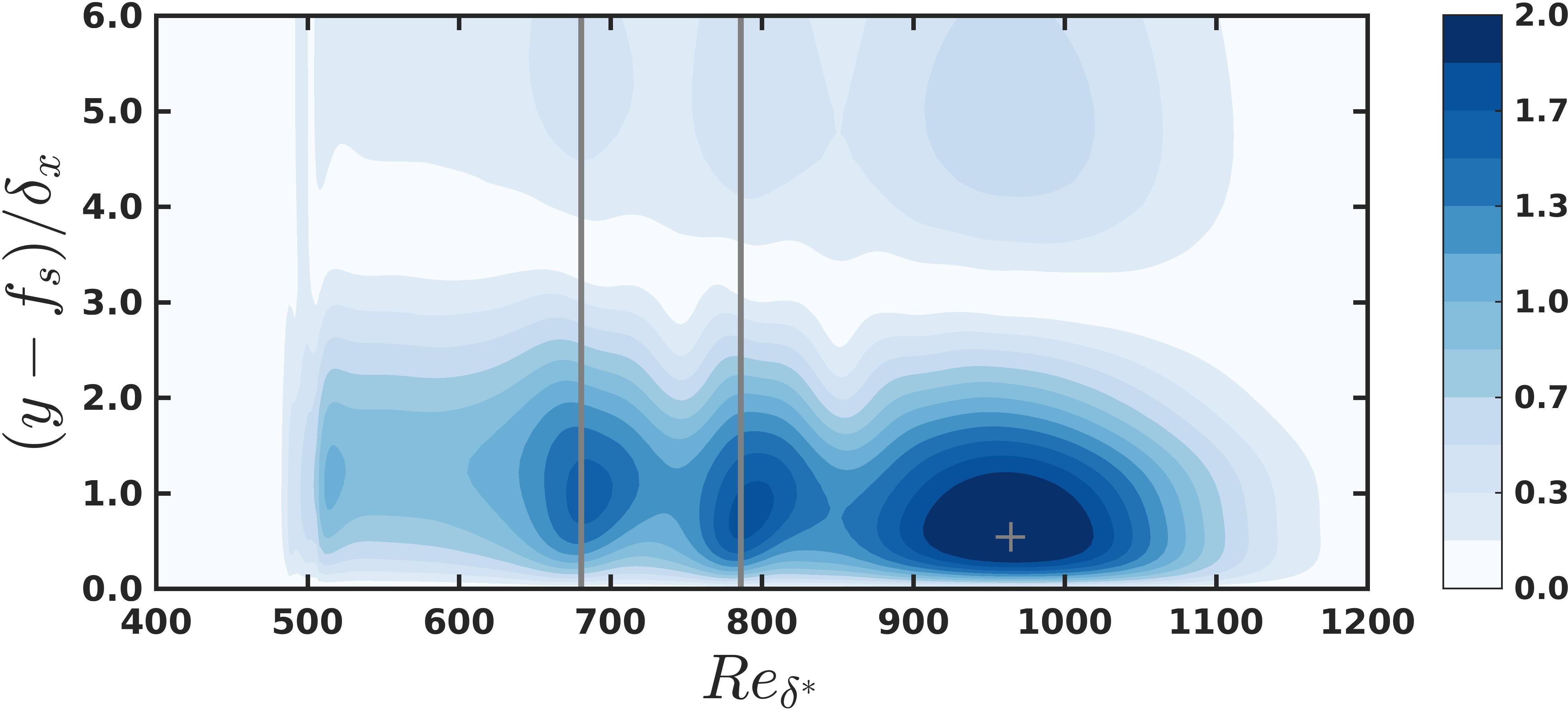}
\put(12,40){(d$_2$)}
\put(40,39){$\hat{h}=30\% \quad \mathcal{F}=150$}
\end{overpic}
}
  \centerline{
\begin{overpic}[height=0.15\textheight]{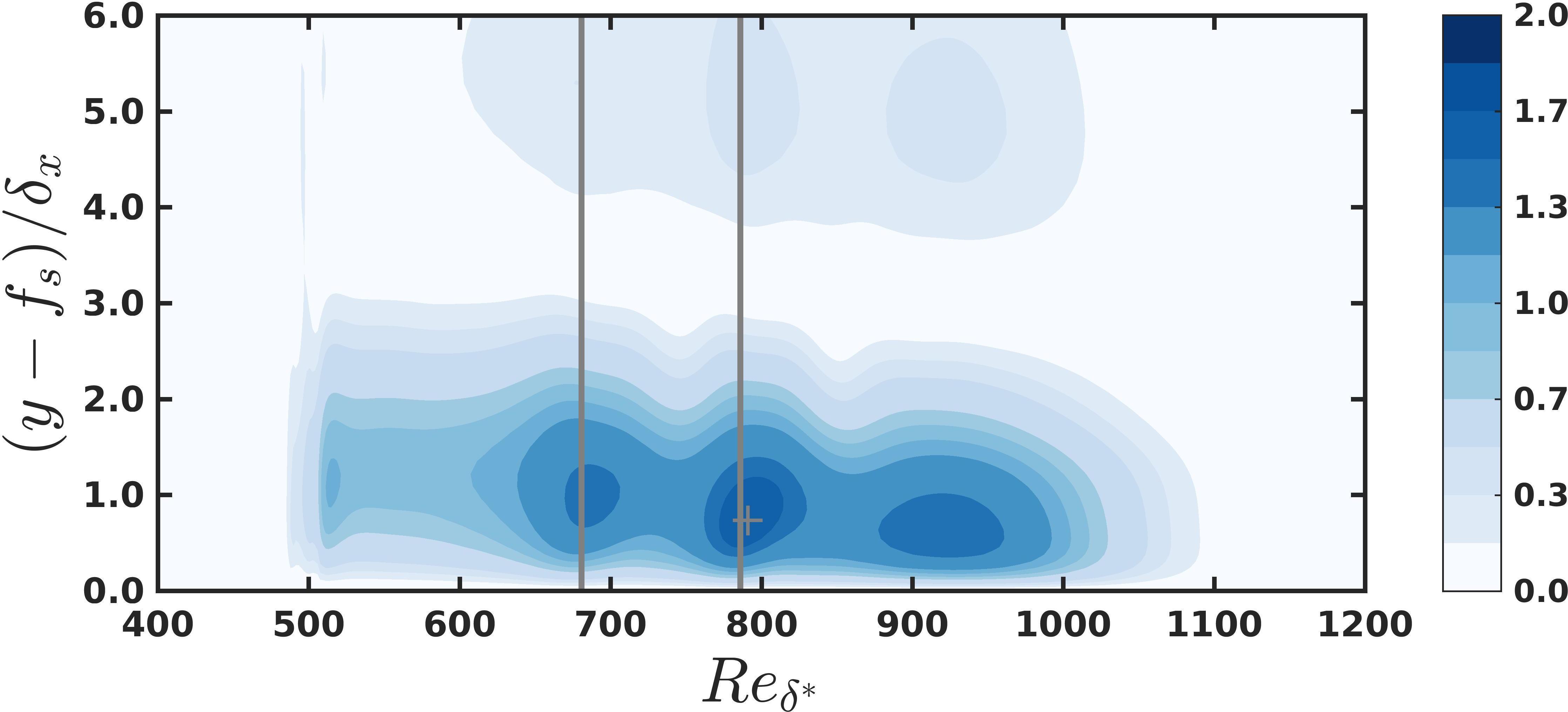}
\put(12,40){(c$_3$)}
\put(40,39){$\hat{h}=20\% \quad \mathcal{F}=160$}
\end{overpic}
\begin{overpic}[height=0.15\textheight]{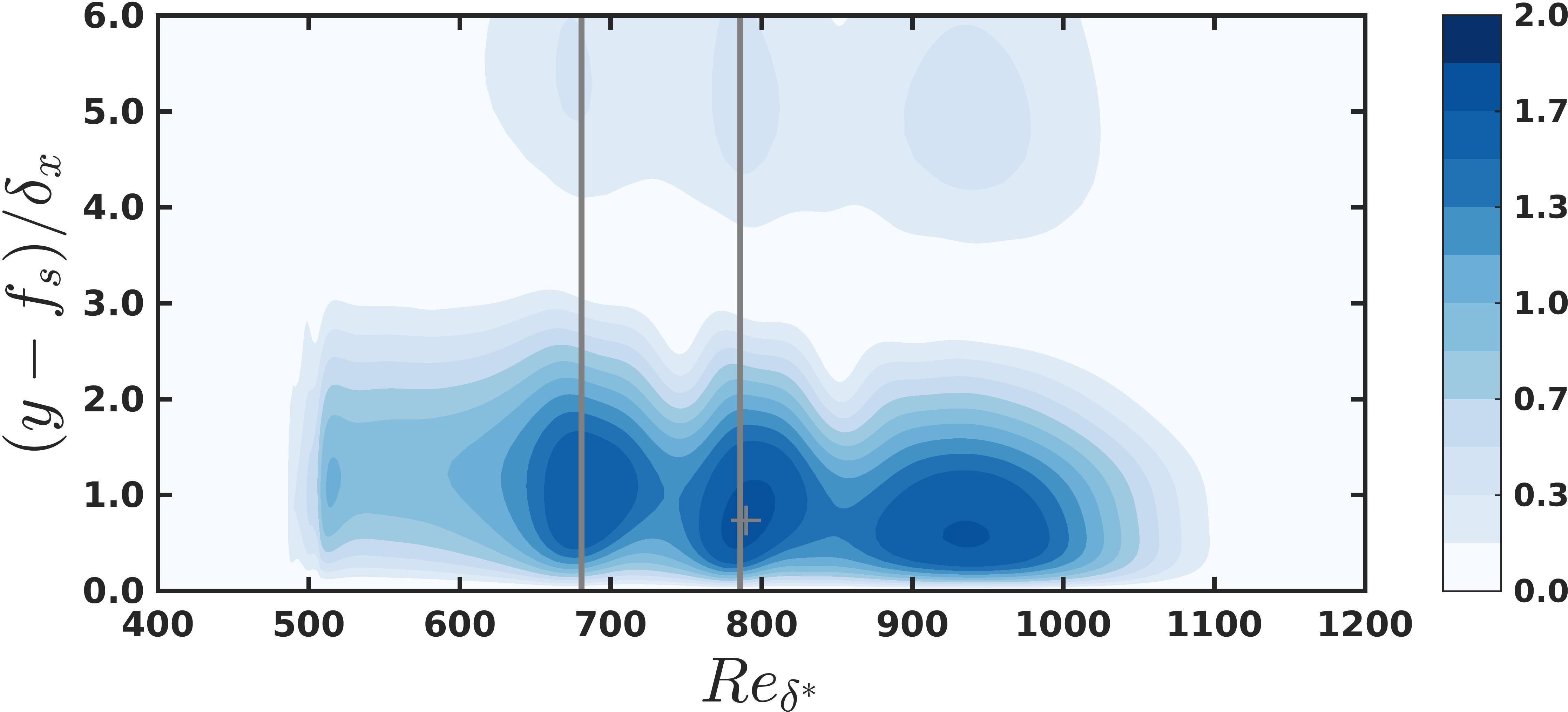}
\put(12,40){(d$_3$)}
\put(40,39){$\hat{h}=30\% \quad \mathcal{F}=160$}
\end{overpic}
}

 \caption{ Contour plots of $|\tilde{u}|/A_0$ and the physical parameters corresponding to (a$_\#$), (b$_\#$), (c$_\#$) and (d$_\#$) are from Case A, B, C  and D in Table \ref{tab:071114} where $\#$ denotes the number 1, 2 or 3, which corresponds to frequency $\mathcal{F}_\#$. The step positions are determined by $\Rey_{\delta_*^{\rm c_1}}$ and  $\Rey_{\delta_*^{\rm c_2}}$.  The grey cross + indicates the location of the maximum amplitude of the TS-wave and the vertical grey line represents the location of the forward facing smooth step.}
\label{fig:071114mf2}
\end{figure}

The result of our linear analysis reveal distinct behavior from the results for a sharp step ($\gamma=1$) of height $h/\delta^*=0.235$ ($\hat{h}=8.24\%$) given by \cite{worner2003} where they 
claim that a forward facing sharp step showed a stabilising effect without a local destablisation regime despite a separation bubble being reported in front of the step. Recently, through further numerical calculations, \cite{edelmann2015} observed that, for subsonic Mach numbers and large height scale ($h/\delta^*=0.94$ or $\hat{h}=33\%$), two separation bubbles, in front and  on top of the step, were observed and strong amplification of the disturbances was found in front of and behind the step. The results from the linear analysis and, more generally the results from the DNS presented in \S 3.4, underline the attractiveness of replacing a forward facing sharp step ($\gamma=1$ with large $\hat{h}$  by a smooth one $\gamma < 1$ because the  smooth step with the same $\hat{h}$ does not lead to a separation bubble. We attribute the discrepancy between the current findings and the work of \cite{edelmann2015} to the step-induced separation bubble that has strong destabilising effect on the TS mode. Furthermore, for a forward-facing smooth step of fixed height, the position of global maximum value of a contour varies with respect to the given frequency, which is consistent with the relative position of the step with respect to the position of the upper branch when changing frequency in the neutral curve diagram (\ref{fig:071114}(c1-c3) for example). For two isolated smooth steps, a similar  phenomenon is observed in figure \ref{fig:071114mf2}(c1-c3) except that two distinct local maxima are observed around each smooth step for large $\hat{h}$. From figures \ref{fig:071114} and  \ref{fig:071114mf2}, it can be concluded that both for single- and two-step configuration  global maximum values of $|\tilde{u}|/A_0$ depended on frequencies, $\hat{h}$ and smoothness.

\begin{figure}
  \centerline{
\begin{overpic}[width=0.5\textwidth]{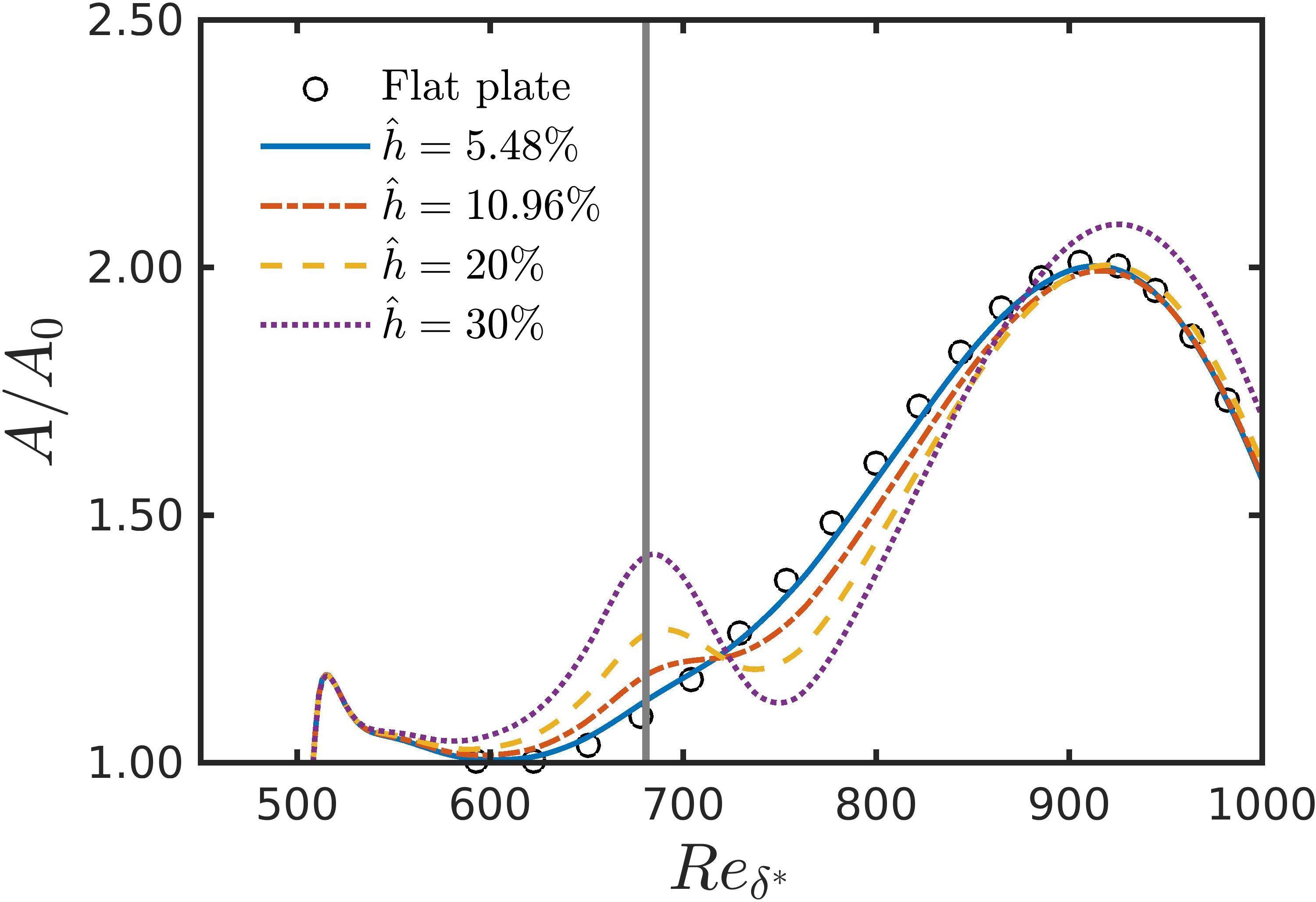}
\put(86,60){(a$_1$)}
\end{overpic}
\begin{overpic}[width=0.5\textwidth]{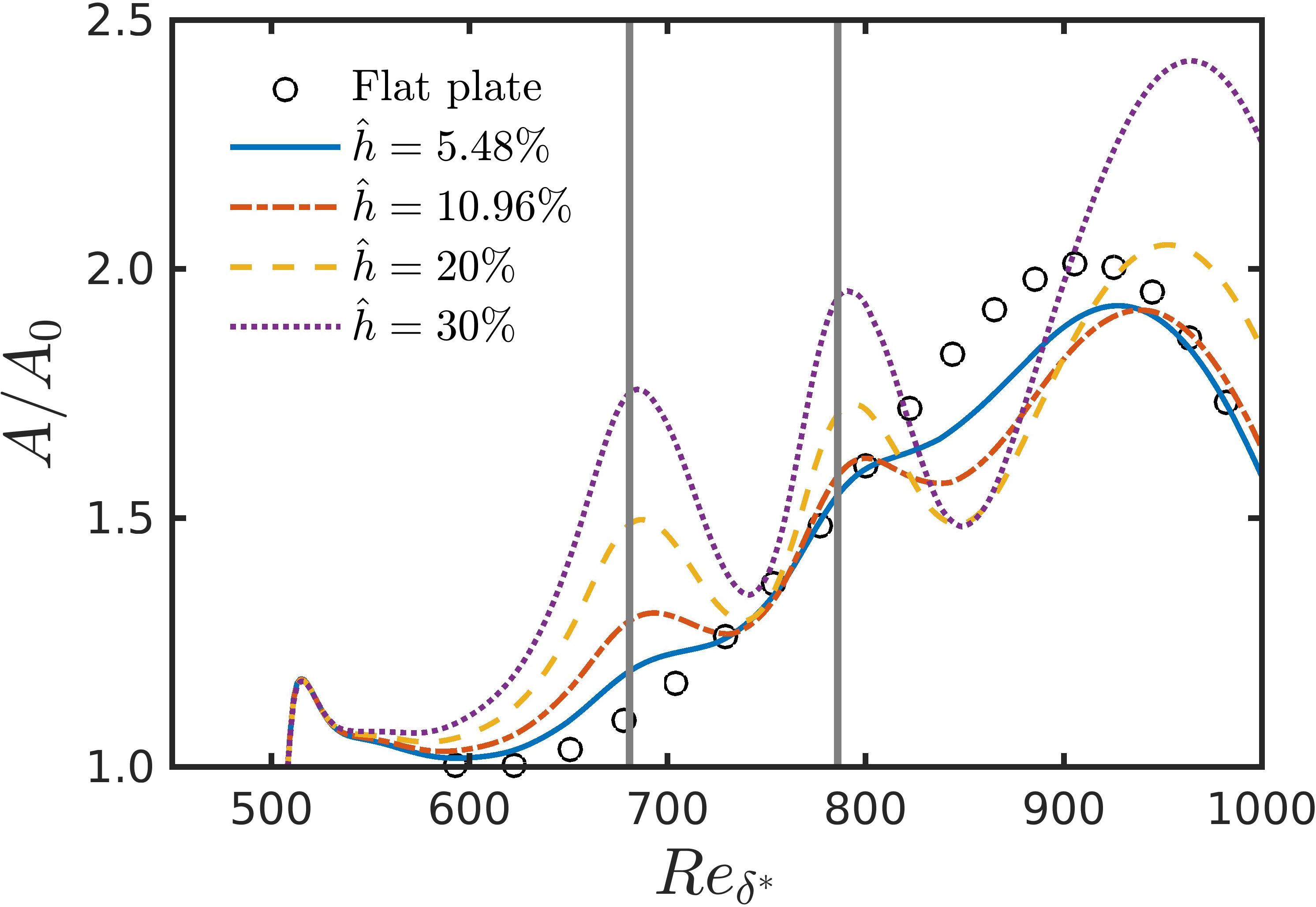}
\put(86,60){(b$_1$)}
\end{overpic}
}\vskip1mm
  \centerline{
\begin{overpic}[width=0.5\textwidth]{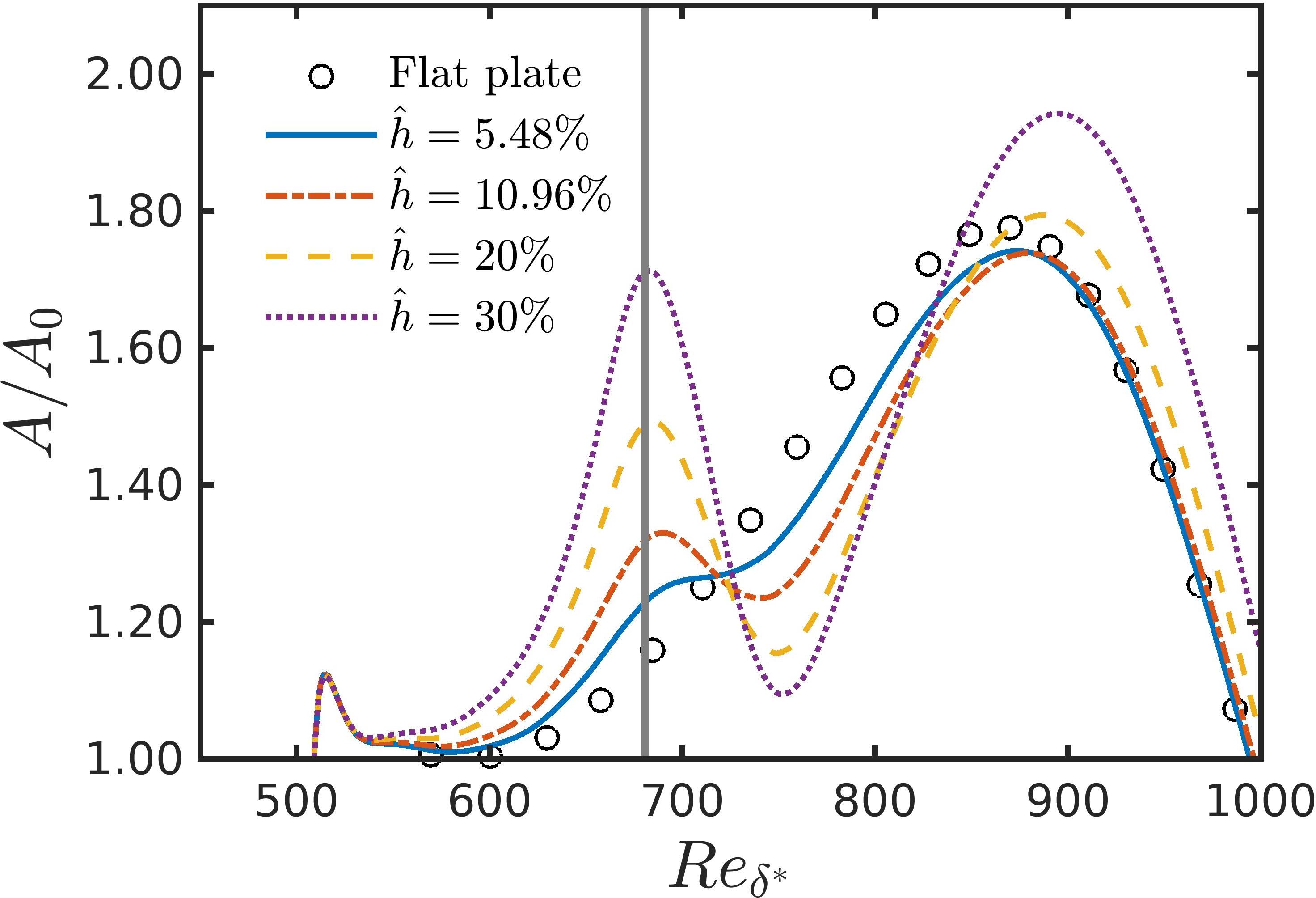}
\put(86,60){(a$_2$)}
\end{overpic}
\begin{overpic}[width=0.5\textwidth]{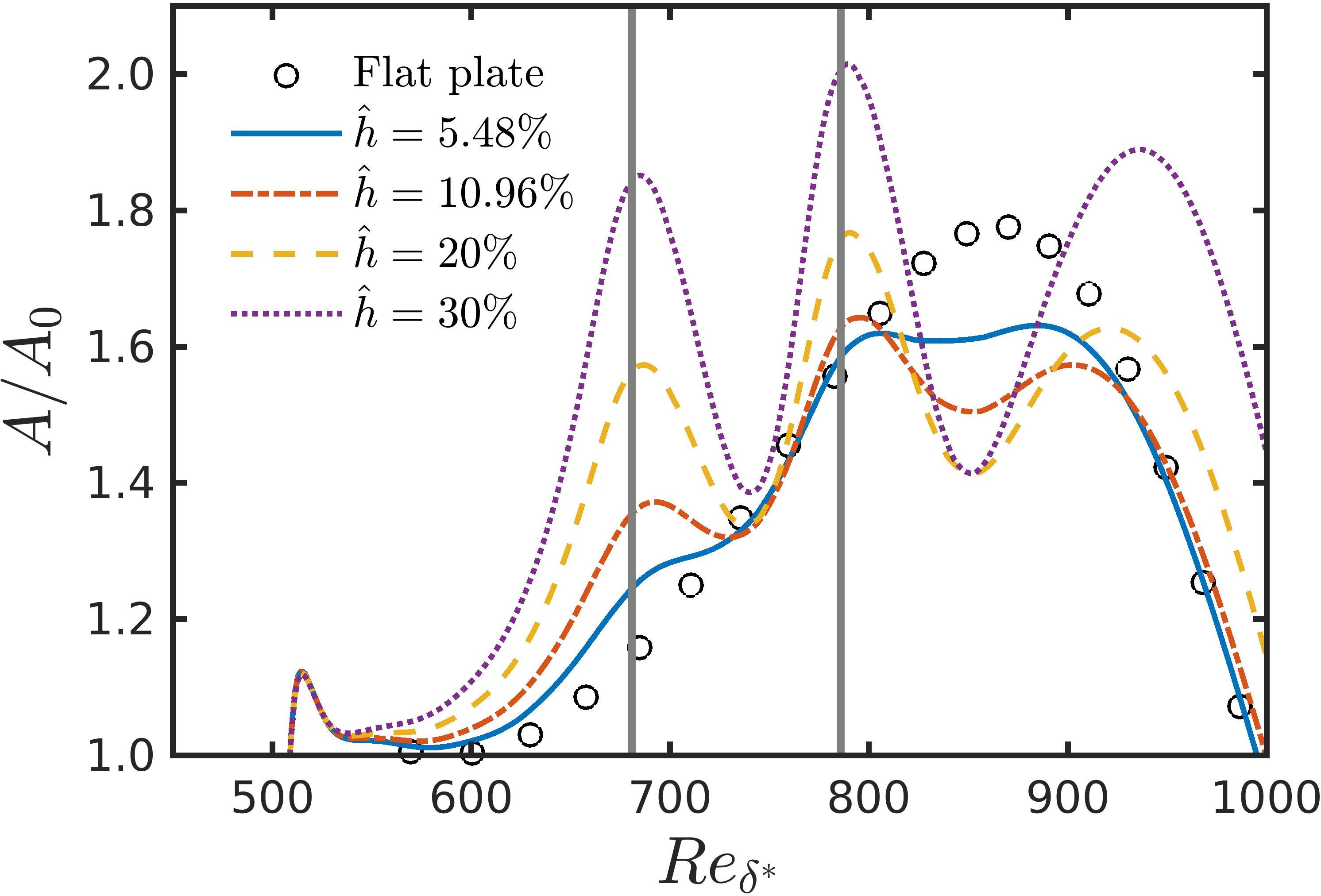}
\put(86,60){(b$_2$)}
\end{overpic}
}\vskip1mm
  \centerline{
\begin{overpic}[width=0.5\textwidth]{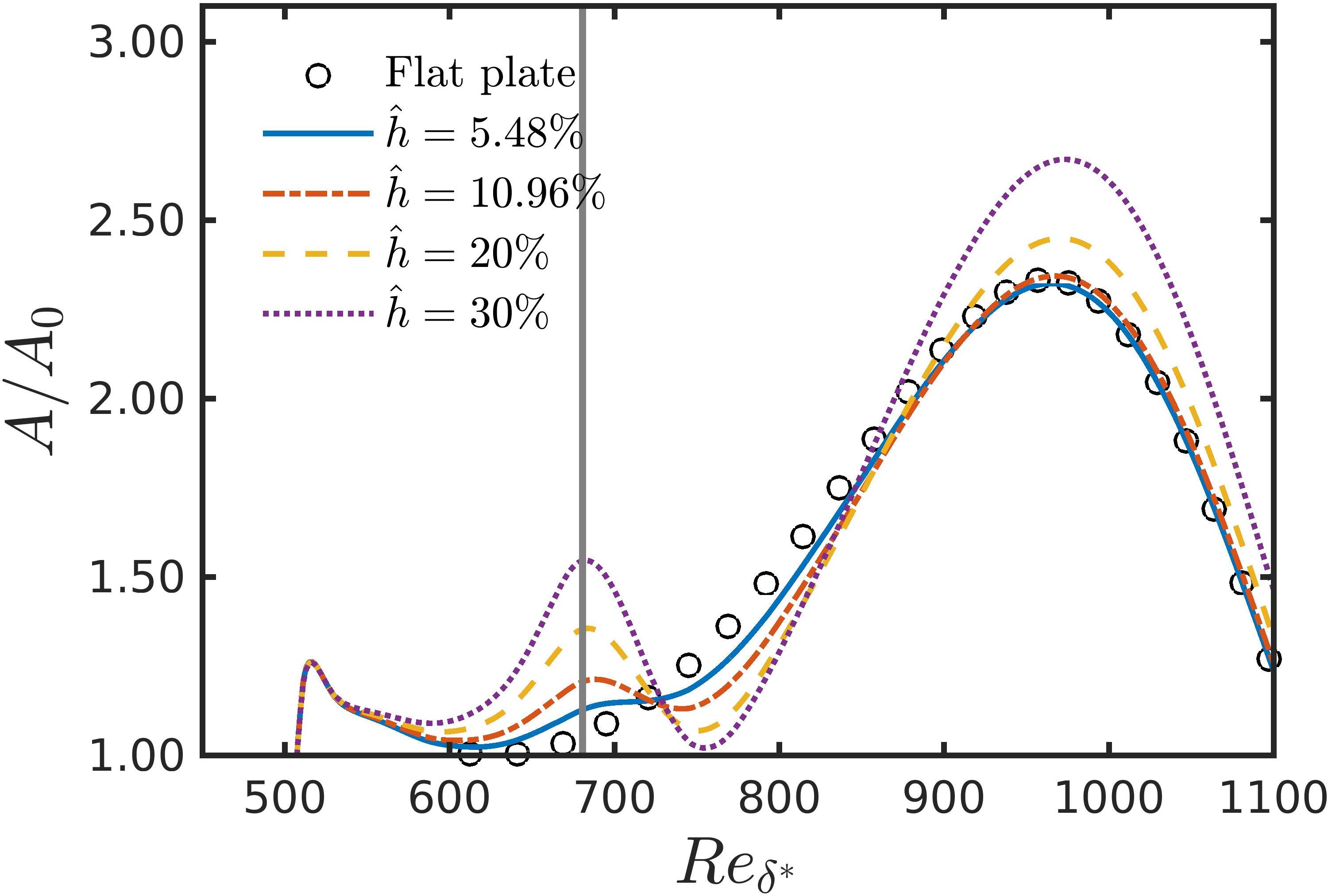}
\put(86,60){(a$_3$)}
\end{overpic}
\begin{overpic}[width=0.5\textwidth]{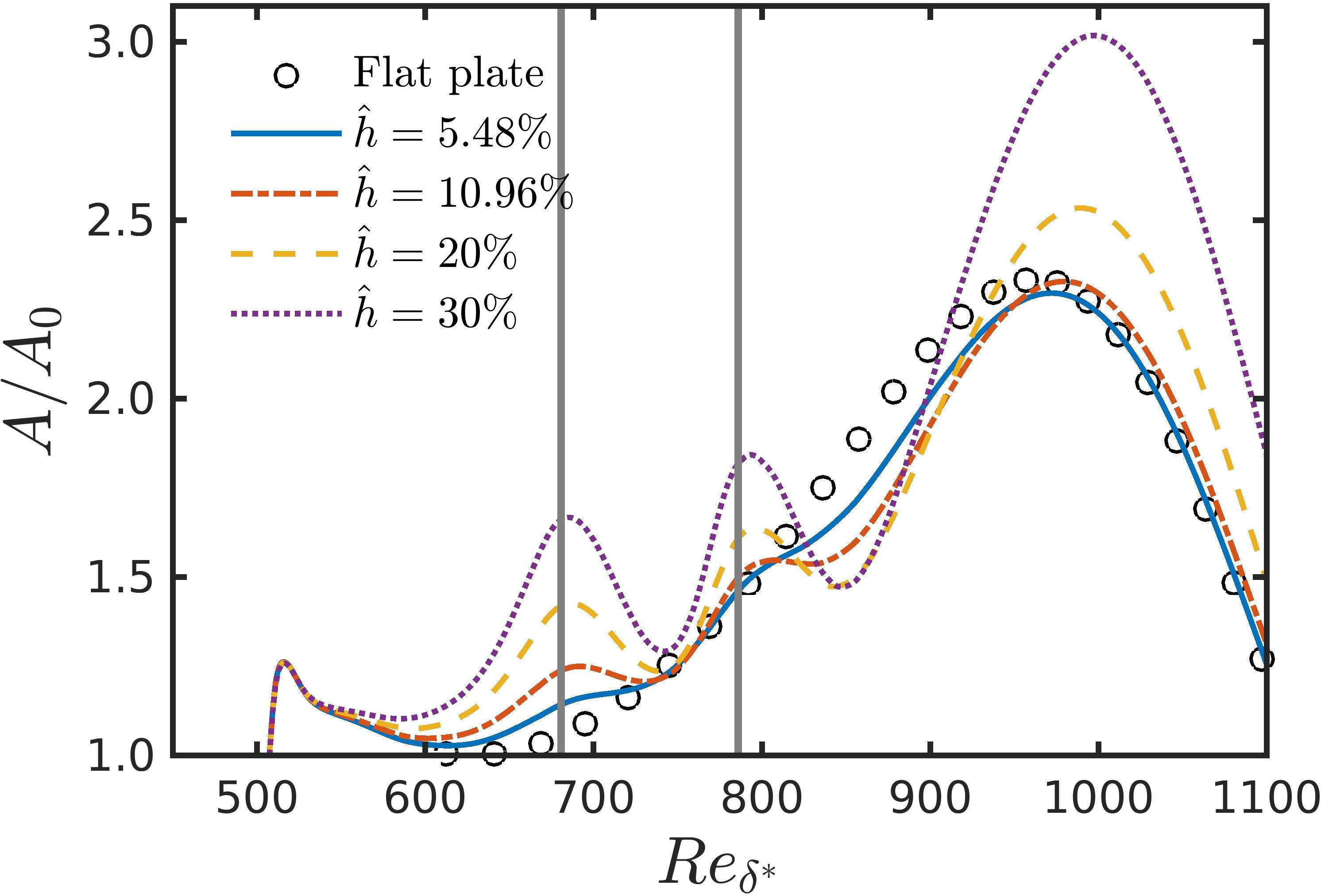}
\put(86,60){(b$_3$)}
\end{overpic}
}
 \caption{ The physical parameters corresponding to A, B,C and D are from Case A, B, C and D in Table \ref{tab:071114}. (a$_\#$) and (b$_\#$), respectively,  denote single step and two steps where where $\#$ denotes the number 1, 2 or 3, which corresponds to frequency $\mathcal{F}_\#$ in Table \ref{tab:071114}. The vertical grey lines represents the location of the forward facing smooth step.}
\label{fig:071114ts}
\end{figure}

\begin{figure}
  \centerline{
\begin{overpic}[width=0.96\textwidth]{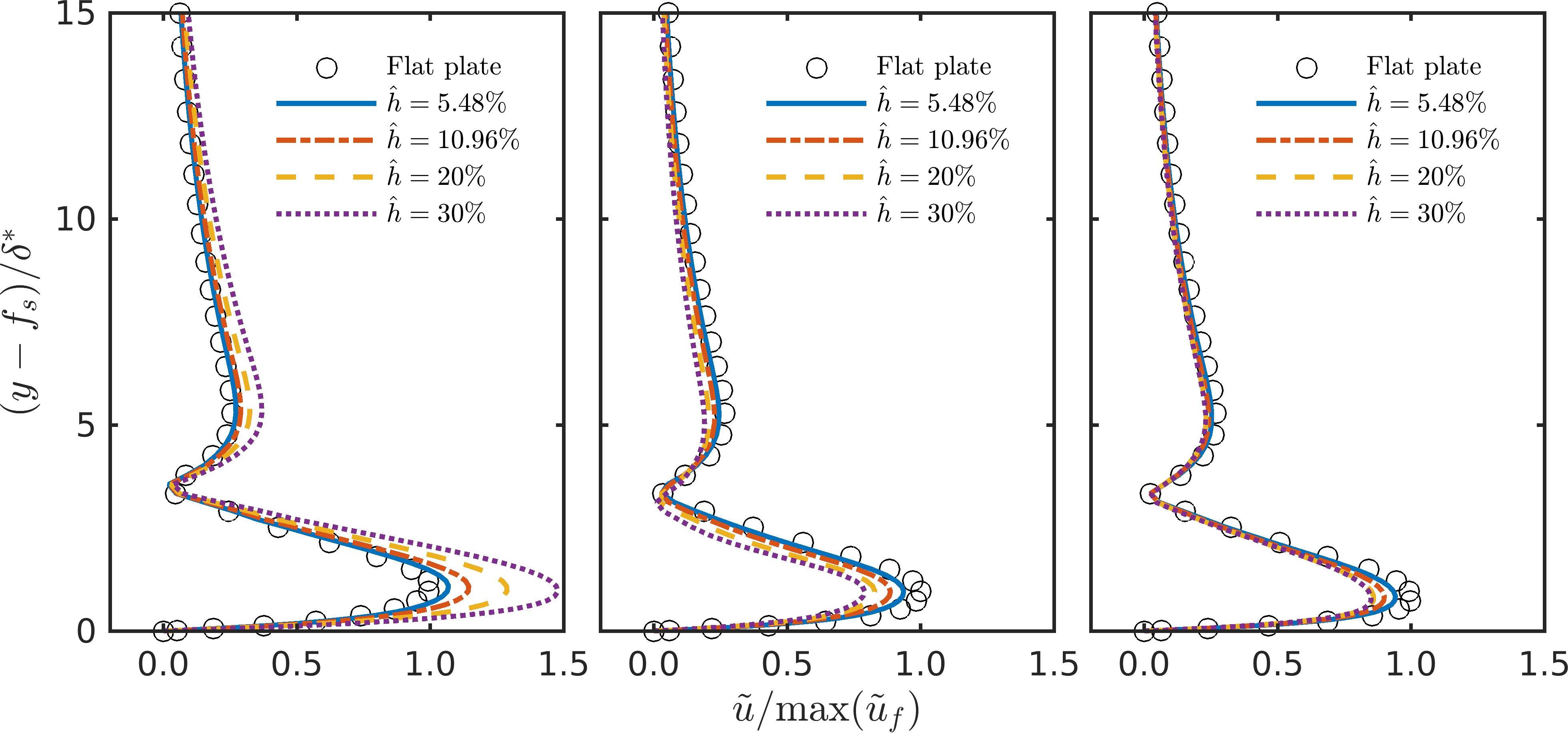}
\end{overpic}
}
  \centerline{
\begin{overpic}[width=0.96\textwidth]{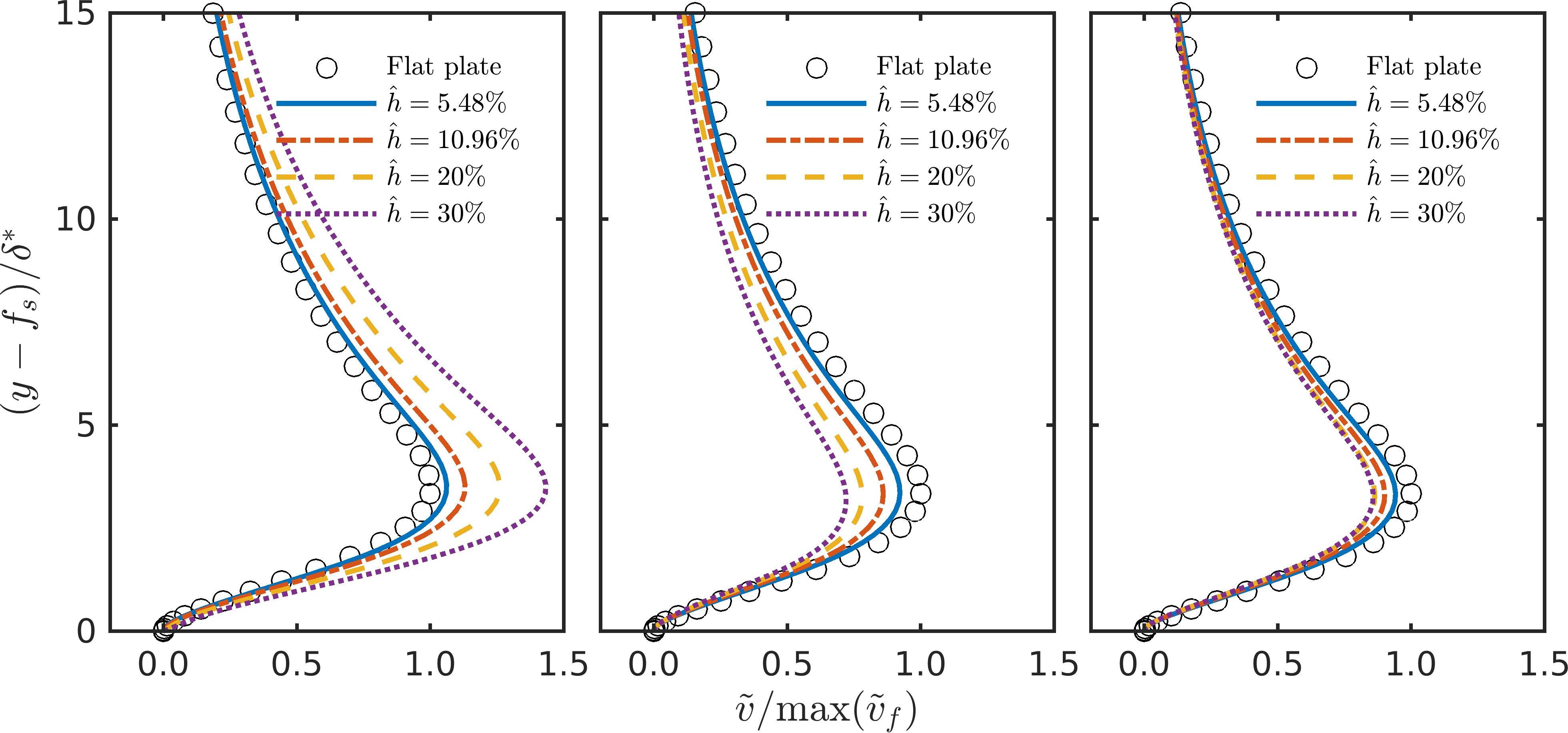}
\end{overpic}
}
 \caption{ Comparison of the TS modes at different location over a single smooth step. The parameters corresponding to A, B, C and D in the figures are from Case A, B, C and D in Table \ref{tab:071114} and $\mathcal{F}=150$. (Left-Top) and (Left-Bottom): $\Rey_\delta^*=680$; (Mid-Top) and (Mid-Bottom): $\Rey_\delta^*=750$; (Right-Top) and (Right-Bottom): $\Rey_\delta^*=800$. $\tilde{u}$ and $\tilde{v}$ are, receptively, normalised by ${\rm max}(\tilde{u}_f)$ and ${\rm max}(\tilde{v}_f)$ from the corresponding flat plat boundary.}
\label{fig:071114tsp}
\end{figure}

By extracting maximum values of the envelope of the TS mode at each streamwise location, the $A/A_0$ profiles for each case are shown in figure \ref{fig:071114ts}.  We notice that when height scale of one single smooth step is not high enough, it does significantly energise the TS wave and the energised region is limited to the vicinity of the the step.  From figure \ref{fig:071114ts}(a$_1$-a$_{3}$), we observe that when $\hat{h}<20\%$, the TS waves are stabilised downstream for small $\hat{h}$. In figure \ref{fig:071114tsp}, a comparison of the TS modes at three different locations are shown, further elucidating the stabilisation effect. We find that for a single smooth step, when frequencies are changed, behaviours of the TS modes are different from each other. When $\hat{h}<20\%$, for all three frequencies,  the global maximum value of the TS mode does not exceed the global maximum values of  the TS modes in the corresponding flat plate boundary layers. Analysing  the frequencies in figure \ref{fig:neutral},  when $\hat{h}\ge20\%$,  the most amplified TS modes correspond to the lowest frequency $\mathcal{F}=140$. The reason for this lies in that for a fixed step position, choosing a low frequency indicates that the step is moved towards the lower branch of the neutral stability curve of a flat plate boundary layer. This is only a heuristic explanation, since large step heights have a significant impact on the position of the lower branch, which can lead to deformation of the shape of the neutral stability curve and moving the neutral position upstream with respect to that of a flat plate boundary. This non-local phenomenon can be seen from a comparison between the TS mode in a flat plate boundary and the TS mode  in a boundary layer over a large height scale smooth step as illustrated in figure \ref{fig:071114ts}(a$_1$-a$_{3}$).  

From figure \ref{fig:071114ts}(a$_1$-a$_{3}$),  assuming the local growth  (or destabilisation) of the TS wave does not trigger any nonlinear phenomena, a smooth step with a low height scale is not harmful for the TS wave with a high frequency ($140 \leq \mathcal{F} \leq160$). In fact, to some extent, a boundary layer can benefit from a smooth step since the net instability can be reduced. In figure \ref{fig:071114ts}(b$_1$-b$_{3}$), with the same frequencies, envelopes of the TS waves over two isolated smooth steps are shown. The position of the two smooth steps can be found from Table \ref{tab:071114} as schematically illustrated in figure \ref{fig:neutral}(b). For the frequencies considered here, the second steps still lie in the unstable regime of the neutral stability curve. We observe that, surprisingly,  the second steps do not locally lead to further amplification of the TS waves when $\hat{h}<20\%$; in contrast, the amplitudes of the TS waves are further damped compared  to the amplitude of the TS waves amplitudes over both the single step boundary layers and a flat plate boundary. A destabilising of the TS mode is only introduced by large height scale steps.

From the results of the linear stability analysis presented above, we deduce that stabilisation or destabilisation behaviours of smooth steps strongly depends on the smoothness parameter $\gamma$ as well as the height. In a suitable range of parameters the influence of smooth steps on a boundary layer can be dominated by stabilisation effect. To conclude, in the unstable regime of the neutral stability curve, which is typically near the leading edge, if a step-like structure is inevitable, a smooth step will have less of destabilising effect on the TS waves than a sharp one.

\begin{table}
  \begin{center}
\def~{\hphantom{0}}
  \begin{tabular}{ccccccccccccc}
 Case &$\Rey_{\delta^*_{i}}$   &$\Rey_{\delta^*_{\rm c_1}}$&$\Rey_{\delta^*_{\rm c_2}}$& $\mathcal{F}_1$& $\mathcal{F}_2$& $\mathcal{F}_3$ &$\hat{h} \%$ & $\hat{d}$ & $\gamma\times 10^4$  & $L_x/\delta_{99}$ & $L_y/\delta_{99}$    \\[3pt]
      A & 320 &680&786&140&150&160 &5.48 &4&0.74 & 250 & 30 \\
       B&---&---&---&---&---&---&10.96&---& 2.99 &---&---\\
       C&---&---&---&---&---&---&20.00&---&9.97 &---&---\\
       D&---&---&---&---&---&---&30.00&---&22.44 &---&---
  \end{tabular}
  \caption{Parameters for smooth steps where $\Rey_{\delta^*_i}$, $\Rey_{\delta^*_{\rm c_1}}$  and $\Rey_{\delta^*_{\rm c_2}}$ are, respectively,  the inlet Reynolds number, the Reynolds number at the centre of the  first step and  the Reynolds number at the centre of the second step . $\mathcal{F}$ denotes the non-dimensional perturbation frequency.  $L_x$ and $L_y$ denote streamwise extent and height of the domain for which the 2D base flow field obtained  was independent of domain size.}
  \label{tab:071114}
  \end{center}
\end{table}

\subsection{Linear analysis of smooth steps at lower excitation frequencies}
\begin{figure}
  \centerline{
\begin{overpic}[width=0.48\textwidth]{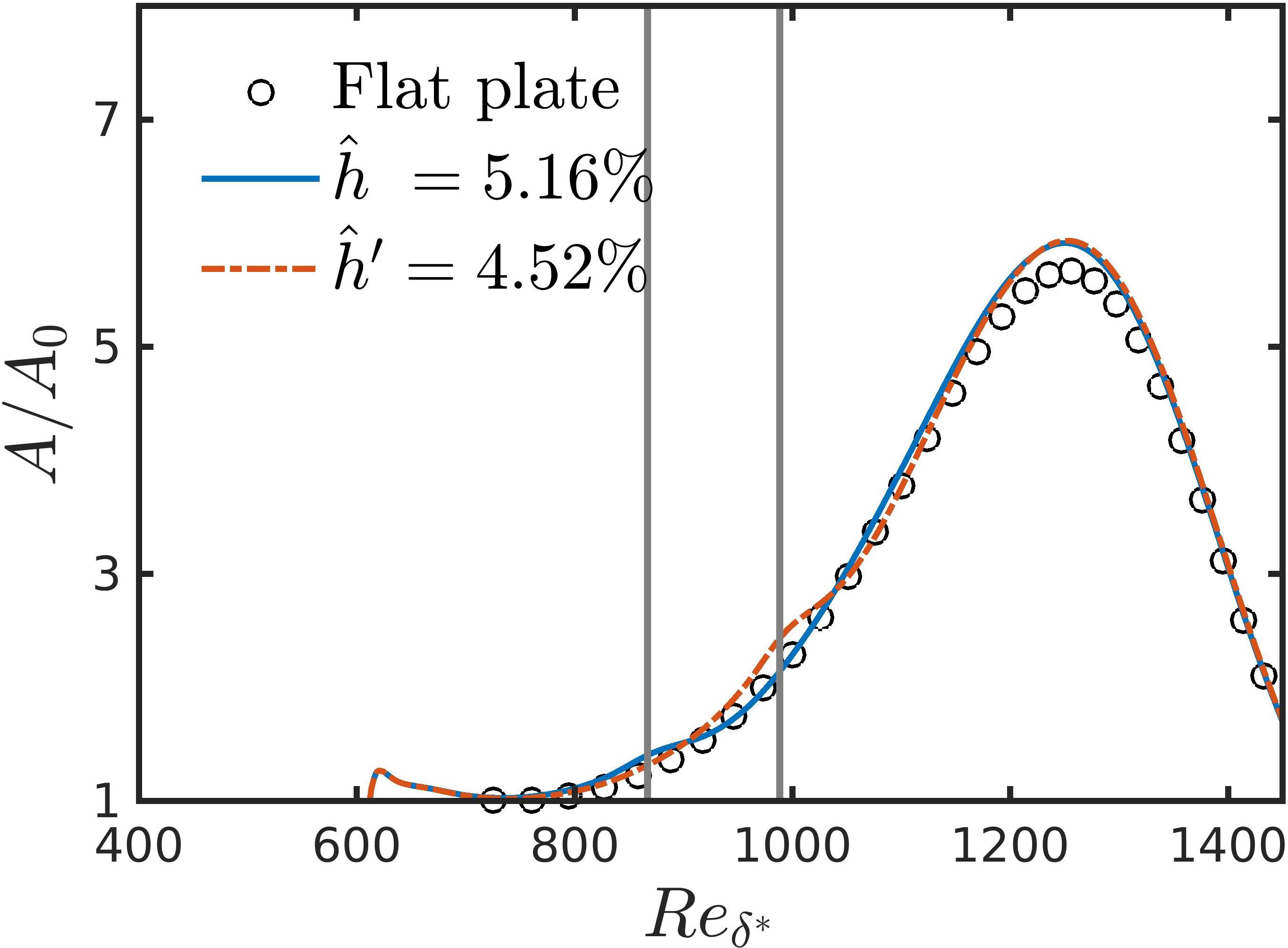}
\put(90,65){(a)}
\end{overpic}
\begin{overpic}[width=0.48\textwidth]{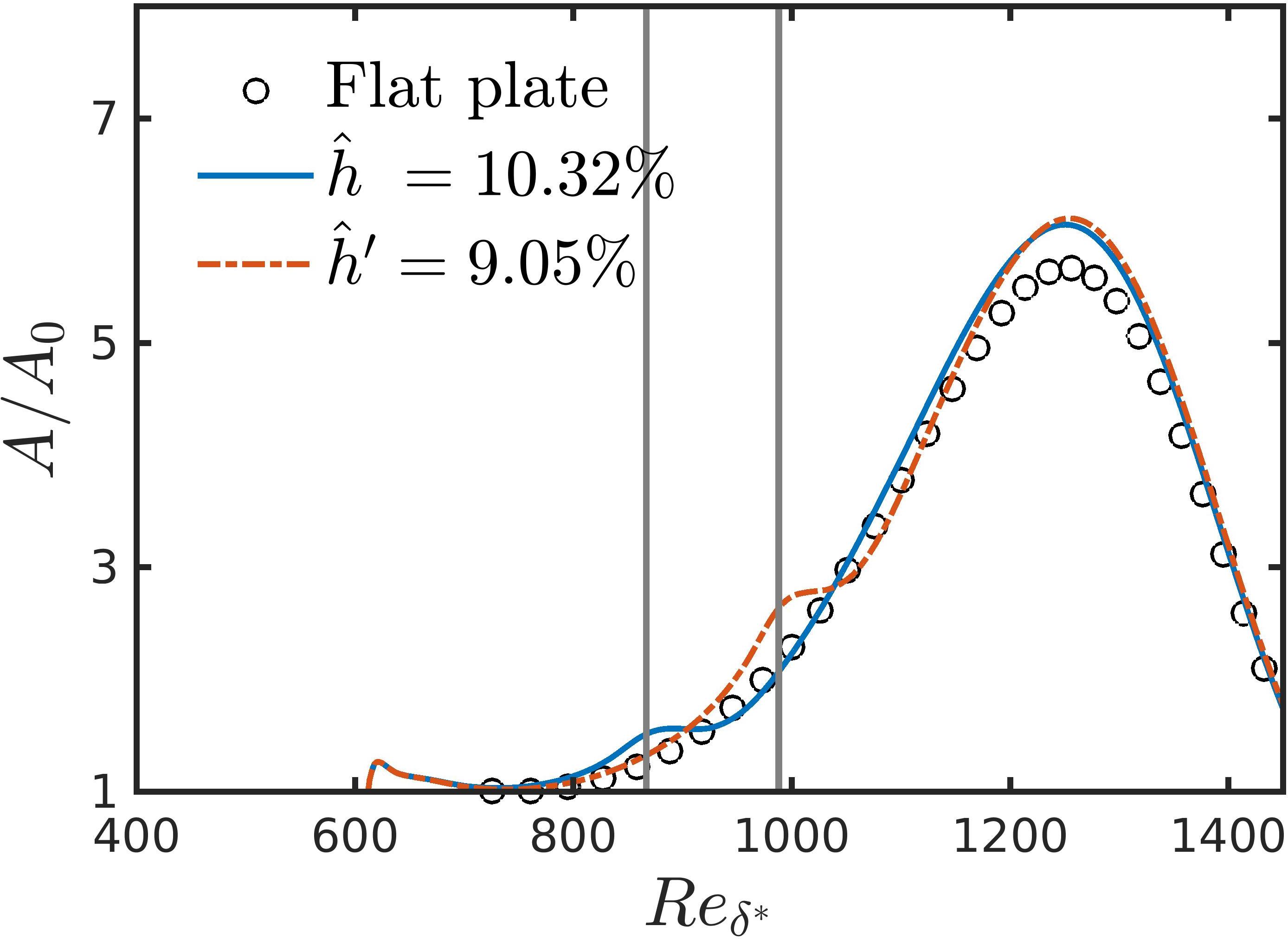}
\put(90,65){(b)}
\end{overpic}
}
  \centerline{
\begin{overpic}[width=0.48\textwidth]{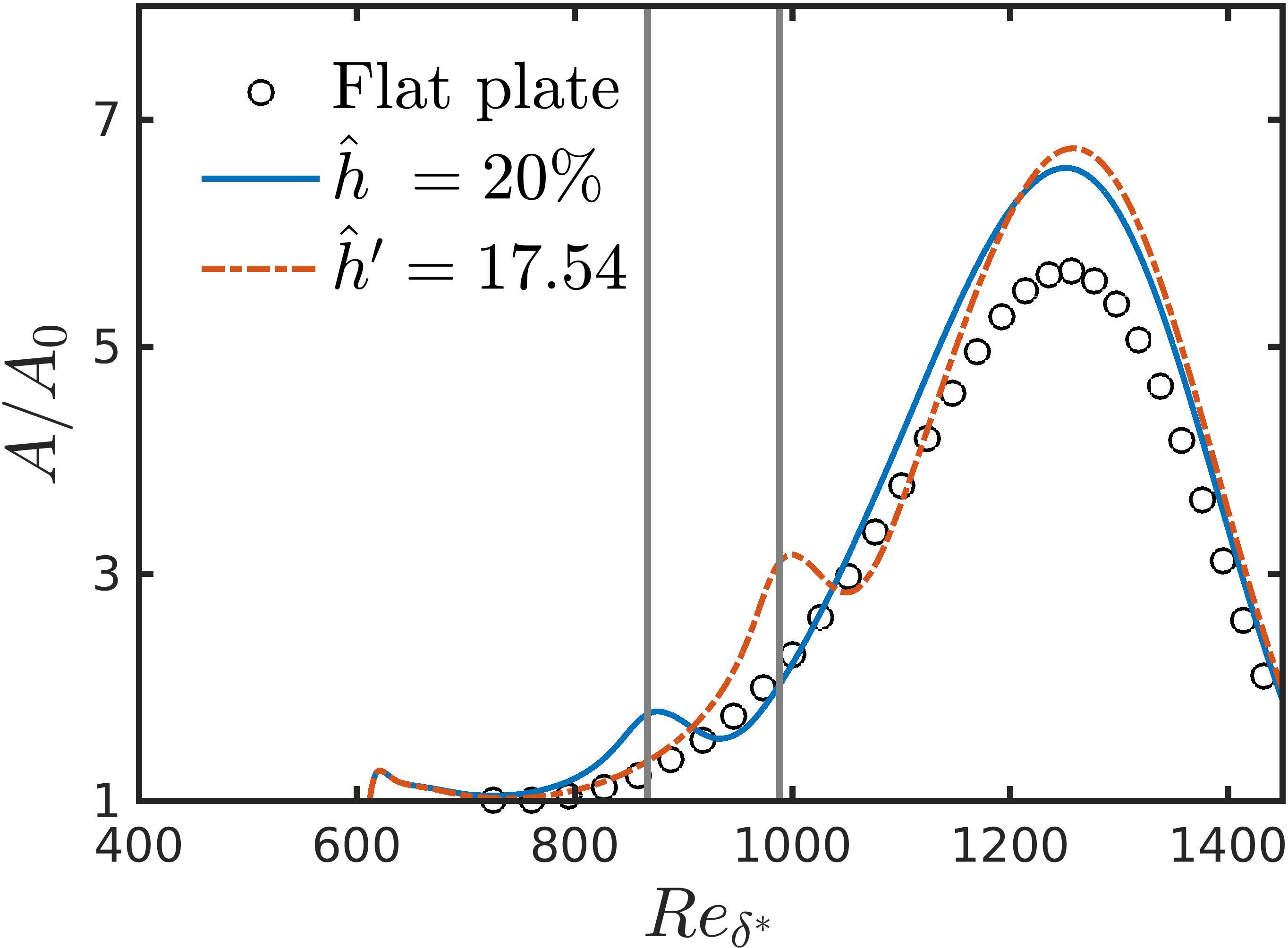}
\put(90,65){(c)}
\end{overpic}
\begin{overpic}[width=0.48\textwidth]{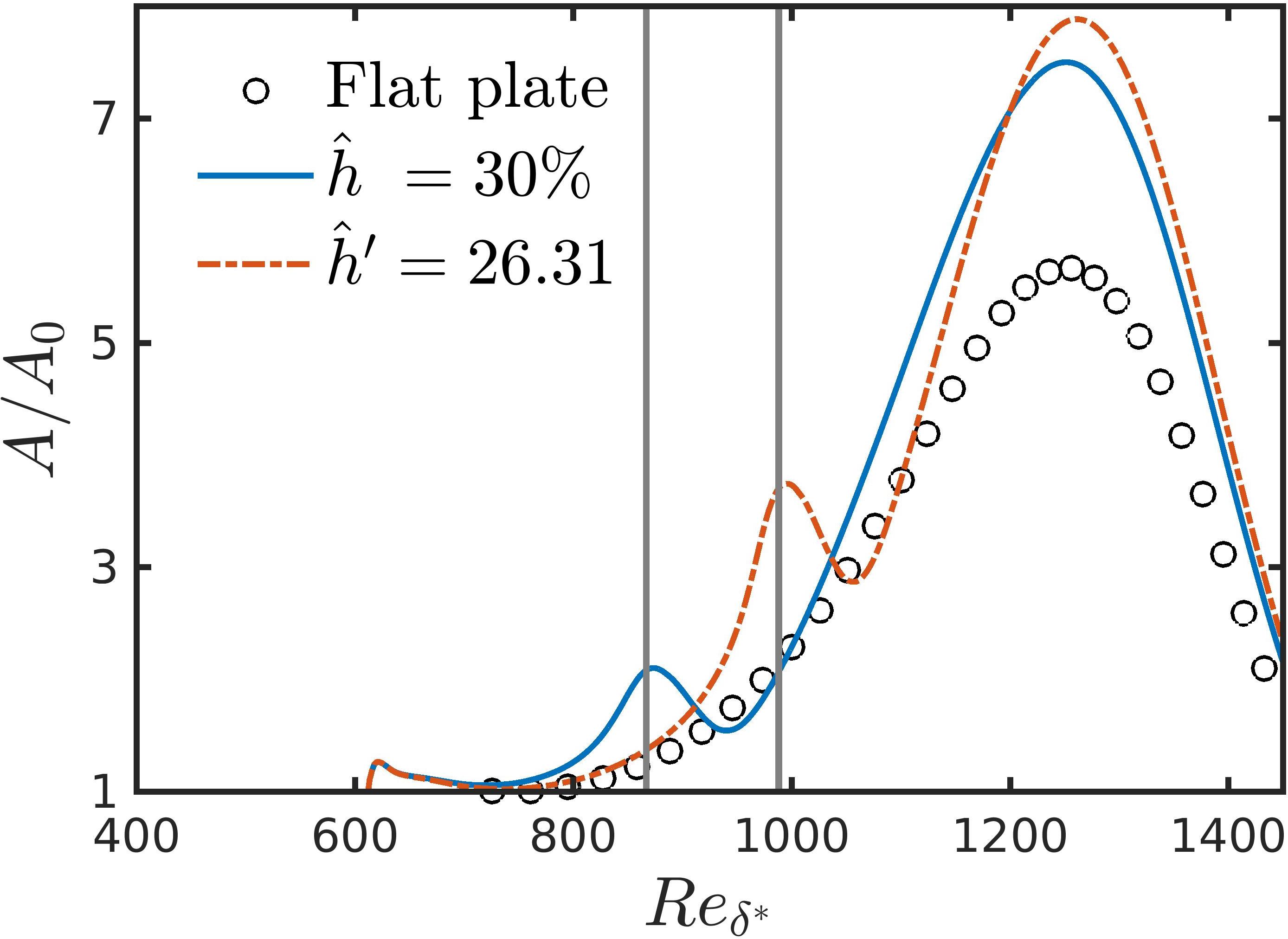}
\put(90,65){(d)}
\end{overpic}
}
 \caption{ Comparison for steps at different locations and the parameters for (a)-(d)  are given in Table \ref{tab:111215}. (A-D) one  single step at $\Rey_{\delta^*_{\rm c}}$; (A$^\prime$-D$^\prime$) one single step at $\Rey_{\delta^*_{\rm c^{\prime}}}$. The vertical grey line represents the location of the forward facing smooth step.}
\label{fig:111215comp}
\end{figure}

In the previous section,  the investigations of the effect of a smooth forward facing step on the stability of a boundary layer  focused on the high frequencies $\mathcal{F}\in\{140,150,160\}$. The unstable regions corresponding to these frequencies are relatively narrow compared with lower frequencies. For example for $\mathcal{F}=160$ the unstable region of the TS-mode on the flat plate ranges from $\Rey_{\delta^*}=580$ to $\Rey_{\delta^*}=830$, whereas for a perturbation frequency of $\mathcal{F}=100$ the unstable region ranges from $\Rey_{\delta^*}=700$ to $\Rey_{\delta^*}=1100$. In this section, we  consider the TS wave with a frequency $\mathcal{F}=100$ in a boundary layer over a single smooth step and the cases with the parameters given in Table \ref{tab:111215}. Physically, the size of steps at $\Rey_{\delta_*^{\rm c^\prime}}$ is kept the same as that of steps at $\Rey_{\delta_*^{\rm c}}$. We are now interested in assessing the effect of the position of the step on the excitation of the TS mode. In figure \ref{fig:111215comp}, the comparison of the TS envelopes in the boundary layers is given. Clearly, figure \ref{fig:111215comp}(a) indicates that small $\hat{h}$ cannot induce significant destabilisation. If $\hat{h}$ is increased as seen in figure \ref{fig:111215comp}(b)-(d), a destabilisation effect emerges and larger $\hat{h}$ give rise to larger global maximum amplitude of the TS wave. Meanwhile, figures \ref{fig:111215comp}(c)-(d) clearly indicate that moving the location of the smooth step downstream, far from the upper branch of the neutral stability curve, it induces more amplification than that of a more upstream location. 

\begin{figure}
  \centerline{
\begin{overpic}[width=0.48\textwidth]{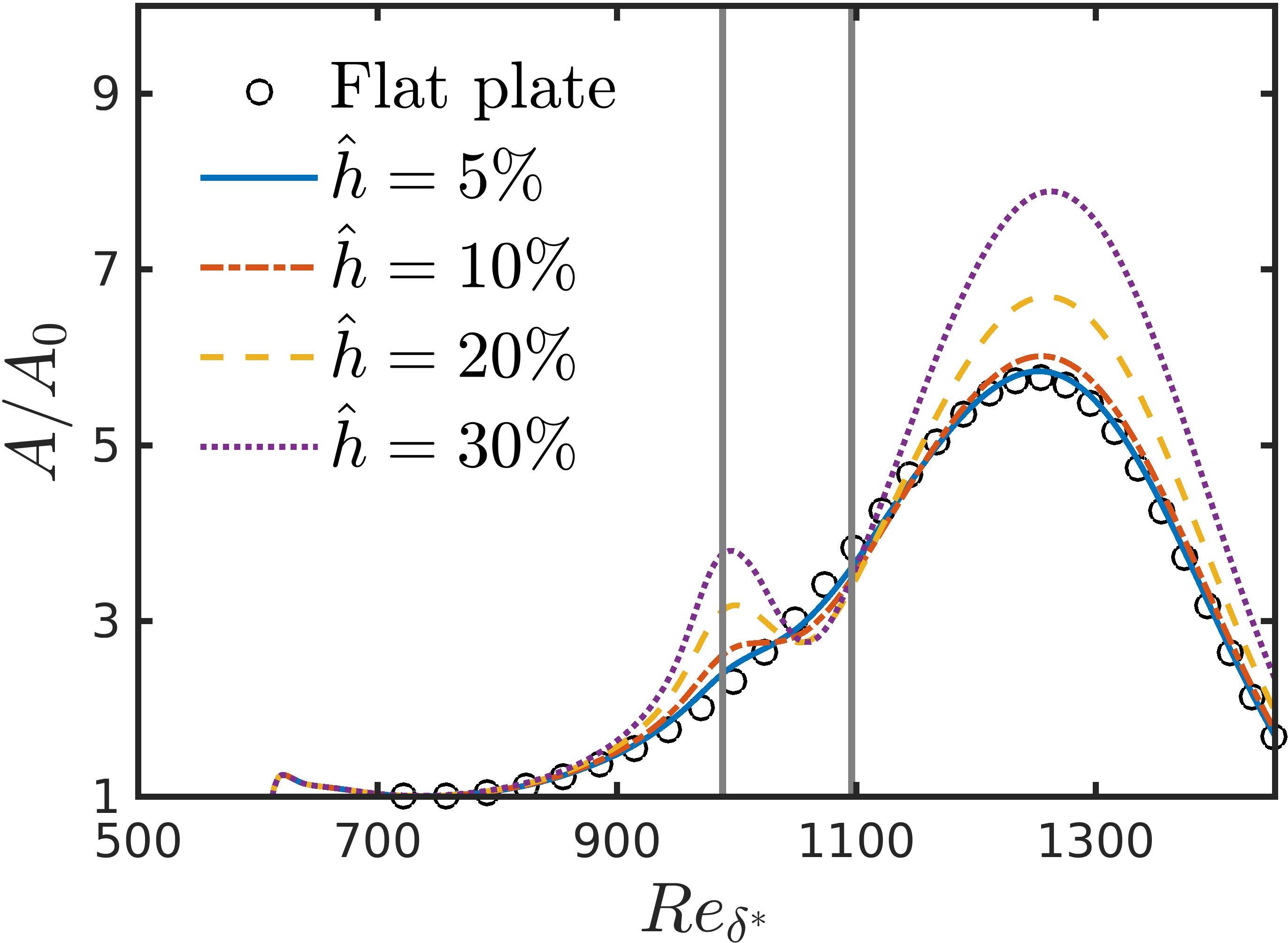}
\put(90,65){(a)}
\end{overpic}
\begin{overpic}[width=0.48\textwidth]{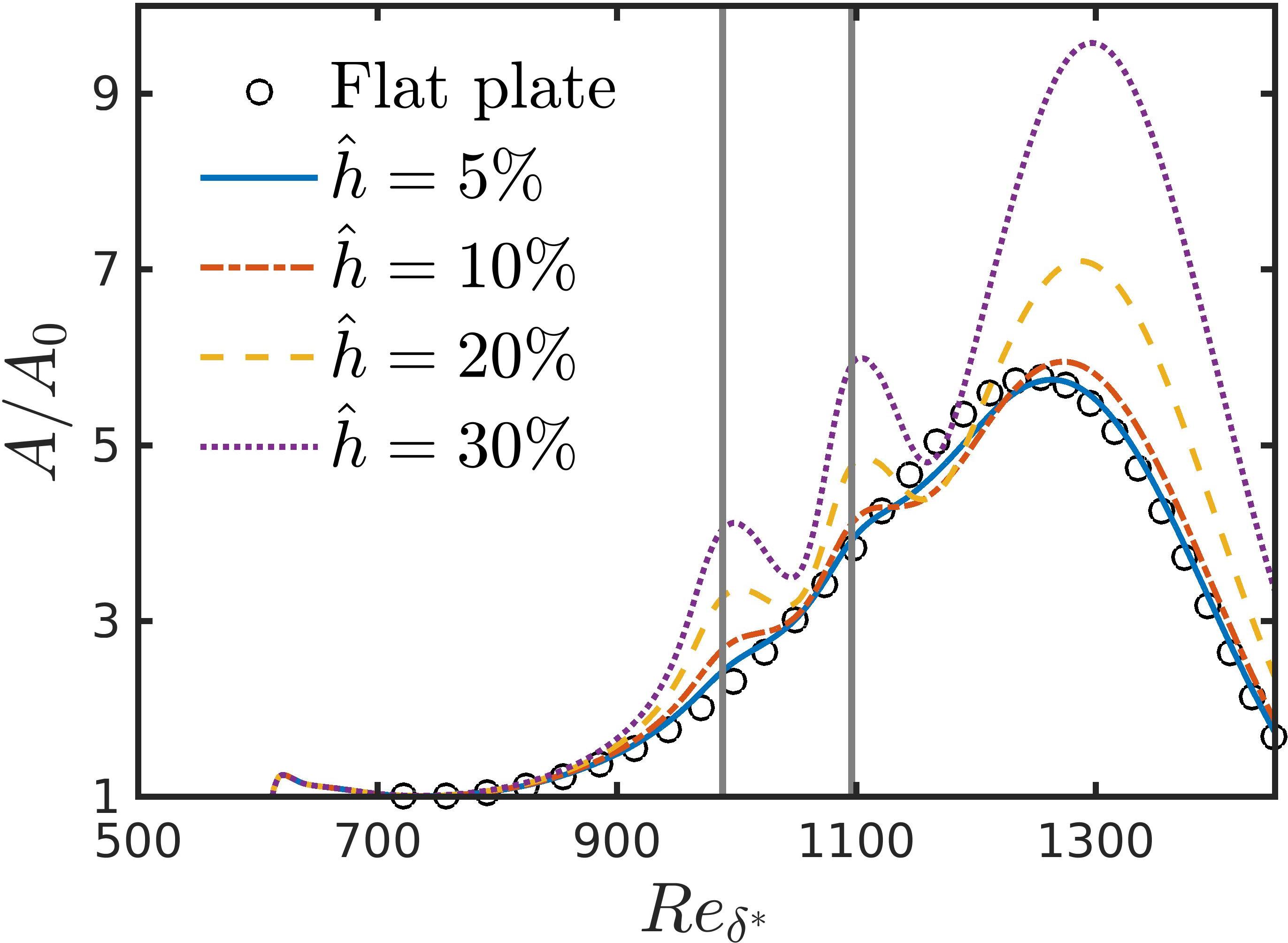}
\put(90,65){(b)}
\end{overpic}
}
 \caption{ The physical parameters corresponding to A, B,C and D are from Case A, B, C and D in Table \ref{tab:121106}. (a) one  single step at $\Rey_{\delta_*^{\rm c_1}}$; (b) two steps at $\Rey_{\delta_*^{\rm c_1}}$ and $\Rey_{\delta_*^{\rm c_2}}$. The vertical grey line represents the location of the forward facing smooth step.}
\label{fig:120216ts}
\end{figure}

This means that when the TS wave has a low frequency, smooth step could play a destabilisation role for the TS wave. From figure \ref{fig:111215comp}, we learn that growth rate of the TS waves in a flat boundary at $\Rey_{\delta^*_{\rm c^{\prime}}}$ is greater than that at $\Rey_{\delta^*_{\rm c}}$. We attribute the greater destabilisation effect induced by the step at $\Rey_{\delta^*_{\rm c^{\prime}}}$ to the greater growth rate at that position  in a flat plate boundary layer. This phenomenon is further considered in figure \ref{fig:120216ts} for $\hat{h}=20$ and $30$  for the same nondimensional scales $\hat{h}$ and $\hat{d}$ at $\Rey_{\delta^*_{\rm c}}=866$ and  $\Rey_{\delta^*_{\rm c^{\prime}}}=988$. The correlation between the position of the step and the TS mode amplification acts as a guideline for choosing the location of a smooth step. For large $\hat{h}$, the effect of a step located in a larger growth rate region gives rise to larger amplification of the TS wave. However, it is worthy to note that when height scale is reduced, there no longer exists a strong destabilisation effect from a single smooth step, regardless of its position within the unstable region.

To compare the effect of one and two steps additional simulations are carried out for two smooth steps located in regions of large growth rate of the TS wave. The amplification $A/A_0$ is plotted as as function of $\Rey_{\delta^*}$ in figure \ref{fig:120216ts}(a) for the single step and \ref{fig:120216ts}(b) for the two steps. We observe that for small $\hat{h}(<20\%)$,  the second steps  do not locally induce more destabilisation effect than the single step. Case A in figure \ref{fig:120216ts}(b) indicates that for small $\hat{h}(<20\%)$,  there exists a week stabilisation effect over the second step. However, a strong destabilisation effect is  observed from $\hat{h}(\ge20\%)$ corresponding to cases C and D in figure \ref{fig:120216ts}(a-b). Both for the single and the two step case, increasing the height of the step moves the location of the maximum amplitude of the TS mode to higher $\Rey_{\delta^*}$.

\begin{table}
  \begin{center}
\def~{\hphantom{0}}
  \begin{tabular}{cccccccccc}
 Case &$\Rey_{\delta^*_{i}}$   &$\Rey_{\delta^*_{\rm c}}$& $\mathcal{F}$ &$\hat{h}\%$ & $\hat{d}$ & $\gamma\times 10^4$  & $L_x/\delta_{99}$ & $L_y/\delta_{99}$    \\[3pt]
       A& 388 &866&100 &5.16 &4&0.66 & 287 & 30 \\
       B&---&---&---&10.32&---& 2.65 &---&---\\
       C&---&---&---&20.00&---&9.97&---&---\\
       D&---&---&---&30.00&---&22.44 &---&---\\ \hline\\[-2mm]
 &$\Rey_{\delta^*_{i}}$   &$\Rey_{\delta^*_{\rm c{^\prime}}}$& $\mathcal{F}$ &$\hat{h}^\prime(=\hat{h}\delta_{99}^{\rm c}/\delta_{99}^{\rm c{^\prime}})\%$ & $\hat{d}^\prime=\hat{d}\delta_{99}^{\rm c}/\delta_{99}^{\rm c{^\prime}}$ & $\gamma\times 10^4$  & $L_x/\delta_{99}$ & $L_y/\delta_{99}$    \\[3pt]
       A$^\prime$&388&988&100&4.52&3.51&0.66&287&30\\
       B$^\prime$&---&---&---&9.05&---&2.65 &---&---\\
       C$^\prime$&---&---&---&17.54&---&9.97 &---&---\\
       D$^\prime$&---&---&---&26.31&---&22.44 &---&---\\
  \end{tabular}
   \caption{Parameters for smooth steps: $\Rey_{\delta^*_i}$ is  the inlet Reynolds number, $\Rey_{\delta^*_{\rm c}}$  and $\Rey_{\delta^*_{\rm c^\prime}}$ indicate two different locations with respect to two same-size single steps. $\mathcal{F}$ denotes the non-dimensional perturbation frequency.  }
  \label{tab:111215}
  \end{center}
\end{table}

\begin{table}
  \begin{center}
\def~{\hphantom{0}}
  \begin{tabular}{ccccccccccc}
 Case &$\Rey_{\delta^*_{i}}$   &$\Rey_{\delta^*_{\rm c_1}}$&$\Rey_{\delta^*_{\rm c_2}}$& $\mathcal{F}$ &$\hat{h}\%$ & $\hat{d}$ & $\gamma\times 10^4$  & $L_x/\delta_{99}$ & $L_y/\delta_{99}$    \\[3pt]
      A & 388 &988&1096&100 &5.00 &4&0.58 & 287 & 30 \\
       B&---&---&---&---&10.00&---& 2.30 &---&---\\
       C&---&---&---&---&20.00&---&9.26&---&---\\
       D&---&---&---&---&30.00&---&20.84 &---&---
  \end{tabular}
   \caption{Parameters for smooth steps where $\Rey_{\delta^*_i}$, $\Rey_{\delta^*_{\rm c_1}}$  and $\Rey_{\delta^*_{\rm c_2}}$ are, respectively,  the inlet Reynolds number, the Reynolds number at the centre of the  first step and  the Reynolds number at the centre of the second step. $\mathcal{F}$ denotes the non-dimensional perturbation frequency. }
  \label{tab:121106}
  \end{center}
\end{table}

\subsection{Does a smooth step significantly amplify the low-frequency TS waves ?}
\begin{figure}
  \centerline{
\begin{overpic}[width=0.48\textwidth]{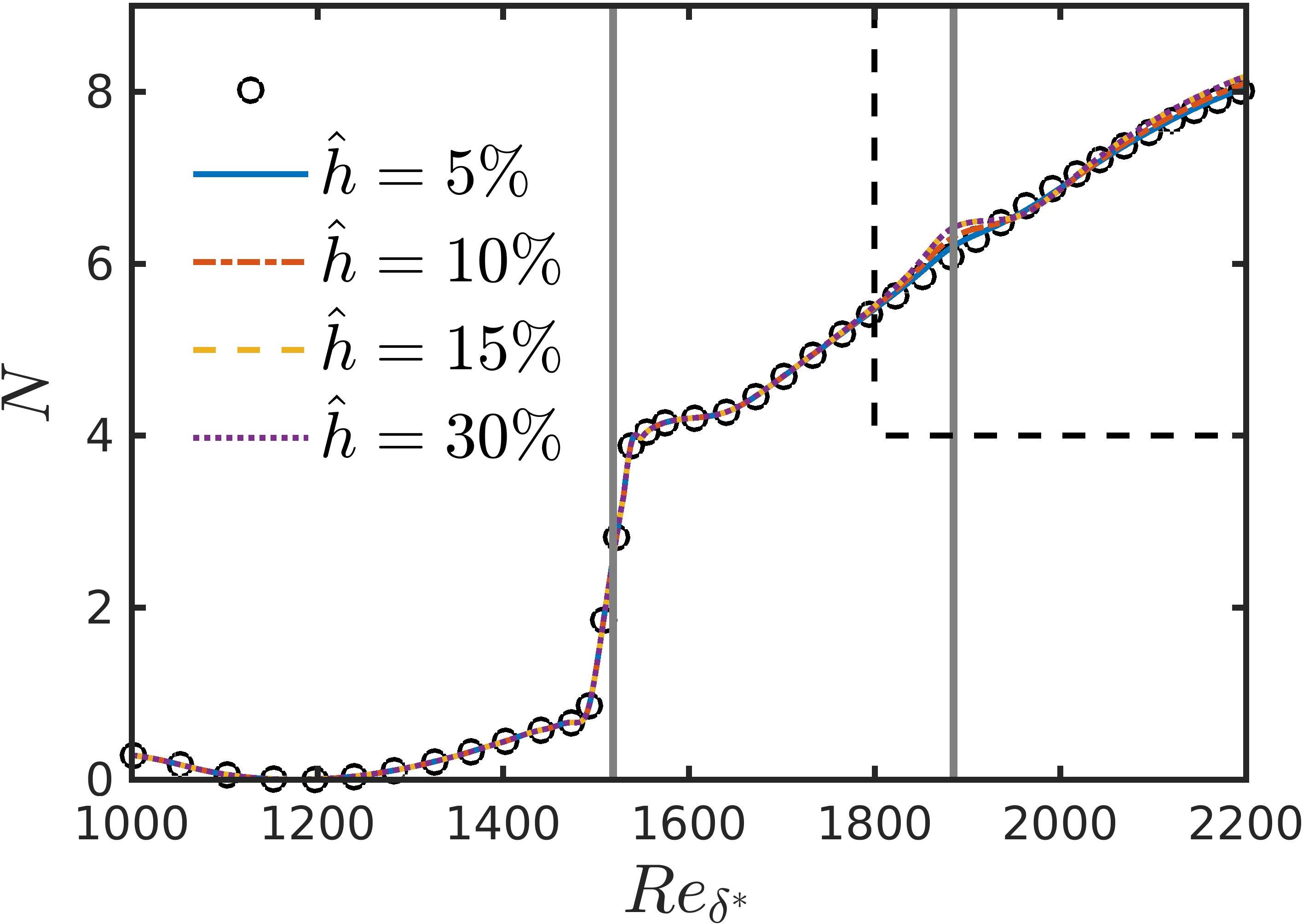}
\end{overpic}
\begin{overpic}[width=0.48\textwidth]{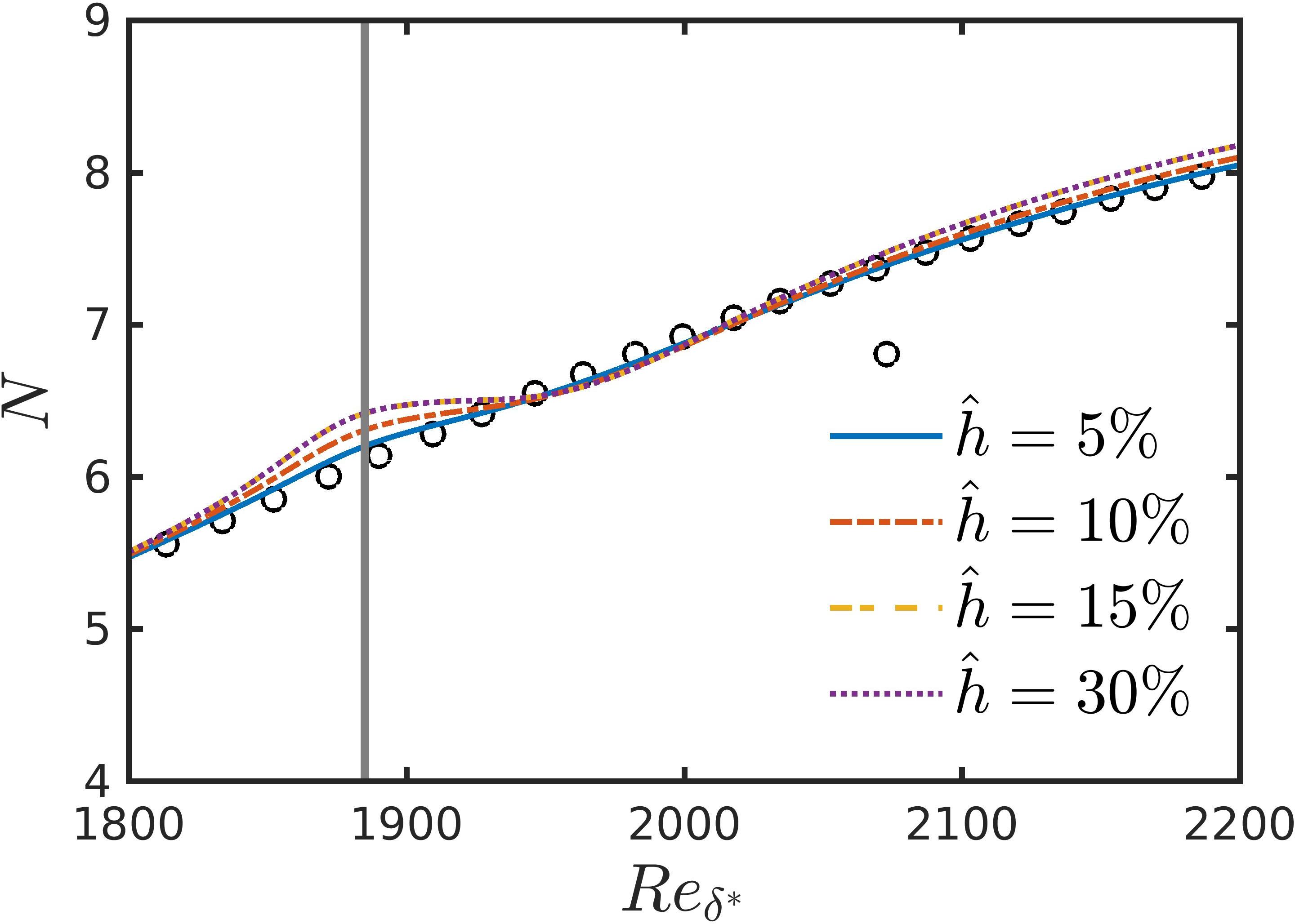}
\end{overpic}
}
 \caption{Effect of a smooth step for the TS wave with a low frequency. $N={\rm ln}\left(A/A_0\right)$. The roughness for amplifying the TS wave is located at the position of the first vertical line and the step is located at the position of the second vertical line.  (Left) Overview of the TS waves' envelopes; (Right) Local view of the TS waves' envelopes around the smooth steps. The parameters are from Table \ref{tab:281215}. The vertical grey line represents the location of the forward facing smooth step.}
\label{fig:281215f}
\end{figure}

The studies by \cite{worner2003,edelmann2015} considered the effect of low frequency TS modes for a long range from the leading edge of the flat plate with $\Rey_{\delta_*}$ (i.e. $\Rey_{\delta_*}=2200 \gg 1500$). In order to further assess  the effect of a forward facing smooth step on the amplification of the TS wave at high $\Rey_{\delta_*}$ regime where the $N$ factor is close to 8, we consider a TS wave with $\mathcal{F}=42$. The excited TS wave is further amplified by the recirculation bubble inside the indentation, located in $\Rey_{\delta^*_r}=1519$, far upstream of the transition criterion position of $N=8$. We then add a smooth forward facing step downstream of the indentation in the unsteady region of the TS mode. This configuration has its practical interest in roughness prompted transition. The indentation is defined by
\begin{equation}\label{roughness}
f_r=\left\{\begin{array}{rl}
-\hat{h}_r\cdot{\rm cos}\left(\pi{X^r}/{\hat{\lambda}}\right)^3, &X^r\in [-\hat{\lambda}/2, \hat{\lambda}/2],\\
0, & X^r\notin[-\hat{\lambda}/2, \hat{\lambda}/2],
\end{array}
\right.
\end{equation}
where $X^r=x-x_c^r/\delta_{99}^r$. The parameters of this computation are given in Table \ref{tab:281215}. Note that the width scale $\hat{\lambda}$ is comparable with the corresponding TS wavelength. A separation bubble is induced in the indentation region and when a base flow undergoes this distortion, the TS wave is strongly amplified. In figure \ref{fig:281215f}(Left), we observe that the TS waves are strongly amplified around the indentation and the N-factor in the computational domain reaches N=8. Downstream of the roughness, for  $\Rey_{\delta_*}>1800$, the smooth forward facing step only has weak, local, stabilising or destabilising effect for all three cases (see figure \ref{fig:281215f}(Right)).

For the case $\hat{h}=5$, the envelope of the TS wave is not significantly modified by the smooth step. Without talking about any possible stabilisation effect, destabilisation effect for such a low frequency is negligible. That is to say, for a TS wave with a low frequency $\mathcal{F}=42$, a smooth step does not significantly impact the N-factor. In contrast, a strong destabilisation of the TS wave by the sharp step ($\gamma=1$) for low frequency is reported by  \cite{edelmann2015} where the separation bubble induces a strong increase in N-factor from $N=4$ to $N=6$ because of the existence of separation bubbles in front of the step. This contrasts with the weak destabilisation influence of a smooth step on the TS wave in a boundary layer where negligible variation of the N-factor can be seen in figure \ref{fig:281215f} even for a large $\hat{h}>20\%$ step height.
\begin{table}
  \begin{center}
\def~{\hphantom{0}}
  \begin{tabular}{cccccccccccc}
 Case &$\Rey_{\delta^*_{i}}$   &$\Rey_{\delta^*_r}$&$\Rey_{\delta^*}$& $\mathcal{F}$ &$\hat{h}_r\%$ & $\hat{h}\%$ & $\hat{\lambda}$ &$\hat{d}$   & $L_x/\delta_{99}$ & $L_y/\delta_{99}$    \\[3pt]
      $\circ$& 596 &1519&1885&42 & 74.77 &0&5.5& 4 & 312 & 30 \\
       A&---&---&---&---&---& 5.00&--- &---&---&---\\
       B&---&---&---&---&---&10.00&---&---&---&---\\
       C&---&---&---&---&---&15.00&--- &---&---&---\\
       D&---&---&---&---&---&30.00&--- &---&---&---
  \end{tabular}
   \caption{Parameters for a wall with an indentation and a smooth step. $\Rey_{\delta^*_i}$, $\Rey_{\delta^*_r}$  and $\Rey_{\delta^*}$ are, respectively,  the inlet Reynolds number, the Reynolds number at the centre of the indentation and  the Reynolds number at the centre of the smooth step. $\mathcal{F}$ denotes the non-dimensional perturbation frequency. $\hat{h}_r$ and $\hat{\lambda}$ are used to define the indentation (\ref{roughness}), which  are normalised by the boundary layer thickness at the centre position of the roughness.}
  \label{tab:281215}
  \end{center}
\end{table}

\subsection{DNS investigation of the effect on K- and H-type transition of two forward facing smooth steps excited by a high frequency TS wave}
\begin{figure}
\vskip0.5cm
\centerline{
\begin{overpic}[width=0.7\textwidth]{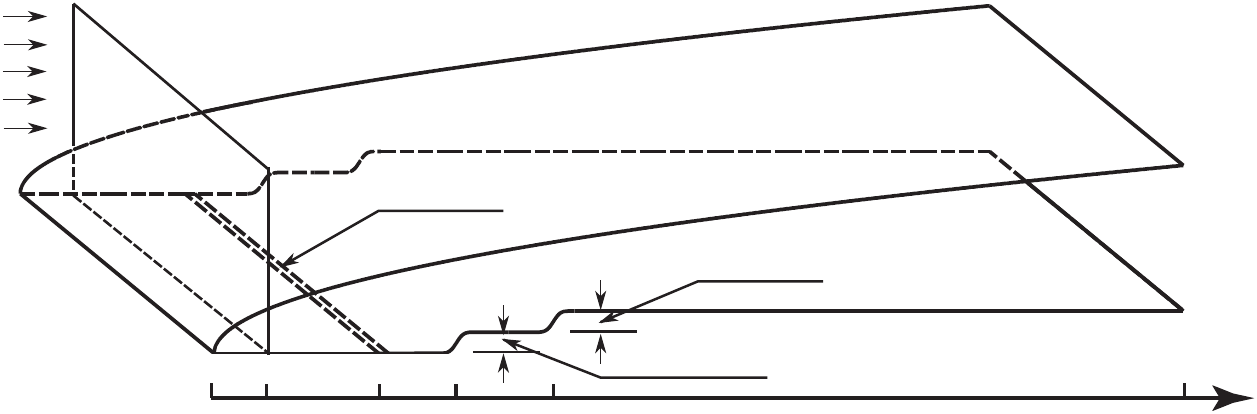}
\put(-10,28){$U_\infty$}
\put( 28,16.5){{ Disturbance strip}}
\put( 63,11){{$h$}}
\put( 57,3.5){{$h$}}
\put( 18,-2){{$\Rey_{\delta_*^i}$}}
\put( 33,-2){{$\Rey_{\delta_*^{c_1}}$}}
\put( 42,-2){{$\Rey_{\delta_*^{c_2}}$}}
\put( 90,-2){{$\Rey_{\delta_*}=1500$}}
\end{overpic}
}
\vskip0.5cm
 \caption{Overview of the computational setup with the Blasius boundary layer profile at the inflow, the disturbance strip and two smooth steps used for the DNS.
}
\label{fig:domain}
\end{figure}
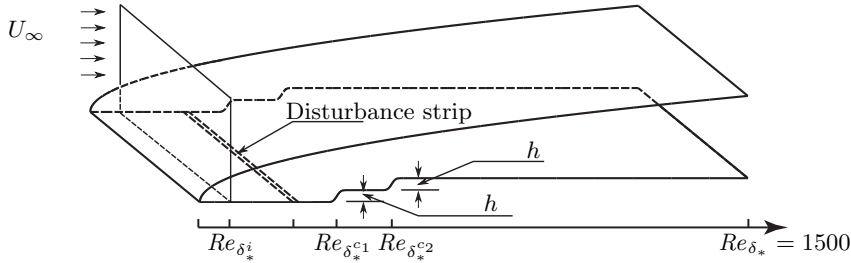

The TS mode is two dimensional, which is why linear stability analysis of the boundary layer can be conducted in two dimension. However secondary instabilities and transition to fully developed flow is a highly three dimensional. To this end further investigation of the influence of smooth steps on two transition scenarios is achieved with a hybrid Fourier-Spectral/{\it hp} discretisation is employed to solve 3D Incompressible Navier-Stokes equations. The spanwise direction was assumed to be periodic and discretised by 80 Fourier modes and the streamwise and wall normal plane was discretized using 5576 elements (quad and triangle) within which a polynomial expansion of degree 7  is imposed. K- and H-type transitions are simulated for the flow over a flat plate and with smooth steps. A Blasius profile is imposed at the inflow and, for both scenarios, a wall-normal velocity along the disturbance strip is prescribed by the blowing and suction boundary condition \citep{huai1997} ,
\begin{equation}
v(x,z,t)=A\cdot f(x)\cdot\sin(\omega_A t)+B\cdot f(x)\cdot g(z)\cdot\sin(\omega_B t),
\end{equation}
where $\omega_A$ and $\omega_B$ are the frequencies of the 2D TS wave and the oblique waves, respectively. 
$A$ and $B$ are the disturbance amplitudes of the fundamental and the oblique waves. The function $f(x)$ is defined by \citep{fasel1990}
\begin{equation}
f(x)=15.1875\xi^5-35.4375\xi^4+20.25\xi^3,
\end{equation}
with the parameter $\xi$
\begin{equation}
\xi=
\left\{\begin{array}{ccc}
\displaystyle\frac{x-x_1}{x_m-x_1} &\mbox{for}& x_1\le x\le x_m\\[1em]
\displaystyle\frac{x_2-x}{x_2-x_m} &\mbox{for}& x_m\le x\le x_2
\end{array}
\right.,
\end{equation}
where $x_m=(x_1+x_2)/2$ and $g(z)=\cos(2\pi z/\lambda_z)$ with the spanwise wavelength  $\lambda_z$. At $x_1$ and $x_2$, $\Rey_{\delta^*_{x_1}}$ and  $\Rey_{\delta^*_{x_2}}$ are equal to $591.37$ and $608.51$, respectively. For K-type transition, the oblique waves have the same frequency as the two dimensional wave ($\omega_A=\omega_B$). Also, for H-type transition, the oblique waves are sub-harmonic ($\omega_B=\omega_A/2$). All parameters used in the investigation are given in Table \ref{tab:3}. Note that the frequency  used for the perturbation  are consistent with the parameters used for linear analysis in Table \ref{tab:071114}. The schematic of the computational domain is illustrated in figure \ref{fig:domain}. It is worth mentioning that, in order to guarantee that nonlinear calculations are converged, all unsteady simulations terminate when the nondimensional time scale $T$ is equal to 20. 

\begin{table}
  \begin{center}
\def~{\hphantom{0}}
  \begin{tabular}{ccccccccccccc}
  $\Rey_{\delta^*_{i}}$  &$\Rey_{\delta^*_{\rm c_1}}$&$\Rey_{\delta^*_{\rm c_2}}$& $\mathcal{F}_A$& $\mathcal{F}_B$ & $A/U_{\infty}$& $B/U_{\infty}$& $\hat{h}\%$  & $L_x/\delta_{99}$ & $L_y/\delta_{99}$ & $L_z/\delta_{99}$ &  $\lambda_z/\delta_{99}$ & $T$\\[3pt]
       320 &680&786&150 &150 &0.5\%&0.03\%&0 & 340 & 30 & 8 &4 & 20 \\
        ---&---&---&---&75 &---&---&---  &---&---&---&---&---\\
      ---&---&---&---&150 &---&---&5.48   &---&---&---&---&---\\ 
       ---&---&---&---&75 &---&---&5.48   &---&---&---&---&---\\ 
       ---&---&---&---&150 &---&---&12.79   &---&---&---&---&---\\ 
       ---&---&---&---&75 &---&---&12.79  &---&---&---&---&---\\ 
  \end{tabular}
  \caption{Parameters used for the DNS simulations. $\mathcal{F}_A$ and $\mathcal{F}_B$ denote non-dimensional perturbation frequencies of the disturbance strip. $A/U_{\infty}$ and $B/U_{\infty}$ are the relative amplitudes of the disturbance amplitudes of the fundamental and oblique waves, respectively. The spanwise $L_z$ extent of the domain is expressed as function of the boundary layer thickness $\delta_{99}$. $T$ is the finial nondimensional  time length scale which we simulate, which is defined by $T=tU_\infty/L$. }
  \label{tab:3}
  \end{center}
\end{table}

\subsubsection{Effect of smooth forward facing steps on K- and H-type transitions}

\begin{figure}
  \centerline{
\begin{overpic}[width=1\textwidth]{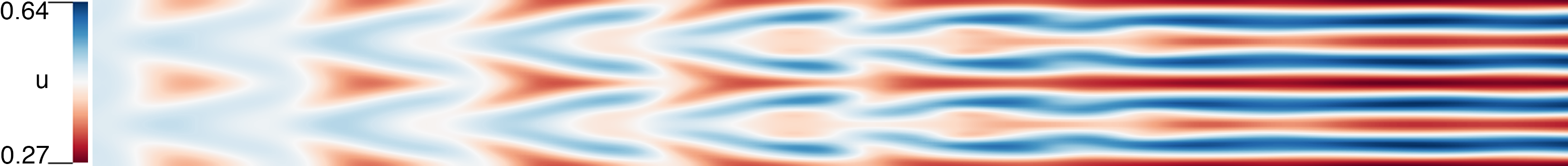}
\put(5.5,8){(a)}
\end{overpic}
}\vskip1em
  \centerline{
\begin{overpic}[width=1\textwidth]{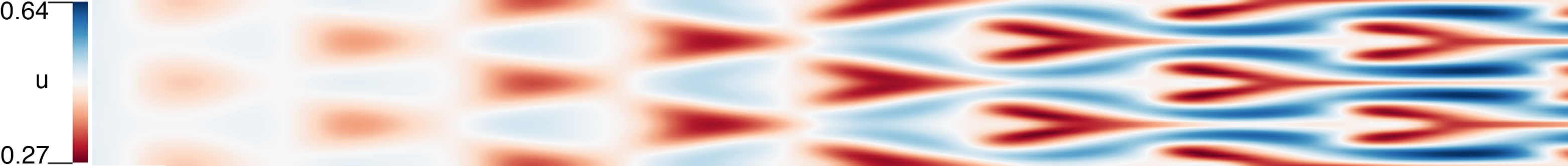}
\put(5.5,8){(b)}
\end{overpic}
}
 \caption{Instantaneous contours of stream-wise velocity in $xz$-plane at height $y=0.6\delta_{99}^{i}$ in $\Rey_{\delta^*}\in[963 , 1111]$ for the K- (a) and H-type (b) transition scenarios for a flat plate $\hat{h}=0$.}
\label{fig:9}
\end{figure}

\begin{figure}
  \centerline{
\begin{overpic}[width=0.45\textwidth]{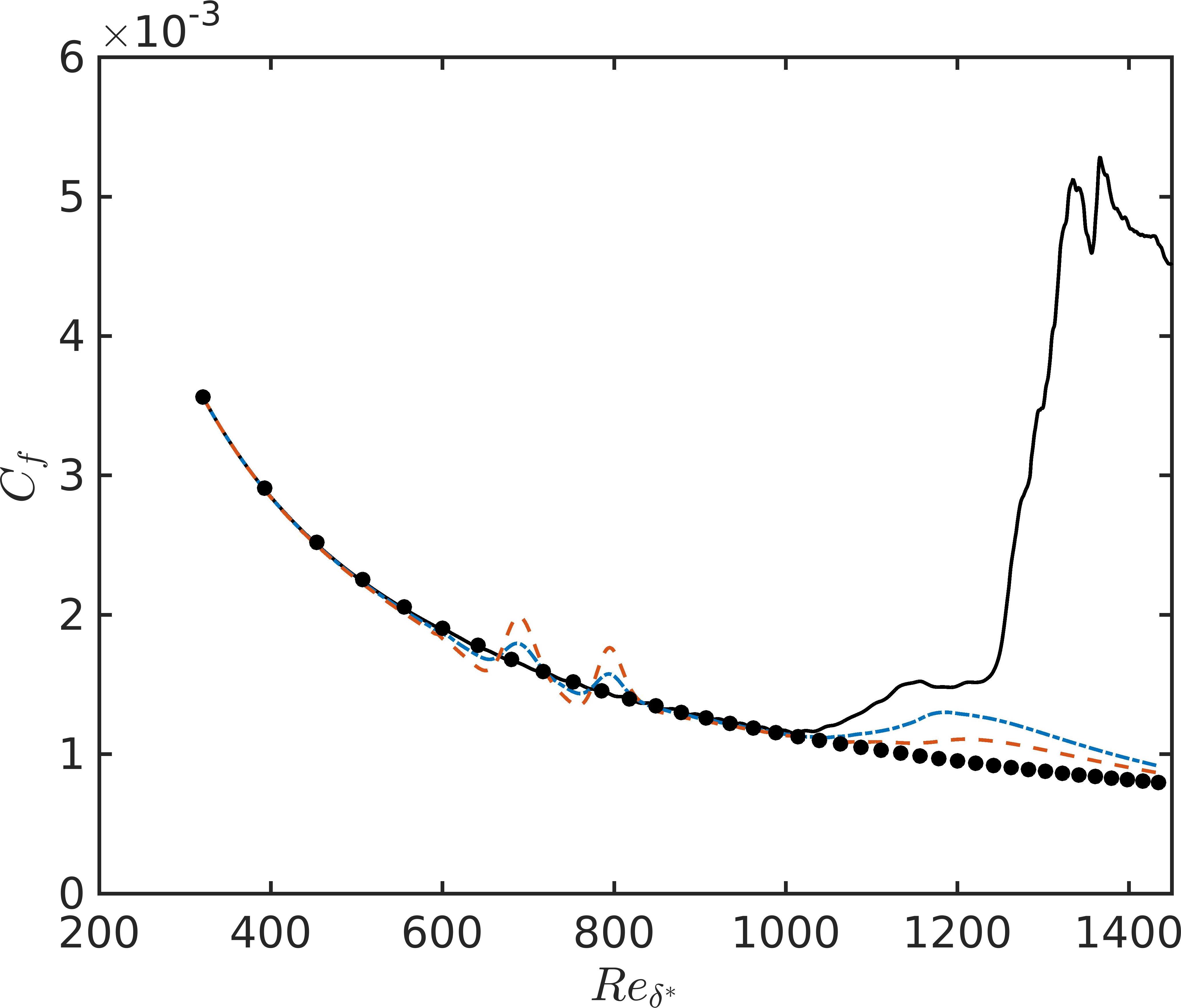}
\put(11,75){(a)}
\end{overpic}
\begin{overpic}[width=0.45\textwidth]{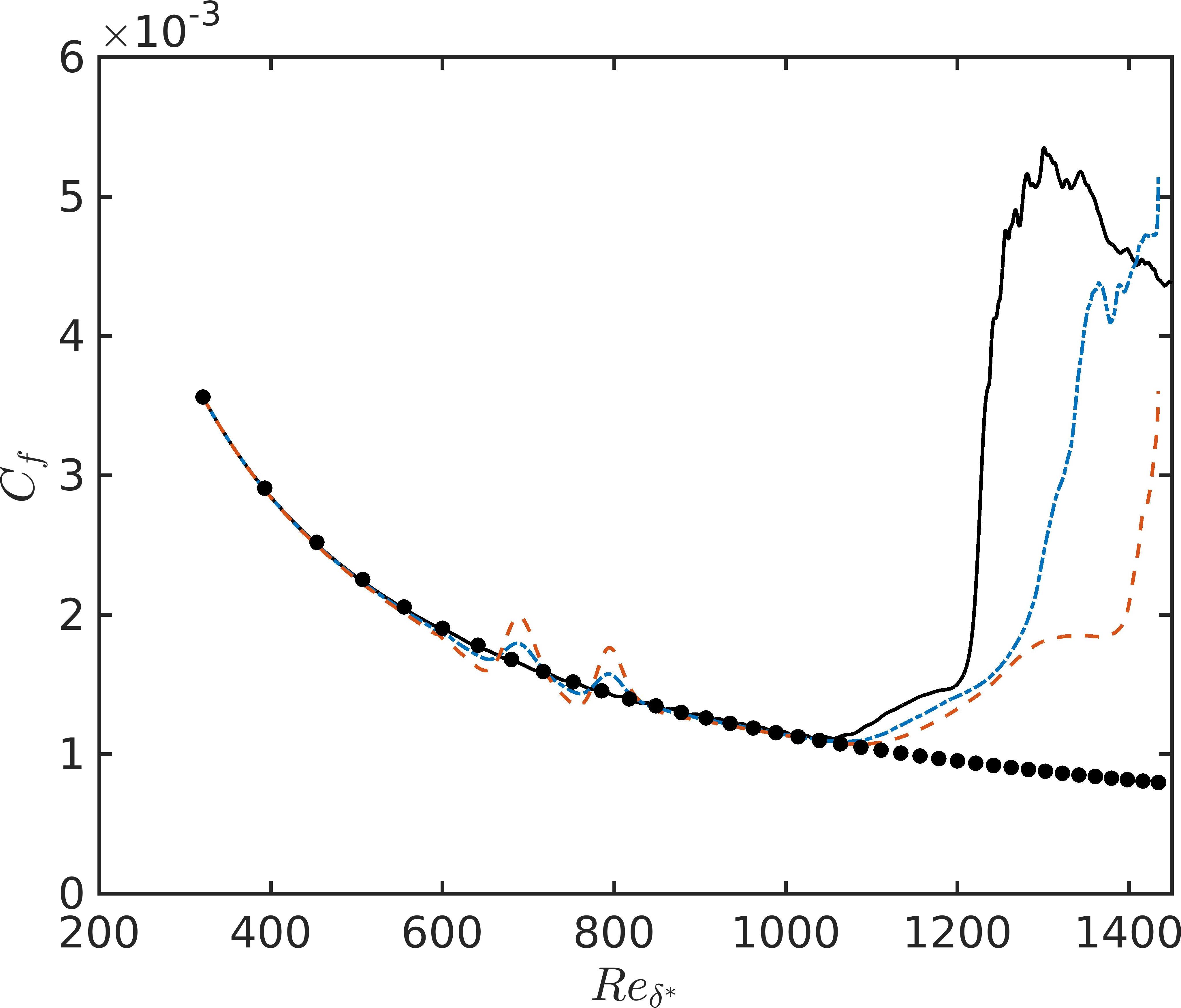}
\put(11,75){(b)}
\end{overpic}
}
 \caption{Comparison of time- and spanwise-averaged skin friction versus streamwise position $\Rey_x$ for K- (a) and H-type (b) transition scenarios for a flat plate (\textemdash) and two of height $\hat{h}=5.48\%$ ($-\cdot-$) and $\hat{h}=12.79\%$ ($- -$). The skin-friction profile of the Blasius boundary layer ($\bullet$) is given for reference.}
\label{fig:8}
\end{figure}

The validation of the DNS results for both K- and H-type transitions is corroborated by recovering the aligned arrangement of the $\Lambda$ vortices  for the K-type transition (see figure \ref{fig:9} (a)) and staggered arrangement for the H-type transition (see figure \ref{fig:9} (b)) as experimentally observed by \cite{berlin1999} for the flat plate boundary layer. Figure \ref{fig:8} shows the evolution of the skin-friction coefficient versus $\Rey_{\delta^*}$ for two different normalized height scales. We observe that, for a flat plate boundary layer, the skin friction coefficient diverges from that of the Blasius boundary layer at the location where $\Lambda$ vortices appear, as illustrated in figure \ref{fig:9}.  The streamwise evolution of the skin-friction (see figure \ref{fig:8}) shows the K-type  transition is fully inhibited by the two smooth steps whereas  the H-type transition is delayed. Additionally, increasing the height $\hat{h}$ ($<20\%$) further reduces the skin friction coefficient $C_f$ in both scenarios. The observation of   these phenomena supports the result of linear analysis. 

\begin{figure}
  \centerline{
\begin{overpic}[width=0.5\textwidth]{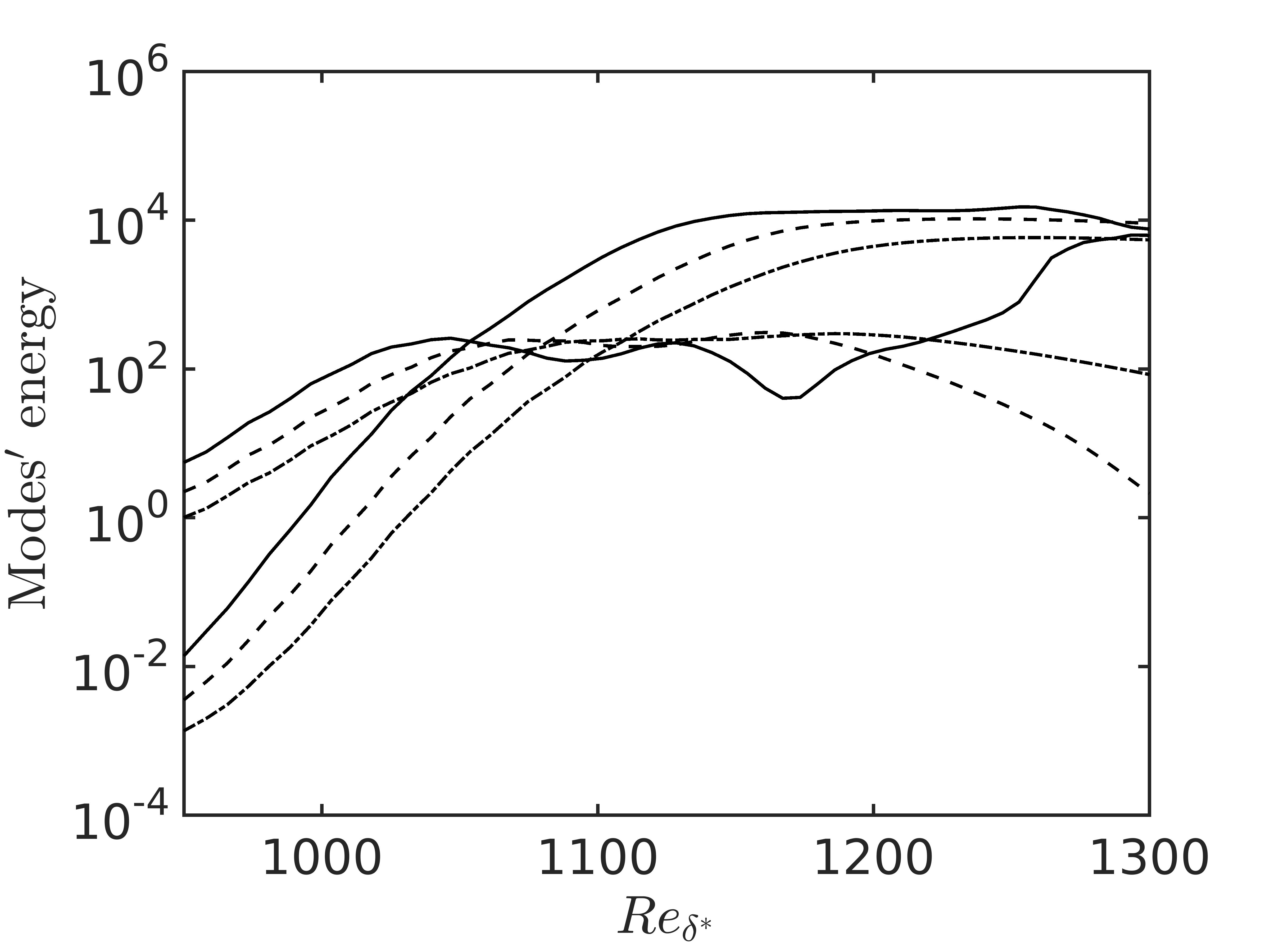}
\put(20,65){(a)}
\put(55,65){(0,2) modes}
\put(55,35){(0,1) modes}
\end{overpic}
\begin{overpic}[width=0.5\textwidth]{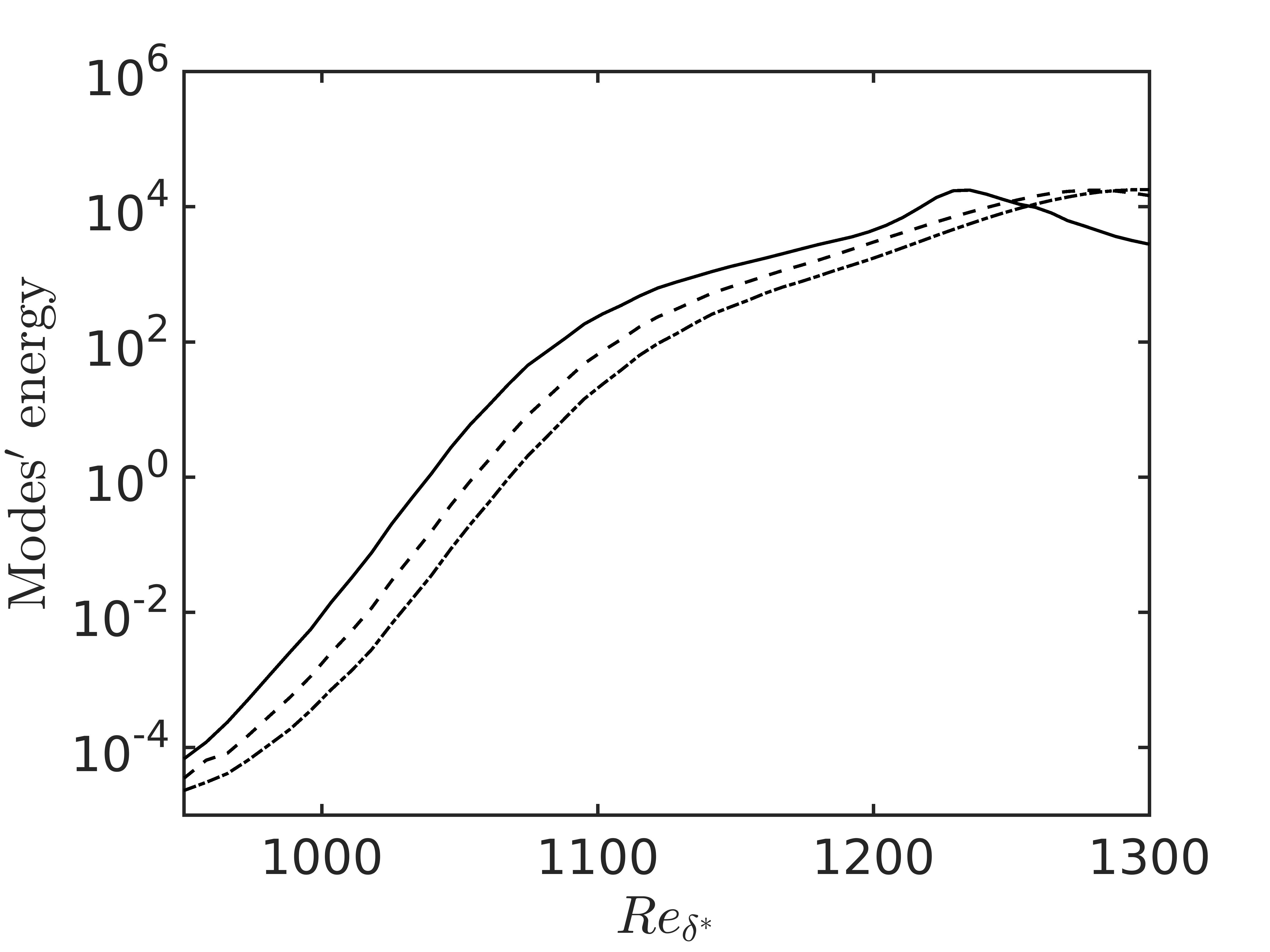}
\put(20,65){(b)}
\put(55,45){(0,2) modes}
\end{overpic}
}
 \caption{Comparison of the energy in modes (0,1) and (0,2) versus  streamwise position $\Rey_{\delta^*}$  over a flat plate ($-$) and two smooth steps with height $\hat{h}=5.48\%$ ($--$) and $\hat{h}=12.79\%$ ($-\cdot -$)  for K- (a) and H-type (b)  transition scenario.}
\label{fig:10b}
\end{figure}
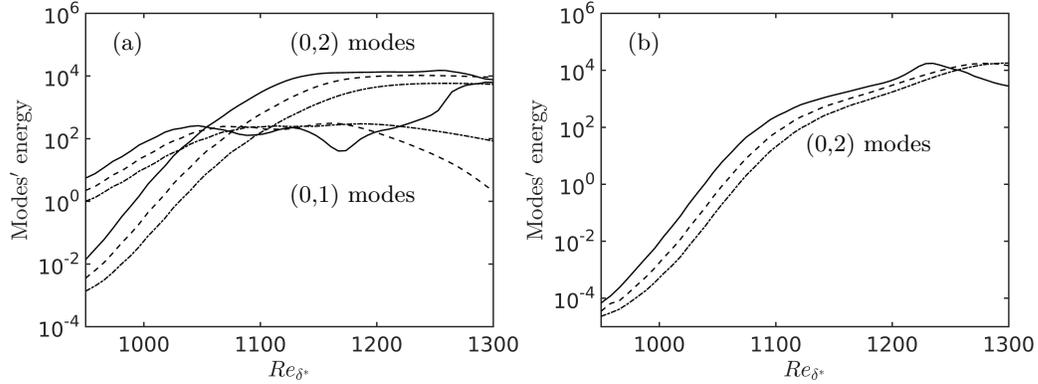

To gain further insight into the different impact on  two  transition scenarios of the two forward facing smooth steps we consider the energy growth of the main modes. We label these modes using the notation $(\omega,\beta)$ \citep{berlin1999}, where $\omega$ and $\beta$ are respectively, the frequency and spanwise wavenumber each normalized by the corresponding fundamental frequency/wavenumber. It has been observed that the K-type  transition scenario has the main initial energy in the $(1,0)$ mode. The $(1,\pm 1)$ mode also generates  the $(0,\pm 2)$ mode  with a small amplitude through non-linear interaction \citep{berlin1999}. At the late stage, the $(0,\pm  2)$ mode can grow to an amplitude comparable to that of the $(0,\pm 1)$ mode. \cite{laurien1989} and \cite{berlin1999} have shown that the initial conditions for the H-type  transition has the main energy in the $(1,0)$ mode with a small amount in the oblique subharmonic $(1/2,\pm1)$ mode. The important mode is the vortex-streak $(0,\pm2)$ mode, which is nonlinearly generated by the subharmonic mode and vital in the transition process.
As illustrated in figure \ref{fig:10b}, the $(0,2)$ mode  plays a significant role in the late stages of  transition for both transition scenarios with the two smooth steps.  For the K-type scenario, in the transition regime, the energy of mode $(0,2 )$ grows and exceeds the energy of mode $(0,1)$. For the flat plate, the energy of mode (0,1) finally grows again until turbulence occurs.  The energy of mode $(0,2)$ with the two smooth steps  is less than that of mode $(0,2)$ for the flat-plate and increasing normalised smooth step height $\hat{h}$ yields stronger reduction of the energy. A similar reduction in energy is also observed for mode (0,1). Furthermore,  the energy of mode (0,2) exceeds that of mode (0,1) in $\Rey_{\delta^*}=1080$ for $\hat{h}=5.48\%$  to $\Rey_{\delta^*}=1100$ for $\hat{h}=12.79\%$ (figure \ref{fig:10b} (a)) and from this points onwards the energy of mode (0,1) decays. Based on the results presented in figure  \ref{fig:10b} (a), we  observe that the spanwise modulation induced by mode (0,1) with energy decaying on the smooth steps leads to the stabilisation of the boundary layer.

\section{Conclusion}

In this paper we introduce the smoothness  of a step $\gamma$ and show that it affects on destabilising effect of step and on an otherwise flat plate boundary layer. Linear stability analyses are conducted with high $140< \mathcal{F}< 160$, medium $\mathcal{F}=100$ and low $\mathcal{F}=42$ frequency forcing with respect to the neutral stability curve of the flat plate boundary layer. One and two step configurations with different heights and smoothness are considered. For the low frequency forcing case an indentation is added upstream of the forward facing smooth step and the effect on the stability of this highly perturbed flow is reported. Finally direct numerical simulation of K- and H-type transition scenarios are undertaken for a high-frequency forcing case to confirm the results from the linear analysis.

The net effect of a smooth forward facing step on the stability of the TS-mode depends on height, width and flatness. From an application point of view, small height smooth steps are safe for TS waves of frequency ranging from $\mathcal{F} \in [42, 160 ]$.  For $\hat{h}<20\%$ both the single and two forward facing smooth step configurations lead to a stabilising effect of the TS-mode of up to 5\% and 10\% respectively. For $\hat{h}>20\%$ the smooth step $\gamma <1$ has a weaker destabilising effect than previous reports show for a  sharp step $\gamma =1$. Further investigation indicates that for the TS wave with a low frequency, small height smooth steps can be safe and don't significantly amplify the TS wave. This result contrasts with that of a sharp forward facing step and we attribute the  large amplification of the TS wave by a sharp step to separation bubbles. This effect is similar to the destabilisation effect induced by the separation bubbles in an indentation or behind a hump or a bump \citep{gao2011,park2013a,xu2016a}. From a global stability point of view, a smooth step, even with a large height, cannot lead to a destabilisation  in front of or over the step because of the absence of recirculation bubbles meaning no global instability can be introduced. The results obtained by DNS, for a high-frequency forcing $\mathcal{F}=150$, support the conclusion that smooth steps can have non-negligible and positive impact on the stability of the boundary layer. For $\hat{h}=5.48$ and $\hat{h}=12.79$ the transition to fully developed turbulent state is even delayed for the H-type transition scenario and suppressed for the K-type scenario.

More generally these results suggest  a smooth forward facing step can have a significant influence on the stability of the boundary layer.

\section*{Acknowledgements}
This research was performed in the Laminar Flow Control Centre (LFC-UK) at Imperial College London. 
The Centre is supported by EPSRC, Airbus Global Innovations under grant EP/I037946/1. The authors would also like to acknowledge support from the United Kingdom Turbulence Consortium
(UKTC) under grant EP/L000261/1 as well as from the Engineering and Physical
Sciences Research Council (EPSRC) for access to ARCHER UK National
Supercomputing Service (http://www.archer.ac.uk). SJS additionally acknowledges Royal Academy of Engineering support under their research chair scheme. The authors thank Olga Evstafyeva  for a number of helpful suggestions for the manuscript.


\bibliographystyle{jfm}

\bibliography{jfm-huixu}

\end{document}